\documentclass[leqno,fleqn,11pt]{article}%
\usepackage{MathSpad,makeidx,ifthen,latexsym}
\usepackage{a4wide,amssymb,euler,beton,graphicx,hyperref}
\def\mathit#1{#1}

\newcommand{\rightsymdivision}{{\setminus}{\hspace{-1.5mm}}{\setminus}}
\newcommand{\leftsymdivision}{/{\hspace{-1.5mm}}/}
\def\tuberight{{\unitlength=1ex\begin{picture}(1.4,1)(0,0)\put(0,0.7){\line(1,0){1}}\put(0,0.3){\line(1,0){1}}\put(1,0.7){\line(0,1){0.2}}\put(1,0.3){\line(0,-1){0.2}}\end{picture}}}
\newcommand{\bigsqcap}{{\sqcap}}
\DeclareSymbolFont{cmsygroup}{OMS}{cmsy}{m}{n}
\DeclareSymbolFont{cmrgroup}{OT1}{cmr}{m}{n}
\DeclareMathSymbol{\sim}{\mathrel}{cmsygroup}{"18}
\DeclareMathSymbol{=}{\mathrel}{cmrgroup}{"3D}
\DeclareMathSymbol{=}{\mathrel}{cmrgroup}{"3D}
\DeclareMathSymbol{[}{\mathopen}{cmrgroup}{"5B}
\DeclareMathSymbol{]}{\mathclose}{cmrgroup}{"5D}
\DeclareMathSymbol{(}{\mathopen}{cmrgroup}{"28}
\DeclareMathSymbol{)}{\mathclose}{cmrgroup}{"29}

\begin{document}
\title{Diagonals and Block-Ordered Relations}
\author{Roland Backhouse\footnote{University of Nottingham, UK;
corresponding author} 
 and Ed Voermans\footnote{Independent Researcher, The Netherlands}}
\date{\today}
\maketitle
\begin{abstract}More than 70  years ago, Jaques Riguet suggested the existence of   an ``analogie frappante''
(striking analogy) between so-called ``relations de Ferrers'' and a class of difunctional relations,
members of which we call ``diagonals''.   Inspired by his suggestion, we formulate an  ``analogie
frappante'' linking  the notion of a block-ordered relation and the notion of the diagonal of a relation.  
We formulate several novel properties of the core/index  of a diagonal, and use these properties to
rephrase our ``analogie frappante''.  Loosely speaking, we show that a block-ordered relation  is a
provisional ordering   up to isomorphism and reduction to its  core.  (Our theorems make this informal
statement precise.) Unlike Riguet  (and others who follow his example),  we avoid almost entirely the
use of nested complements  to express  and reason about properties of these notions:   we  use factors (aka
residuals) instead.  The only (and inevitable) exception to this is to show that our definition of a  ``staircase''
relation is equivalent to Riguet's definition of a ``relation de Ferrers''.    Our ``analogie frappante'' also
makes it obvious that a  ``staircase'' relation is not necessarily block-ordered, in spite of the mental
picture of such a relation presented by Riguet.\end{abstract}

%\newpage
%\tableofcontents
%\newpage

\section{Introduction}\label{Indices:Introduction}

More than seventy years ago, in a series of publications \cite{Riguet48,Riguet50,Riguet51},  Jacques Riguet introduced
the notions of a ``relation difonctionelle'' and  ``relations de Ferrers''.   In
\cite{Riguet51} he remarked on an ``analogie frappante'' between these two notions via what he referred to
as the ``diff\a'{e}rence''  of a given relation.   Riguet \cite{Riguet51} states the following theorem:

\begin{quote}Pour que $R$ soit une relation de Ferrers, il
faut et il suffit que $R$ soit r\a'{e}union de rectangles dont les projections de m\^{e}me nom sont totalement
ordonn\a'{e}es par inclusion et tels que si la premi\a`{e}re projection de l'un des rectangles est contenue dans 
la premi\a`{e}re projection d'un autre rectangle, la seconde projection du second est contenue dans 
la seconde projection du premier.
\end{quote}

\begin{figure}[h]
\centering \includegraphics{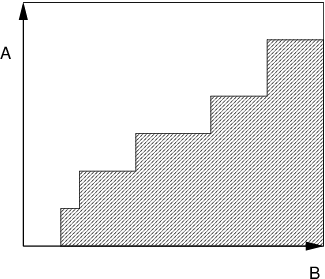}
 
\caption{Mental Picture of a  Staircase Relation}\label{fig:staircase}
\end{figure}

For those unable to read French, the theorem states a necessary and sufficient condition for a relation to
be ``de Ferrers'' in terms of totally ordered rectangles  (``rectangles $\ldots$ totalement
ordonn\a'{e}es'').   The theorem clearly begs the question what is the definition of a ``relation de Ferrers''.  We
postpone answering this question until section \ref{sec:Staircase Relations}.  The reason for doing so is that Riguet gives both  a formal
definition and a mental picture ---a picture like the one  in fig.\ \ref{fig:staircase} of what we call a ``staircase
relation''--- but it is far from obvious how Riguet's definition and the mental picture are related.

Riguet does not give a proof of the theorem and his  ``analogie frappante'' between  ``relations de Ferrers'' and
difunctional relations is unclear.  The proof of Riguet's theorem  is quite 
straightforward: see \cite{RCB2020}.   However, clarifying the ``analogie frappante'' is  more difficult.  

The work presented here initially began as an attempt to properly understand Riguet's work and to
rectify errors in extant literature.  We introduce, in section \ref{difunctional.Diagonal},  the ``diagonal'' of a
relation and, in section \ref{Block-Ordered Relations},  the notion  of a ``block-ordered relation''.  The
``diagonal'' of a relation is what Riguet referred to as the ``diff\a'{e}rence''; it is a difunctional relation.  We
formulate an ``analogie frappante'' (specifically, theorem \ref{BD.diagdom}) linking block-ordered relations
and diagonals.   We also formalise  the notion of a ``staircase'' relation:    as shown in \cite{RCB2020},  our
definition  of a ``staircase'' relation and Riguet's  definition of a  ``relation de Ferrers'' are
equivalent\footnote{The name ``staircase'' is more informative than ``relation de Ferrers'' and our
definition uses factors rather than complementation  (from which Riguet's terminology ``diff\a'{e}rence'' is
derived),   which is why we prefer to deviate from Riguet's presentation}.    Ostensibly a ``staircase''
relation is ``block-ordered'' where the ``blocks'' are totally ordered.  But, as we observed in our
initial investigations, this is not the case: as we show in section \ref{sec:Staircase Relations}, the less-than
relation on real numbers is a ``staircase'' relation but is not ``block-ordered''.   

In addition to our  ``analogie frappante'', a major contribution of our work is the application of the notions
of a core/index of a relation.  These notions and their properties are briefly summarised in section 
\ref{Indices General}.  Informally,  a core\footnote{We say ``a'' core because a relation typically has many
cores; all cores of a relation are, however, isomorphic.} of a relation captures its essential properties.  For example, a core
of a difunctional relation is an isomorphism.  An index of a relation is a core of the relation that has the
same type as the relation.   
A fuller account of their properties is given in \cite{VB2022,VB2023a}.  In this paper, we
state and prove a number of properties of the core/index of the diagonal of a relation.  These properties
are relevant to our analogie frappante: theorem \ref{BD.diagdom} gives a method of testing whether or not a
relation is block-ordered in terms of the diagonal of the relation, and 
theorem \ref{diagonal.core.domain} reformulates the test in terms of the core of the relation.

During the course of our investigation, we became aware of very similar work by Winter \cite{Winter2004}. 
Winter's notion of ``relations of order shape'' is identical to our notion of ``block-ordered relation''. 
Winter also introduces the notion of the ``diagonal'' of a relation --- but does not give the notion a name
and does not attribute the notion to Riguet \cite{Riguet51}.  (He does cite \cite{Riguet51} and he denotes the
``diagonal'' of a relation $R$ by $R^{d}$, his definition of which is identical to the definition of the  ``diff\a'{e}rence'' of $R$
given by Riguet.)  Consequently, there is some overlap between our work and Winter's work.  We believe
the overlap is justified because we formulate the notion of a ``diagonal'' in terms of factors (``residuals'' in
the terminology used by Winter) and we avoid the use of complements altogether --- with the exception
of section \ref{sec:Staircase Relations} where we observe the equivalence of the notion of a ``staircase''
relation with Riguet's notion of a ``relation de Ferrers''.  (Winter does express some properties in terms of
factors but, following in Riguet's footsteps,  his calculations invariably use complements, which he
denotes by an overbar.)  Winter also observes that not all ``staircase'' relations are ``block-ordered'' but
does not give any concrete example.  See the concluding section for further discussion of this aspect of
Winter's paper. 

In order to make this paper relatively self-contained,  sections \ref{Point-freeAxiomatisation} and  \ref{Domains} 
summarise the axioms of (point-free) relation algebra and some basic concepts.  Typically, proofs of 
properties stated in these sections  are omitted.  Exceptions to this rule concern properties that we deem
less familiar to many readers (for example, factorisation properties of functional relations   in section
\ref{Functionality}).   Section \ref{Indices General} introduces the notions of a core/index of a relation.   For proofs
of properties stated in this section see \cite{VB2022,VB2023a}.   

\section{Axioms of Point-free Relation Algebra}\label{Point-freeAxiomatisation}

In this section, we give a brief summary of the axioms of point-free relation algebra.  For full details
of the axioms, see \cite{BACKHOUSE2022100730}.    

\subsection{Summary}\label{Indices:The AxiomSystemSummary}

Point-free relation algebra comprises three layers with interfaces between the layers plus additional axioms
peculiar to relations.  The axiom system is typed.  For types $A$ and $B$,  \setms{0.15em}$A{\sim}B$ denotes a set;  the elements of the set are called
\emph{(heterogeneous) relations of type}  $A{\sim}B$.  Elements of type $A{\sim}A$, for some  type  $A$, are called \emph{homogeneous
relations}.  

The first layer axiomatises the properties of a partially ordered  set.  We postulate that, for each pair of
types $A$ and $B$,  $A{\sim}B$ forms  a complete, universally distributive lattice.     We use the symbol ``${\subseteq}$'' for the ordering relation, and  ``${\cup}$''
and ``${\cap}$'' for the supremum and infimum operators.   We assume that this notation is familiar to the
reader, allowing us to skip a more detailed account of its  properties.   However, we use   ${\MPplatbottom}$
for the least element of the ordering  (rather than the conventional $\emptyset$)  and ${\MPplattop}$ for the greatest element.  In keeping with the
conventional practice of overloading the  symbol ``$\emptyset$'', both these symbols are overloaded.  

It is important to note that there is no axiom stating that a relation is a set, and there is no corresponding
notion of membership.   The lack of a notion of membership distinguishes point-free relation algebra from
pointwise algebra.

The second layer adds a composition operator.  If $R$ is a relation of type $A{\sim}B$ and $S$ is a relation of type
$B{\sim}C$, the composition of $R$ and $S$ is a relation of type $A{\sim}C$ which we denote by $R{\MPcomp}S$.     Composition is
associative and, for each type $A$, there is an identity relation which we denote by $I_{A}$.  We often overload
the notation for the identity relation, writing just $I$.  Occasionally, for greater clarity, we do supply the
type information.

The interface between the first and second layers defines a relation algebra to be an instance of a \emph{regular
algebra} \cite{B2004a} (also  called a \emph{standard Kleene algebra}, or \textbf{S}-\emph{algebra} \cite{CON71b}).   For this paper, 
the most important aspect of this interface is the existence and properties of the factor operators.  These
are introduced in section \ref{Barbosa:Basic Structures}.  Also, ${\MPplatbottom}$ is a zero of composition: for all $R$,
 ${\MPplatbottom}{\MPcomp}R\ms{1}{=}\ms{1}{\MPplatbottom}\ms{1}{=}\ms{1}R{\MPcomp}{\MPplatbottom}$.  

The completeness axiom in the first layer  allows the reflexive-transitive closure $R^{*}$ of
each element  $R$  of type  $A{\sim}A$, for some type $A$,  to be defined.  

The third layer is the introduction of a \emph{converse} operator.   If $R$ is a relation of type $A{\sim}B$, the converse of
$R$, which we denote by $R^{\MPrev}$ (pronounced $R$ ``wok'') is a relation of type $B{\sim}A$. 
The interface with the first  layer is that converse is simultaneously  the lower and upper adjoint in a Galois
connection, and thus  a poset isomorphism (in particular,  ${\MPplatbottom}^{\MPrev}\ms{1}{=}\ms{1}{\MPplatbottom}$ and
${\MPplattop}^{\MPrev}\ms{1}{=}\ms{1}{\MPplattop}$);   the interface with the
second layer is formed by the two rules   $I^{\MPrev}\ms{1}{=}\ms{1}I$ and, for all relations $R$ and $S$ of appropriate type, 
 $(R{\MPcomp}S)^{\MPrev}\ms{2}{=}\ms{2}S^{\MPrev}\ms{1}{\MPcomp}\ms{1}R^{\MPrev}$. 

Additional axioms characterise properties peculiar to relations.
The modularity rule (aka Dedekind's rule \cite{Riguet48}) is that, for all relations $R$, $S$ and $T$,\begin{equation}\label{dedekind.rule}
R{\MPcomp}S\ms{1}{\cap}\ms{1}T\ms{3}{\subseteq}\ms{3}R\ms{1}{\MPcomp}\ms{1}(S\ms{3}{\cap}\ms{3}R^{\MPrev}\ms{1}{\MPcomp}\ms{1}T)~~.
\end{equation}The dual property, obtained by exploiting properties of the converse operator, is, for all relations $R$, $S$ and $T$,\begin{equation}\label{dedekind.rule.conv}
S{\MPcomp}R\ms{1}{\cap}\ms{1}T\ms{3}{\subseteq}\ms{3}(S\ms{3}{\cap}\ms{3}T\ms{1}{\MPcomp}\ms{1}R^{\MPrev})\ms{1}{\MPcomp}\ms{1}R~~.
\end{equation}The modularity rule is 
necessary to the derivation of some of the properties we state without proof (for example, the
properties of the domain operators given in section \ref{domain.ops}).  Another rule is  the \emph{cone rule}:\begin{equation}\label{cone.rule}
{\left\langle\forall{}R\ms{4}{:}{:}\ms{4}{\MPplattop}{\MPcomp}R{\MPcomp}{\MPplattop}\ms{2}{=}\ms{2}{\MPplattop}\ms{5}{\equiv}\ms{5}R\ms{1}{\neq}\ms{1}{\MPplatbottom}\right\rangle}~~.
\end{equation}The cone rule limits consideration to ``unary'' relation algebras:  constructing the cartesian product of two
relation algebras to form a relation algebra (whereby the operators are defined pointwise) does not yield
an algebra satisfying the cone rule.

\subsection{Factors}\label{Barbosa:Basic Structures}

If $R$ is a relation of type $A{\sim}B$ and $S$ is a relation of type $A{\sim}C$, the relation $R{\setminus}S$ of type $B{\sim}C$ is defined by
the Galois connection, for all $T$ (of type $B{\sim}C$),\begin{equation}\label{under}
T\ms{2}{\subseteq}\ms{2}R{\setminus}S\ms{5}{\equiv}\ms{5}R{\MPcomp}T\ms{2}{\subseteq}\ms{2}S~~.
\end{equation}Similarly,  if $R$ is a relation of type $A{\sim}B$ and $S$ is a relation of type $C{\sim}B$, the relation $R{/}S$ of type $A{\sim}C$ is defined by
the Galois connection, for all $T$,\begin{equation}\label{over}
T\ms{2}{\subseteq}\ms{2}R{/}S\ms{5}{\equiv}\ms{5}T{\MPcomp}S\ms{2}{\subseteq}\ms{2}R~~.
\end{equation}The existence of factors is a property of a regular algebra; in relation algebra, factors are also known as 
``residuals''.    Factors enjoy a rich theory which underlies many of our calculations.  

The relations $R{\setminus}R$ (of type $B{\sim}B$ if $R$ has type $A{\sim}B$) and $R{/}R$ (of type $A{\sim}A$ if $R$ has type $A{\sim}B$) play a
central r\^{o}le in section \ref{sec:Staircase Relations}.  As is easily verified, both are \emph{preorders}.  That is, both are \emph{transitive}:\begin{displaymath}R{\setminus}R\ms{1}{\MPcomp}\ms{1}R{\setminus}R\ms{4}{\subseteq}\ms{4}R{\setminus}R\ms{9}{\wedge}\ms{9}R{/}R\ms{1}{\MPcomp}\ms{1}R{/}R\ms{4}{\subseteq}\ms{4}R{/}R\end{displaymath}and both are \emph{reflexive}:\begin{displaymath}I\ms{3}{\subseteq}\ms{3}R{\setminus}R\ms{8}{\wedge}\ms{8}I\ms{3}{\subseteq}\ms{3}R{/}R~~.\end{displaymath}(The notation ``$I$'' is overloaded in the above equation.  In the left conjunct, it
denotes the identity relation of type $B{\sim}B$ and, in the right conjunct, it
denotes the identity relation of type $A{\sim}A$, assuming $R$ has type $A{\sim}B$.  We often overload constants in
this way. Note, however, that we do not attempt to combine the two inclusions into one.)
In addition, for all $R$,\begin{equation}\label{factor.eq}
R\ms{1}{\MPcomp}\ms{1}R{\setminus}R\ms{6}{=}\ms{6}R\ms{6}{=}\ms{6}R{/}R\ms{1}{\MPcomp}\ms{1}R~~,
\end{equation}\begin{equation}\label{preorder.1}
R{/}(R{\setminus}R)\ms{4}{=}\ms{4}R\ms{4}{=}\ms{4}(R{/}R){\setminus}R~~,
\end{equation}\begin{equation}\label{preorder.2}
(R{\setminus}R){/}(R{\setminus}R)\ms{4}{=}\ms{4}R{\setminus}R\ms{4}{=}\ms{4}(R{\setminus}R){\setminus}(R{\setminus}R)\mbox{~~and}
\end{equation}\begin{equation}\label{preorder.3}
(R{/}R){\setminus}(R{/}R)\ms{4}{=}\ms{4}R{/}R\ms{4}{=}\ms{4}(R{/}R){/}(R{/}R)~~.
\end{equation}In fact, we don't use (\ref{preorder.1}) directly; its relevance is as the initial step in proving the leftmost 
equations of (\ref{preorder.2}) and (\ref{preorder.3}).  We choose not to exploit the associativity of the over and
under operators in (\ref{preorder.2}) and (\ref{preorder.3}) ---by writing, for example, $(R{\setminus}R){/}(R{\setminus}R)$ as
$R{\setminus}R{/}(R{\setminus}R)$--- in order to emphasise their r\^{o}le as properties of the
preorders  $R{\setminus}R$  and $R{/}R$.

In relation algebra (as opposed to regular algebra) it is possible to eliminate the factor operators
altogether because they can be expressed in terms of complements and converse.  The rules for doing so
are  as follows: for all $R$, $S$ and $T$,  \begin{equation}\label{underovernot}
R{\setminus}S\ms{4}{=}\ms{4}{\neg}(R^{\MPrev}\ms{2}{\MPcomp}\ms{2}{\neg}S)\ms{7}{\wedge}\ms{7}S{/}T\ms{4}{=}\ms{4}{\neg}({\neg}S\ms{2}{\MPcomp}\ms{2}T^{\MPrev})~~.
\end{equation}\begin{equation}\label{underover.to.not}
R{\setminus}S{/}T\ms{4}{=}\ms{4}{\neg}(R^{\MPrev}\ms{1}{\MPcomp}\ms{1}{\neg}S\ms{1}{\MPcomp}\ms{1}T^{\MPrev})~~.
\end{equation} Although the  elimination of factors  is highly
undesirable,  we are obliged to introduce complements in order to compare our work with that of Riguet
in section \ref{sec:Staircase Relations} on staircase relations.
 
Property (\ref{factor.eq}) exemplifies how much easier calculations with factors can be compared to
calculations that combine complements with converses.  The property is very easy to spot and apply. 
Expressed using  (\ref{underover.to.not}), it is equivalent to \begin{displaymath}R\ms{1}{\MPcomp}\ms{1}{\neg}(R^{\MPrev}\ms{1}{\MPcomp}\ms{1}{\neg}R)\ms{6}{=}\ms{6}R\ms{6}{=}\ms{6}{\neg}({\neg}R\ms{1}{\MPcomp}\ms{1}R^{\MPrev})\ms{1}{\MPcomp}\ms{1}R~~.\end{displaymath}In this form, the property is difficult to spot and its correct application is difficult to check.

It is useful to record the distributivity properties of converse over the factor operators:
\begin{Lemma}\label{factor.props}{\rm \ \ \ For all $R$ and $S$,\begin{equation}\label{factor.props.under}
R^{\MPrev}\ms{1}{\setminus}\ms{1}S^{\MPrev}\ms{4}{=}\ms{4}(S{/}R)^{\MPrev}\ms{4}{=}\ms{4}{\neg}R\ms{1}{/}\ms{1}{\neg}S~~.
\end{equation}Symmetrically,\begin{equation}\label{factor.props.over}
R^{\MPrev}\ms{1}{/}\ms{1}S^{\MPrev}\ms{4}{=}\ms{4}(S{\setminus}R)^{\MPrev}\ms{4}{=}\ms{4}{\neg}R\ms{1}{\setminus}\ms{1}{\neg}S~~.
\end{equation}Also,\begin{equation}\label{factor.props.both}
(R{\setminus}S{/}T)^{\MPrev}\ms{4}{=}\ms{4}T^{\MPrev}\ms{1}{\setminus}\ms{1}S^{\MPrev}\ms{1}{/}\ms{1}R^{\MPrev}~~.
\end{equation}\vspace{-7mm}
}%
\end{Lemma}%
\MPendBox

In  (\ref{factor.props.under}) and (\ref{factor.props.over}), the inclusion of terms involving complements
 is only  relevant in section \ref{sec:Staircase Relations}.  

\section{Some Definitions}\label{Domains}

This section introduces a number of concepts which have been studied in detail elsewhere.  For the most
part, their properties are stated without proof.  An exception to this is in section \ref{Functionality} where we
combine the Galois connections defining factors with the Galois connection defining functionality.

\subsection{Coreflexives}\label{Coreflexives}

In point-free relation algebra, ``coreflexives'' of a given type   represent sets of elements of that type.  A
\emph{coreflexive of type} $A$ is a relation $p$ such that $p\ms{1}{\subseteq}\ms{1}I_{A}$.  Frequently used properties are that, for all
coreflexives $p$,  \begin{displaymath}p\ms{2}{=}\ms{2}p^{\MPrev}\ms{2}{=}\ms{2}p{\MPcomp}p\end{displaymath}and, for all coreflexives $p$ and $q$, \begin{displaymath}p{\MPcomp}q\ms{3}{=}\ms{3}p\ms{1}{\cap}\ms{1}q\ms{3}{=}\ms{3}q{\MPcomp}p~~.\end{displaymath}(The proof of these properties relies on the modularity rule.)  In the literature, coreflexives have several
different names, usually depending on the application area in question.  Examples are ``monotype'',
``pid'' (short for ``partial identity'')  and ``test''.

\subsection{The Domain Operators}\label{domain.ops}

The ``domain operators'' (see eg. \cite{BAC92b})  play a dominant and unavoidable role.  We
exploit their properties frequently in calculations, so much so that we assume great familiarity with
them.  
\begin{Definition}[Domain Operators]\label{lr.squares}{\rm \ \ \ Given  relation $R$ of type $A{\sim}B$,  
the \emph{left domain} $R{\MPldom{}}$  of   $R$ is a relation of type $A$ defined by the equation \begin{displaymath}R{\MPldom{}}\ms{6}{=}\ms{6}I_{A}\ms{3}{\cap}\ms{3}R\ms{1}{\MPcomp}\ms{1}R^{\MPrev}\end{displaymath}and the \emph{right  domain} $R{\MPrdom{}}$  of    $R$is a relation of type $B$  is defined by the equation \begin{displaymath}R{\MPrdom{}}\ms{6}{=}\ms{6}I_{B}\ms{3}{\cap}\ms{3}R^{\MPrev}\ms{1}{\MPcomp}\ms{1}R~~.\end{displaymath}\vspace{-7mm}
}
\MPendBox\end{Definition}

The name ``domain operator'' is chosen because of the fundamental properties: for all $R$ and all
coreflexives $p$, \begin{equation}\label{rdom.is.least}
R\ms{1}{=}\ms{1}R{\MPcomp}p\ms{6}{\equiv}\ms{6}R{\MPrdom{}}\ms{2}{=}\ms{2}R{\MPrdom{}}\ms{1}{\MPcomp}\ms{1}p
\end{equation}and \begin{equation}\label{ldom.is.least}
R\ms{1}{=}\ms{1}p{\MPcomp}R\ms{6}{\equiv}\ms{6}R{\MPldom{}}\ms{2}{=}\ms{2}p\ms{1}{\MPcomp}\ms{1}R{\MPldom{}}~~.
\end{equation}A simple, often-used consequence of (\ref{rdom.is.least}) and (\ref{ldom.is.least}) is the property:\begin{equation}\label{ldom.and.rdom}
R{\MPldom{}}\ms{1}{\MPcomp}\ms{1}R\ms{4}{=}\ms{4}R\ms{4}{=}\ms{4}R\ms{1}{\MPcomp}\ms{1}R{\MPrdom{}}~~.
\end{equation}
As is the case for factors, the domain operators enjoy a rich theory which underlies many of our 
calculations but we omit the details here.  

\subsection{Pers and Per Domains}\label{Per Domains}

Given relations $R$ of type $A{\sim}B$  and $S$ of type $A{\sim}C$,  the symmetric
\emph{right-division}  is the  relation $R\rightsymdivision{}S$  of type $B{\sim}C$  defined in terms of \emph{right} factors as  \begin{equation}\label{double.under.def}
R\rightsymdivision{}S\ms{5}{=}\ms{5}R{\setminus}S\ms{2}{\cap}\ms{2}(S{\setminus}R)^{\MPrev}~~.
\end{equation}Dually, given relations $R$ of type $B{\sim}A$ and $S$ of type $C{\sim}A$,  the
symmetric \emph{left-division} is the  relation $R\leftsymdivision{}S$  of type $B{\sim}C$ defined in terms of left factors as \begin{equation}\label{double.over.def}
R\leftsymdivision{}S\ms{5}{=}\ms{5}R{/}S\ms{3}{\cap}\ms{3}(S{/}R)^{\MPrev}~~.
\end{equation}The relation $R\rightsymdivision{}R$ is an equivalence relation.  
 Voermans \cite{Vo99} calls it the ``greatest 
right domain'' of $R$. Riguet \cite{Riguet48} calls $R\rightsymdivision{}R$ the ``noyau''  of $R$ (but defines it using nested
complements).   Others  (see   \cite{Ol17} for references) call it the  ``kernel'' of $R$.  

As remarked elsewhere \cite{Ol17},  the \emph{symmetric left-division} inherits a number of (left) cancellation
properties from the properties of factorisation in terms of which it is defined.  In this section the focus is on 
the left and right ``per domains'' introduced by Voermans \cite{Vo99}.
\begin{Definition}[Right and Left Per Domains]\label{perdoms}{\rm \ \ \ The \emph{right per domain} of relation $R$, denoted
$R{\MPperrdom{}}$,   is defined by the equation\begin{equation}\label{per.rightdomain}
R{\MPperrdom{}}\ms{4}{=}\ms{4}R{\MPrdom{}}\ms{1}{\MPcomp}\ms{1}R\rightsymdivision{}R~~.
\end{equation}Dually, the \emph{left per domain} of relation $R$, denoted $R{\MPperldom{}}$, is defined  by the equation\begin{equation}\label{per.leftdomain}
R{\MPperldom{}}\ms{4}{=}\ms{4}R\leftsymdivision{}R\ms{1}{\MPcomp}\ms{1}R{\MPldom{}}~~.
\end{equation}\vspace{-7mm}
}
\MPendBox\end{Definition}

In order to prove additional
properties,   it is useful to record the left and right domains of  the   relation $R\rightsymdivision{}R\ms{1}{\MPcomp}\ms{1}R{\MPrdom{}}$.   
\begin{Lemma}\label{symleftdiv.dom}{\rm \ \ \ For all $R$,\begin{displaymath}(R\rightsymdivision{}R\ms{1}{\MPcomp}\ms{1}R{\MPrdom{}}){\MPrdom{}}\ms{4}{=}\ms{4}R{\MPrdom{}}\ms{4}{=}\ms{4}(R{\MPrdom{}}\ms{1}{\MPcomp}\ms{1}R\rightsymdivision{}R){\MPldom{}}~~,\end{displaymath}\begin{displaymath}(R\rightsymdivision{}R\ms{1}{\MPcomp}\ms{1}R{\MPrdom{}}){\MPldom{}}\ms{4}{=}\ms{4}R{\MPrdom{}}\ms{4}{=}\ms{4}(R{\MPrdom{}}\ms{1}{\MPcomp}\ms{1}R\rightsymdivision{}R){\MPrdom{}}~~,\end{displaymath}\begin{displaymath}R\rightsymdivision{}R\ms{1}{\MPcomp}\ms{1}R{\MPrdom{}}\ms{3}{=}\ms{3}R{\MPrdom{}}\ms{1}{\MPcomp}\ms{1}R\rightsymdivision{}R\ms{1}{\MPcomp}\ms{1}R{\MPrdom{}}\ms{3}{=}\ms{3}R{\MPrdom{}}\ms{1}{\MPcomp}\ms{1}R\rightsymdivision{}R~~.\end{displaymath}\vspace{-9mm}
}%
\end{Lemma}%
\MPendBox

The left and right per domains are ``pers'' where ``per'' is an abbreviation of ``partial equivalence
relation''.  
\begin{Definition}[Partial Equivalence Relation (per)]\label{def:per}{\rm \ \ \ A relation  is a \emph{partial equivalence
relation} iff  it is symmetric and transitive. That is, $R$ is a partial equivalence relation iff \begin{displaymath}R\ms{1}{=}\ms{1}R^{\MPrev}\ms{5}{\wedge}\ms{5}R{\MPcomp}R\ms{1}{\subseteq}\ms{1}R~~.\end{displaymath}Henceforth we abbreviate partial equivalence relation to \emph{per}.
}
\MPendBox\end{Definition}

That $R{\MPperldom{}}$ and $R{\MPperrdom{}}$ are indeed pers is a direct consequence of the symmetry and transitivity of $R\rightsymdivision{}R$.  

The left and right per domains are called ``domains'' because,  like the coreflexive domains, we have the
properties: for all relations $R$ and pers $P$,\begin{equation}\label{per.rdom.is.least}
R\ms{1}{=}\ms{1}R{\MPcomp}P\ms{6}{\equiv}\ms{6}R{\MPperrdom{}}\ms{2}{=}\ms{2}R{\MPperrdom{}}\ms{1}{\MPcomp}\ms{1}P\mbox{~~, and}
\end{equation}\begin{equation}\label{per.ldom.is.least}
R\ms{1}{=}\ms{1}P{\MPcomp}R\ms{6}{\equiv}\ms{6}R{\MPperldom{}}\ms{2}{=}\ms{2}P\ms{1}{\MPcomp}\ms{1}R{\MPperldom{}}~~.
\end{equation}The right per domain  $R{\MPperrdom{}}$ can be defined equivalently by the equation \begin{equation}\label{per.rightdomain.equiv}
R{\MPperrdom{}}\ms{4}{=}\ms{4}R\rightsymdivision{}R\ms{1}{\MPcomp}\ms{1}R{\MPrdom{}}~~.
\end{equation}Moreover, \begin{equation}\label{per.rightdomain.doms}
(R{\MPperrdom{}}){\MPldom{}}\ms{5}{=}\ms{5}R{\MPrdom{}}\ms{5}{=}\ms{5}(R{\MPperrdom{}}){\MPrdom{}}~~.
\end{equation} (See \cite{RCB2020} for the proofs of these properties.)  Symmetrical properties hold of $R{\MPperldom{}}$.

For further properties of pers and per domains, see \cite{Vo99}.

\subsection{Functionality}\label{Functionality}

In this section, we present  a number of lesser-known properties of ``functional'' relations.   A relation $R$ of type $A{\sim}B$
is said to be  \emph{left}-\emph{functional} iff $R\ms{1}{\MPcomp}\ms{1}R^{\MPrev}\ms{2}{=}\ms{2}R{\MPldom{}}$.    Equivalently,   $R$ is \emph{left-functional}  iff $R\ms{1}{\MPcomp}\ms{1}R^{\MPrev}\ms{2}{\subseteq}\ms{2}I_{A}$.
It is said to be  \emph{right-functional}  iff  $R^{\MPrev}\ms{1}{\MPcomp}\ms{1}R\ms{2}{=}\ms{2}R{\MPrdom{}}$ (equivalently,  $R^{\MPrev}\ms{1}{\MPcomp}\ms{1}R\ms{2}{\subseteq}\ms{2}I_{B}$).   A relation $R$
is said to be a \emph{bijection} iff it is both left- and right-functional.  

Rather than left-  and right-functional, the more common terminology is ``functional'' and ``injective''
but publications differ on which of left- or right-functional is ``functional'' or ``injective''.  We choose to
abbreviate ``left-functional'' to \emph{functional} and to use the term \emph{injective} instead of right-functional. 
Typically, we use  $f$ and $g$ to denote functional relations, and Greek letters to denote bijections (although
the latter is not always the case).  Other authors make the opposite choice.

The properties we present here stem from 
 the observation that functionality can be defined via a Galois connection. 
Specifically, the relation   $f$ is (left-)functional iff, for all relations $R$ and $S$ (of appropriate type), \begin{equation}\label{functional.GC}
f{\MPcomp}R\ms{2}{\subseteq}\ms{2}S\ms{8}{\equiv}\ms{8}f{\MPrdom{}}\ms{1}{\MPcomp}\ms{1}R\ms{3}{\subseteq}\ms{3}f^{\MPrev}\ms{1}{\MPcomp}\ms{1}S~~.
\end{equation}It is a simple exercise to show that (\ref{functional.GC}) is equivalent to the property $f\ms{1}{\MPcomp}\ms{1}f^{\MPrev}\ms{2}{\subseteq}\ms{2}I$.  (Although
(\ref{functional.GC}) doesn't immediately fit the standard definition of a Galois connection, it can be turned
into standard form by restricting the range of the dummy $R$ to relations that satisfy $f{\MPrdom{}}\ms{1}{\MPcomp}\ms{1}R\ms{2}{=}\ms{2}R$, i.e.\ 
relations $R$ such that $R{\MPldom{}}\ms{1}{\subseteq}\ms{1}f{\MPrdom{}}$.)

The converse-dual of (\ref{functional.GC}) is also used frequently:  $g$ is functional iff, for all $R$ and $S$, \begin{equation}\label{functional.GC.conv}
R\ms{1}{\MPcomp}\ms{1}g^{\MPrev}\ms{3}{\subseteq}\ms{3}S\ms{8}{\equiv}\ms{8}R\ms{1}{\MPcomp}\ms{1}g{\MPrdom{}}\ms{3}{\subseteq}\ms{3}S{\MPcomp}g~~.
\end{equation}
Comparing the Galois connections defining the over and under operators 
%***(see section \ref{BD:Factors})***
with the Galois connection defining functionality (see  (\ref{functional.GC})) suggests a formal relationship
between ``division'' by a functional relation and composition with the relation's converse.  The precise
form of this relationship is given by the following lemma. 

\begin{Lemma}\label{fun.under.R}{\rm \ \ \ For all $R$ and all functional relations $f$,\begin{displaymath}f{\MPrdom{}}\ms{1}{\MPcomp}\ms{1}f{\setminus}R\ms{5}{=}\ms{5}f^{\MPrev}\ms{1}{\MPcomp}\ms{1}R~~.\end{displaymath}
}%
\end{Lemma}%
{\bf Proof}~~~We use the anti-symmetry of the subset relation.  First,
\begin{mpdisplay}{0.15em}{6.5mm}{0mm}{2}
	$f^{\MPrev}\ms{1}{\MPcomp}\ms{1}R\ms{4}{\subseteq}\ms{4}f{\MPrdom{}}\ms{1}{\MPcomp}\ms{1}f{\setminus}R$\push\-\\
	$=$	\>	\>$\{$	\>\+\+\+domains\-\-$~~~ \}$\pop\\
	$f{\MPrdom{}}\ms{1}{\MPcomp}\ms{1}f^{\MPrev}\ms{1}{\MPcomp}\ms{1}R\ms{4}{\subseteq}\ms{4}f{\MPrdom{}}\ms{1}{\MPcomp}\ms{1}f{\setminus}R$\push\-\\
	$\Leftarrow$	\>	\>$\{$	\>\+\+\+monotonicity\-\-$~~~ \}$\pop\\
	$f^{\MPrev}\ms{1}{\MPcomp}\ms{1}R\ms{4}{\subseteq}\ms{4}f{\setminus}R$\push\-\\
	$=$	\>	\>$\{$	\>\+\+\+factors\-\-$~~~ \}$\pop\\
	$f\ms{1}{\MPcomp}\ms{1}f^{\MPrev}\ms{1}{\MPcomp}\ms{1}R\ms{4}{\subseteq}\ms{4}R$\push\-\\
	$\Leftarrow$	\>	\>$\{$	\>\+\+\+definition and monotonicity\-\-$~~~ \}$\pop\\
	$f$ is functional~~~.
\end{mpdisplay}
Second,
\begin{mpdisplay}{0.15em}{6.5mm}{0mm}{2}
	$f{\MPrdom{}}\ms{1}{\MPcomp}\ms{1}f{\setminus}R\ms{4}{\subseteq}\ms{4}f^{\MPrev}\ms{1}{\MPcomp}\ms{1}R$\push\-\\
	$\Leftarrow$	\>	\>$\{$	\>\+\+\+$f{\MPrdom{}}\ms{2}{\subseteq}\ms{2}f^{\MPrev}\ms{1}{\MPcomp}\ms{1}f$; monotonicity and transitivity\-\-$~~~ \}$\pop\\
	$f^{\MPrev}\ms{1}{\MPcomp}\ms{1}f\ms{1}{\MPcomp}\ms{1}f{\setminus}R\ms{4}{\subseteq}\ms{4}f^{\MPrev}\ms{1}{\MPcomp}\ms{1}R$\push\-\\
	$\Leftarrow$	\>	\>$\{$	\>\+\+\+monotonicity\-\-$~~~ \}$\pop\\
	$f\ms{1}{\MPcomp}\ms{1}f{\setminus}R\ms{4}{\subseteq}\ms{4}R$\push\-\\
	$=$	\>	\>$\{$	\>\+\+\+cancellation\-\-$~~~ \}$\pop\\
	$\mathsf{true}~~.$
\end{mpdisplay}
\vspace{-7mm}
\MPendBox

Two  lemmas  that will be needed later now follow.   Lemma \ref{fwok.elim}
 allows the converse of a functional relation (i.e.\ an injective relation) to be cancelled, whilst lemma 
\ref{f.factor.dist} expresses a distributivity property.

\begin{Lemma}\label{fwok.elim}{\rm \ \ \ For all $R$ and all functional relations $f$,\begin{displaymath}f{\MPldom{}}\ms{2}{\MPcomp}\ms{2}f^{\MPrev}\ms{1}{\setminus}\ms{1}(f^{\MPrev}\ms{1}{\MPcomp}\ms{1}R)\ms{6}{=}\ms{6}f{\MPldom{}}\ms{1}{\MPcomp}\ms{1}R~~.\end{displaymath}
}%
\end{Lemma}%
{\bf Proof}~~~
\begin{mpdisplay}{0.15em}{6.5mm}{0mm}{2}
	$f{\MPldom{}}\ms{2}{\MPcomp}\ms{2}f^{\MPrev}\ms{1}{\setminus}\ms{1}(f^{\MPrev}\ms{1}{\MPcomp}\ms{1}R)$\push\-\\
	$=$	\>	\>$\{$	\>\+\+\+assumption:  $f$ is functional\-\-$~~~ \}$\pop\\
	$f\ms{2}{\MPcomp}\ms{2}f^{\MPrev}\ms{2}{\MPcomp}\ms{2}f^{\MPrev}\ms{1}{\setminus}\ms{1}(f^{\MPrev}\ms{1}{\MPcomp}\ms{1}R)$\push\-\\
	$\subseteq$	\>	\>$\{$	\>\+\+\+cancellation\-\-$~~~ \}$\pop\\
	$f\ms{1}{\MPcomp}\ms{1}f^{\MPrev}\ms{1}{\MPcomp}\ms{1}R$\push\-\\
	$=$	\>	\>$\{$	\>\+\+\+assumption:  $f$ is functional\-\-$~~~ \}$\pop\\
	$f{\MPldom{}}\ms{1}{\MPcomp}\ms{1}R~~.$
\end{mpdisplay}
Also,
\begin{mpdisplay}{0.15em}{6.5mm}{0mm}{2}
	$f{\MPldom{}}\ms{1}{\MPcomp}\ms{1}R\ms{3}{\subseteq}\ms{3}f{\MPldom{}}\ms{2}{\MPcomp}\ms{2}f^{\MPrev}\ms{1}{\setminus}\ms{1}(f^{\MPrev}\ms{1}{\MPcomp}\ms{1}R)$\push\-\\
	$\Leftarrow$	\>	\>$\{$	\>\+\+\+monotonicity\-\-$~~~ \}$\pop\\
	$R\ms{2}{\subseteq}\ms{2}f^{\MPrev}\ms{1}{\setminus}\ms{1}(f^{\MPrev}\ms{1}{\MPcomp}\ms{1}R)$\push\-\\
	$=$	\>	\>$\{$	\>\+\+\+factors\-\-$~~~ \}$\pop\\
	$\mathsf{true}~~.$
\end{mpdisplay}
The lemma follows by anti-symmetry of the subset relation.
\MPendBox

\begin{Lemma}\label{f.factor.dist}{\rm \ \ \ For all $R$ and $S$ and all functional relations $f$ ,\begin{displaymath}R{\setminus}(S{\MPcomp}f)\ms{1}{\MPcomp}\ms{1}f{\MPrdom{}}\ms{5}{=}\ms{5}R{\setminus}S\ms{1}{\MPcomp}\ms{1}f~~.\end{displaymath}
}%
\end{Lemma}%
{\bf Proof}~~~
\begin{mpdisplay}{0.15em}{6.5mm}{0mm}{2}
	$R{\setminus}(S{\MPcomp}f)\ms{1}{\MPcomp}\ms{1}f{\MPrdom{}}\ms{4}{\subseteq}\ms{4}R{\setminus}S\ms{1}{\MPcomp}\ms{1}f$\push\-\\
	$\Leftarrow$	\>	\>$\{$	\>\+\+\+$f{\MPrdom{}}\ms{2}{\subseteq}\ms{2}f^{\MPrev}\ms{1}{\MPcomp}\ms{1}f$,  monotonicity\-\-$~~~ \}$\pop\\
	$R{\setminus}(S{\MPcomp}f)\ms{1}{\MPcomp}\ms{1}f^{\MPrev}\ms{4}{\subseteq}\ms{4}R{\setminus}S$\push\-\\
	$=$	\>	\>$\{$	\>\+\+\+factors\-\-$~~~ \}$\pop\\
	$R\ms{1}{\MPcomp}\ms{1}R{\setminus}(S{\MPcomp}f)\ms{1}{\MPcomp}\ms{1}f^{\MPrev}\ms{4}{\subseteq}\ms{4}S$\push\-\\
	$\Leftarrow$	\>	\>$\{$	\>\+\+\+cancellation\-\-$~~~ \}$\pop\\
	$S\ms{1}{\MPcomp}\ms{1}f\ms{1}{\MPcomp}\ms{1}f^{\MPrev}\ms{4}{\subseteq}\ms{4}S$\push\-\\
	$=$	\>	\>$\{$	\>\+\+\+assumption:  $f$ is functional\-\-$~~~ \}$\pop\\
	$\mathsf{true}~~.$
\end{mpdisplay}
Also,
\begin{mpdisplay}{0.15em}{6.5mm}{0mm}{2}
	$R{\setminus}S\ms{1}{\MPcomp}\ms{1}f\ms{4}{\subseteq}\ms{4}R{\setminus}(S{\MPcomp}f)\ms{1}{\MPcomp}\ms{1}f{\MPrdom{}}$\push\-\\
	$\Leftarrow$	\>	\>$\{$	\>\+\+\+monotonicity,  $f\ms{2}{=}\ms{2}f\ms{1}{\MPcomp}\ms{1}f{\MPrdom{}}$\-\-$~~~ \}$\pop\\
	$R{\setminus}S\ms{1}{\MPcomp}\ms{1}f\ms{4}{\subseteq}\ms{4}R{\setminus}(S{\MPcomp}f)$\push\-\\
	$=$	\>	\>$\{$	\>\+\+\+factors and cancellation\-\-$~~~ \}$\pop\\
	$\mathsf{true}~~.$
\end{mpdisplay}
The lemma follows by anti-symmetry of the subset relation.
\MPendBox

The following lemma is crucial to fully understanding Riguet's ``analogie frappante''.
%; see  lemma \ref{BD.provorder}.
(The lemma is complicated by the fact that it
has five free variables.  Simpler, possibly better known, instances can be obtained by instantiating one or
more  of $f$, $g$, $U$ and $W$ to the identity relation.)  
\begin{Lemma}\label{f.under.over.g}{\rm \ \ \ Suppose $f$ and $g$ are functional.  Then, for all $U$, $V$ and $W$,
\begin{mpdisplay}{0.15em}{6.5mm}{0mm}{2}
	\push$~~~\ms{2}$\=\+$f^{\MPrev}\ms{1}{\MPcomp}\ms{1}(g{\MPldom{}}\ms{1}{\MPcomp}\ms{1}U){\setminus}V{/}(W\ms{1}{\MPcomp}\ms{1}f{\MPldom{}})\ms{1}{\MPcomp}\ms{1}g$\push\-\\
	$=$	\>\+\pop$f{\MPrdom{}}\ms{1}{\MPcomp}\ms{1}(g^{\MPrev}\ms{1}{\MPcomp}\ms{1}U\ms{1}{\MPcomp}\ms{1}f){\setminus}(g^{\MPrev}\ms{1}{\MPcomp}\ms{1}V\ms{1}{\MPcomp}\ms{1}f){/}(g^{\MPrev}\ms{1}{\MPcomp}\ms{1}W\ms{1}{\MPcomp}\ms{1}f)\ms{1}{\MPcomp}\ms{1}g{\MPrdom{}}~~.$\-\pop
\end{mpdisplay}
}%
\end{Lemma}%
{\bf Proof}~~~Guided by the assumed functionality of $f$ and $g$, we use the rule of indirect equality.  Specifically,
we  have, for all $R$, $U$, $V$ and $W$,
\begin{mpdisplay}{0.15em}{6.5mm}{0mm}{2}
	$f{\MPrdom{}}\ms{1}{\MPcomp}\ms{1}R\ms{1}{\MPcomp}\ms{1}g{\MPrdom{}}\ms{4}{\subseteq}\ms{4}f^{\MPrev}\ms{1}{\MPcomp}\ms{1}(g{\MPldom{}}\ms{1}{\MPcomp}\ms{1}U){\setminus}V{/}(W\ms{1}{\MPcomp}\ms{1}f{\MPldom{}})\ms{1}{\MPcomp}\ms{1}g$\push\-\\
	$=$	\>	\>$\{$	\>\+\+\+assumption:    $f$ and $g$ are functional,   (\ref{functional.GC}) and (\ref{functional.GC.conv})\-\-$~~~ \}$\pop\\
	$f\ms{1}{\MPcomp}\ms{1}R\ms{1}{\MPcomp}\ms{1}g^{\MPrev}\ms{4}{\subseteq}\ms{4}(g{\MPldom{}}\ms{1}{\MPcomp}\ms{1}U){\setminus}V{/}(W\ms{1}{\MPcomp}\ms{1}f{\MPldom{}})$\push\-\\
	$=$	\>	\>$\{$	\>\+\+\+factors\-\-$~~~ \}$\pop\\
	$g{\MPldom{}}\ms{1}{\MPcomp}\ms{1}U\ms{1}{\MPcomp}\ms{1}f\ms{1}{\MPcomp}\ms{1}R\ms{1}{\MPcomp}\ms{1}g^{\MPrev}\ms{1}{\MPcomp}\ms{1}W\ms{1}{\MPcomp}\ms{1}f{\MPldom{}}\ms{4}{\subseteq}\ms{4}V$\push\-\\
	$=$	\>	\>$\{$	\>\+\+\+assumption:    $f$ and $g$ are functional\\
	i.e.\ ~  $f\ms{1}{\MPcomp}\ms{1}f^{\MPrev}\ms{2}{=}\ms{2}f{\MPldom{}}\ms{3}{\wedge}\ms{3}g\ms{1}{\MPcomp}\ms{1}g^{\MPrev}\ms{2}{=}\ms{2}g{\MPldom{}}$\-\-$~~~ \}$\pop\\
	$g\ms{1}{\MPcomp}\ms{1}g^{\MPrev}\ms{1}{\MPcomp}\ms{1}U\ms{1}{\MPcomp}\ms{1}f\ms{1}{\MPcomp}\ms{1}R\ms{1}{\MPcomp}\ms{1}g^{\MPrev}\ms{1}{\MPcomp}\ms{1}W\ms{1}{\MPcomp}\ms{1}f\ms{1}{\MPcomp}\ms{1}f^{\MPrev}\ms{4}{\subseteq}\ms{4}V$\push\-\\
	$=$	\>	\>$\{$	\>\+\+\+assumption:    $f$ and $g$ are functional, (\ref{functional.GC}) and (\ref{functional.GC.conv})\-\-$~~~ \}$\pop\\
	$g{\MPrdom{}}\ms{1}{\MPcomp}\ms{1}g^{\MPrev}\ms{1}{\MPcomp}\ms{1}U\ms{1}{\MPcomp}\ms{1}f\ms{1}{\MPcomp}\ms{1}R\ms{1}{\MPcomp}\ms{1}g^{\MPrev}\ms{1}{\MPcomp}\ms{1}W\ms{1}{\MPcomp}\ms{1}f\ms{1}{\MPcomp}\ms{1}f{\MPrdom{}}\ms{4}{\subseteq}\ms{4}g^{\MPrev}\ms{1}{\MPcomp}\ms{1}V\ms{1}{\MPcomp}\ms{1}f$\push\-\\
	$=$	\>	\>$\{$	\>\+\+\+domains (four times)\-\-$~~~ \}$\pop\\
	$g^{\MPrev}\ms{1}{\MPcomp}\ms{1}U\ms{1}{\MPcomp}\ms{1}f\ms{1}{\MPcomp}\ms{1}f{\MPrdom{}}\ms{1}{\MPcomp}\ms{1}R\ms{1}{\MPcomp}\ms{1}g{\MPrdom{}}\ms{1}{\MPcomp}\ms{1}g^{\MPrev}\ms{1}{\MPcomp}\ms{1}W\ms{1}{\MPcomp}\ms{1}f\ms{4}{\subseteq}\ms{4}g^{\MPrev}\ms{1}{\MPcomp}\ms{1}V\ms{1}{\MPcomp}\ms{1}f$\push\-\\
	$=$	\>	\>$\{$	\>\+\+\+factors\-\-$~~~ \}$\pop\\
	$f{\MPrdom{}}\ms{1}{\MPcomp}\ms{1}R\ms{1}{\MPcomp}\ms{1}g{\MPrdom{}}\ms{4}{\subseteq}\ms{4}(g^{\MPrev}\ms{1}{\MPcomp}\ms{1}U\ms{1}{\MPcomp}\ms{1}f){\setminus}(g^{\MPrev}\ms{1}{\MPcomp}\ms{1}V\ms{1}{\MPcomp}\ms{1}f){/}(g^{\MPrev}\ms{1}{\MPcomp}\ms{1}W\ms{1}{\MPcomp}\ms{1}f)$\push\-\\
	$=$	\>	\>$\{$	\>\+\+\+$f{\MPrdom{}}$ and $g{\MPrdom{}}$ are coreflexives\-\-$~~~ \}$\pop\\
	$f{\MPrdom{}}\ms{1}{\MPcomp}\ms{1}R\ms{1}{\MPcomp}\ms{1}g{\MPrdom{}}\ms{4}{\subseteq}\ms{4}f{\MPrdom{}}\ms{1}{\MPcomp}\ms{1}(g^{\MPrev}\ms{1}{\MPcomp}\ms{1}U\ms{1}{\MPcomp}\ms{1}f){\setminus}(g^{\MPrev}\ms{1}{\MPcomp}\ms{1}V\ms{1}{\MPcomp}\ms{1}f){/}(g^{\MPrev}\ms{1}{\MPcomp}\ms{1}W\ms{1}{\MPcomp}\ms{1}f)\ms{1}{\MPcomp}\ms{1}g{\MPrdom{}}$
\end{mpdisplay}
The lemma follows by instantiating $R$ to the left and right sides of the claimed equation, simplifying using
domain calculus,  and then applying  the
reflexivity and anti-symmetry of the subset relation.
\MPendBox

The final lemma in this section anticipates the discussion of per domains in section \ref{Isomorphic Relations}.
\begin{Lemma}\label{fwok.elim.gen}{\rm \ \ \ Suppose relations  $R$,     $f$    and $g$  are such that   \begin{displaymath}f\ms{1}{\MPcomp}\ms{1}f^{\MPrev}\ms{3}{=}\ms{3}f{\MPldom{}}\ms{3}{=}\ms{3}R{\MPldom{}}\ms{6}{\wedge}\ms{6}g{\MPldom{}}\ms{3}{=}\ms{3}g\ms{1}{\MPcomp}\ms{1}g^{\MPrev}~~.\end{displaymath}Then, for all $S$,\begin{equation}\label{fwok.elim.gen1}
g{\MPrdom{}}\ms{1}{\MPcomp}\ms{1}(f^{\MPrev}\ms{1}{\MPcomp}\ms{1}R\ms{1}{\MPcomp}\ms{1}g){\setminus}(f^{\MPrev}\ms{1}{\MPcomp}\ms{1}S)\ms{6}{=}\ms{6}g^{\MPrev}\ms{1}{\MPcomp}\ms{1}R{\setminus}S~~.
\end{equation}It follows that \begin{equation}\label{fwok.elim.gen2}
g{\MPrdom{}}\ms{1}{\MPcomp}\ms{1}(f^{\MPrev}\ms{1}{\MPcomp}\ms{1}R\ms{1}{\MPcomp}\ms{1}g){\setminus}(f^{\MPrev}\ms{1}{\MPcomp}\ms{1}R\ms{1}{\MPcomp}\ms{1}g)\ms{1}{\MPcomp}\ms{1}g{\MPrdom{}}\ms{6}{=}\ms{6}g^{\MPrev}\ms{1}{\MPcomp}\ms{1}R{\setminus}R\ms{1}{\MPcomp}\ms{1}g~~.
\end{equation}
}%
\end{Lemma}%
{\bf Proof}~~~The proof of (\ref{fwok.elim.gen1}) is as follows.
\begin{mpdisplay}{0.15em}{6.5mm}{0mm}{2}
	$g{\MPrdom{}}\ms{1}{\MPcomp}\ms{1}(f^{\MPrev}\ms{1}{\MPcomp}\ms{1}R\ms{1}{\MPcomp}\ms{1}g){\setminus}(f^{\MPrev}\ms{1}{\MPcomp}\ms{1}S)$\push\-\\
	$=$	\>	\>$\{$	\>\+\+\+factors:\-\-$~~~ \}$\pop\\
	$g{\MPrdom{}}\ms{1}{\MPcomp}\ms{1}g{\setminus}((f^{\MPrev}\ms{1}{\MPcomp}\ms{1}R){\setminus}(f^{\MPrev}\ms{1}{\MPcomp}\ms{1}S))$\push\-\\
	$=$	\>	\>$\{$	\>\+\+\+lemma \ref{fun.under.R} with $f{,}R\ms{2}{:=}\ms{2}g\ms{1}{,}\ms{1}(f^{\MPrev}\ms{1}{\MPcomp}\ms{1}R){\setminus}(f^{\MPrev}\ms{1}{\MPcomp}\ms{1}S)$\-\-$~~~ \}$\pop\\
	$g^{\MPrev}\ms{1}{\MPcomp}\ms{1}(f^{\MPrev}\ms{1}{\MPcomp}\ms{1}R){\setminus}(f^{\MPrev}\ms{1}{\MPcomp}\ms{1}S)$\push\-\\
	$=$	\>	\>$\{$	\>\+\+\+factors\-\-$~~~ \}$\pop\\
	$g^{\MPrev}\ms{1}{\MPcomp}\ms{1}R{\setminus}(f^{\MPrev}\ms{1}{\setminus}\ms{1}(f^{\MPrev}\ms{1}{\MPcomp}\ms{1}S))$\push\-\\
	$=$	\>	\>$\{$	\>\+\+\+$\left[\ms{2}R{\setminus}S\ms{1}{=}\ms{1}R{\setminus}(R{\MPldom{}}\ms{1}{\MPcomp}\ms{1}S)\ms{2}\right]$ with $R{,}S\ms{3}{:=}\ms{3}R\ms{2}{,}\ms{2}f^{\MPrev}\ms{1}{\setminus}\ms{1}(f^{\MPrev}\ms{1}{\MPcomp}\ms{1}S)$\\
	assumption:  $f{\MPldom{}}\ms{1}{=}\ms{1}R{\MPldom{}}$\-\-$~~~ \}$\pop\\
	$g^{\MPrev}\ms{1}{\MPcomp}\ms{1}R{\setminus}(f{\MPldom{}}\ms{2}{\MPcomp}\ms{2}f^{\MPrev}\ms{1}{\setminus}\ms{1}(f^{\MPrev}\ms{1}{\MPcomp}\ms{1}S))$\push\-\\
	$=$	\>	\>$\{$	\>\+\+\+lemma \ref{fwok.elim} with $f{,}R\ms{1}{:=}\ms{1}f{,}S$\-\-$~~~ \}$\pop\\
	$g^{\MPrev}\ms{1}{\MPcomp}\ms{1}R{\setminus}(f{\MPldom{}}\ms{1}{\MPcomp}\ms{1}S)$\push\-\\
	$=$	\>	\>$\{$	\>\+\+\+assumption:  $f{\MPldom{}}\ms{1}{=}\ms{1}R{\MPldom{}}$,  $\left[\ms{2}R{\setminus}S\ms{1}{=}\ms{1}R{\setminus}(R{\MPldom{}}\ms{1}{\MPcomp}\ms{1}S)\ms{2}\right]$\-\-$~~~ \}$\pop\\
	$g^{\MPrev}\ms{1}{\MPcomp}\ms{1}R{\setminus}S~~.$
\end{mpdisplay}
Now we prove (\ref{fwok.elim.gen2}).
\begin{mpdisplay}{0.15em}{6.5mm}{0mm}{2}
	$g{\MPrdom{}}\ms{1}{\MPcomp}\ms{1}(f^{\MPrev}\ms{1}{\MPcomp}\ms{1}R\ms{1}{\MPcomp}\ms{1}g){\setminus}(f^{\MPrev}\ms{1}{\MPcomp}\ms{1}R\ms{1}{\MPcomp}\ms{1}g)\ms{1}{\MPcomp}\ms{1}g{\MPrdom{}}$\push\-\\
	$=$	\>	\>$\{$	\>\+\+\+(\ref{fwok.elim.gen1}) with $S\ms{1}{:=}\ms{1}R{\MPcomp}g$\-\-$~~~ \}$\pop\\
	$g^{\MPrev}\ms{1}{\MPcomp}\ms{1}R{\setminus}(R{\MPcomp}g)\ms{1}{\MPcomp}\ms{1}g{\MPrdom{}}$\push\-\\
	$=$	\>	\>$\{$	\>\+\+\+lemma \ref{f.factor.dist}\-\-$~~~ \}$\pop\\
	$g^{\MPrev}\ms{1}{\MPcomp}\ms{1}R{\setminus}R\ms{1}{\MPcomp}\ms{1}g~~.$
\end{mpdisplay}
\vspace{-7mm}
\MPendBox

\subsection{Difunctions}\label{bd:Difunctions}

Formally,  relation $R$ is \emph{difunctional} iff \begin{equation}\label{difunctional.def}
R\ms{1}{\MPcomp}\ms{1}R^{\MPrev}\ms{1}{\MPcomp}\ms{1}R\ms{3}{\subseteq}\ms{3}R~~.
\end{equation}As for pers,  there are several equivalent definitions of ``difunctional'', as formulated below.    

\begin{Theorem}\label{difunctional.strongdef}{\rm \ \ \ For all $R$, the following statements are all equivalent.
\begin{description}
\item[(i)~~]$R$ is difunctional (i.e.\  $R\ms{1}{\MPcomp}\ms{1}R^{\MPrev}\ms{1}{\MPcomp}\ms{1}R\ms{3}{\subseteq}\ms{3}R$)~~, 
\item[(ii)~]$R\ms{3}{=}\ms{3}R\ms{1}{\MPcomp}\ms{1}R^{\MPrev}\ms{1}{\MPcomp}\ms{1}R$~~, 
\item[(iii)]$R{\MPrdom{}}\ms{1}{\MPcomp}\ms{1}R{\setminus}R\ms{3}{=}\ms{3}R^{\MPrev}\ms{1}{\MPcomp}\ms{1}R$~~, 
\item[(iv)~]$R{\MPperrdom{}}\ms{3}{=}\ms{3}R^{\MPrev}\ms{1}{\MPcomp}\ms{1}R$~~, 
\item[(v)~~]$R{/}R\ms{1}{\MPcomp}\ms{1}R{\MPldom{}}\ms{3}{=}\ms{3}R\ms{1}{\MPcomp}\ms{1}R^{\MPrev}$~~, 
\item[(vi)~]$R{\MPperldom{}}\ms{3}{=}\ms{3}R\ms{1}{\MPcomp}\ms{1}R^{\MPrev}$~~, 
\item[(vii)]$R\ms{3}{=}\ms{3}R\ms{1}{\cap}\ms{1}(R{\setminus}R{/}R)^{\MPrev}$~~. 
 
\end{description}
\vspace{-7mm}
}
\MPendBox\end{Theorem}

\subsection{Provisional Orderings}\label{bd:orderings}

There are various well-known  notions of ordering: preorder, partial and linear (aka total) ordering.  For
our purposes all of these are too strict --- the fact  is that, in practice,   relations are rarely  ``total'' (for example, not
everyone has a sibling).  So, in this section, we introduce the notion of a ``provisional
ordering''.  The adjective ``provisional'' has been chosen because the notion ``provides'' just what we need. 
For later use, we state a number of properties but without proof.  All proofs can be found in the
companion working document \cite{VB2022}.

The standard definition of an ordering is an anti-symmetric  preorder whereby a preorder is required to
be reflexive and transitive.  It is the reflexivity requirement that is too strict for our purposes.  
So, with the intention of weakening  the standard definition of a preorder to requiring reflexivity of a
relation  over  some superset of its left and right domains, we propose the following definition.  
\begin{Definition}\label{provisionalpreorder}{\rm \ \ \ Suppose  $T$ is a  homogeneous relation.     Then $T$  is said to be a 
\emph{provisional  preorder}   if \begin{displaymath}T{\MPldom{}}\ms{2}{\subseteq}\ms{2}T\ms{6}{\wedge}\ms{6}T{\MPrdom{}}\ms{2}{\subseteq}\ms{2}T\ms{6}{\wedge}\ms{6}T{\MPcomp}T\ms{1}{\subseteq}\ms{1}T~~.\end{displaymath}\vspace{-7mm}
}
\MPendBox\end{Definition}

\begin{figure}[h]
\centering \includegraphics{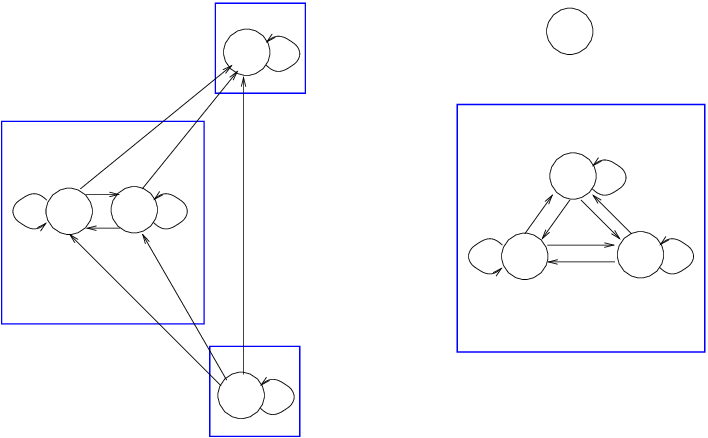}
 
\caption{A Provisional Preorder}\label{ProvisionalPreorder}
\end{figure}

Fig.\ \ref{ProvisionalPreorder}  depicts a provisional preorder on a set of eight elements as a directed graph.  
The blue squares should be ignored for the moment.  (See the discussion following lemma
 \ref{ppreorder.perdomain}.)  Note that the relation depicted is not a preorder
because it is not reflexive:  the top-right node depicts an element that is not in the left or right domain of
the relation.  

An immediate consequence of the definition is:
\begin{Lemma}\label{provisionalpreorder.domains}{\rm \ \ \ If $T$ is a provisional  preorder then $T{\MPldom{}}\ms{2}{=}\ms{2}T{\MPrdom{}}$.
\vspace{-7mm}
}%
\end{Lemma}%
\MPendBox

A trivial property that is nevertheless used frequently:
\begin{Lemma}\label{provisionalpreorder.conv}{\rm \ \ \  $T$  is  a provisional  preorder equivales $T^{\MPrev}$ 
is  a provisional  preorder.
\vspace{-7mm}
}%
\end{Lemma}%
\MPendBox

A preorder is a provisional preorder with left (equally right) domain equal to the identity relation.  In
other words, a preorder is a total provisional preorder.  It is easy to show that, for any relation $R$, the
relations $R{\setminus}R$ and $R{/}R$ are preorders.  It is also easy to show that $R$ is a preorder if and only if $R\ms{1}{=}\ms{1}R{\setminus}R$
(or equivalently if and only if $R\ms{1}{=}\ms{1}R{/}R$).  These properties generalise to provisional preorders. 

\begin{Lemma}\label{R/R.provisional}{\rm \ \ \ For all relations $R$, the relations $R{\MPrdom{}}\ms{1}{\MPcomp}\ms{1}R{\setminus}R$ and $R{/}R\ms{1}{\MPcomp}\ms{1}R{\MPldom{}}$ are provisional 
preorders.
\vspace{-7mm}
}%
\end{Lemma}%
\MPendBox 

{} 
\begin{Lemma}\label{ppreorder.left}{\rm \ \ \  $T$ is a provisional  preorder equivales\begin{displaymath}T\ms{6}{=}\ms{6}T{\MPldom{}}\ms{1}{\MPcomp}\ms{1}T{\setminus}T\ms{6}{=}\ms{6}T{/}T\ms{1}{\MPcomp}\ms{1}T{\MPrdom{}}\ms{6}{=}\ms{6}T{\MPldom{}}\ms{1}{\MPcomp}\ms{1}T{\setminus}T{/}T\ms{1}{\MPcomp}\ms{1}T{\MPrdom{}}~~.\end{displaymath}\vspace{-9mm}
}%
\end{Lemma}%
\MPendBox

Lemma \ref{ppreorder.left} is sometimes used in a form where the domains are replaced by per
domains.  
\begin{Lemma}\label{ppreorder.perleft}{\rm \ \ \ Suppose $T$ is a provisional  preorder.  Then\begin{displaymath}T\ms{6}{=}\ms{6}T{\MPperldom{}}\ms{1}{\MPcomp}\ms{1}T{\setminus}T\ms{6}{=}\ms{6}T{/}T\ms{1}{\MPcomp}\ms{1}T{\MPperrdom{}}\ms{6}{=}\ms{6}T{\MPperldom{}}\ms{1}{\MPcomp}\ms{1}T{\setminus}T{/}T\ms{1}{\MPcomp}\ms{1}T{\MPperrdom{}}~~.\end{displaymath}\vspace{-9mm}
}%
\end{Lemma}%
\MPendBox 

\begin{Lemma}\label{ppreorder.perdomain}{\rm \ \ \ Suppose $T$ is a provisional  preorder.  Then\begin{displaymath}T{\MPperldom{}}\ms{6}{=}\ms{6}T\ms{1}{\cap}\ms{1}T^{\MPrev}\ms{6}{=}\ms{6}T{\MPperrdom{}}~~.\end{displaymath}Hence $T\ms{1}{\cap}\ms{1}T^{\MPrev}$ is a per.
\vspace{-9mm}
}%
\end{Lemma}%
\MPendBox

Referring back to fig.\ \ref{ProvisionalPreorder}, the blue squares depict the equivalence classes of the
symmetric closure of a  provisional preorder.  As remarked earlier, the depicted relation is not a
preorder; correspondingly, the blue squares depict a truly \emph{partial} equivalence relation.  

We assume the reader is familiar with the notions of  an ordering and a linear (or total)
ordering.   We now extend these notions to provisional orderings.   (The at-most relation on the
integers is both anti-symmetric and linear.  The at-most relation restricted to some arbitrary 
subset of the  integers is an  example of a linear provisional ordering according to the definition below.)
\begin{Definition}\label{p.ordering}{\rm \ \ \ Suppose $T$  is a homogeneous relation  of type 
$A{\sim}A$, for some $A$.     Then $T$ is said to be \emph{provisionally anti-symmetric} if  \begin{displaymath}T\ms{1}{\cap}\ms{1}T^{\MPrev}\ms{3}{\subseteq}\ms{3}I_{A}~~.\end{displaymath}Also,  $T$  is said to be a  \emph{provisional  ordering}   if  $T$ is provisionally anti-symmetric and
 $T$ is a provisional preorder.  Finally, $T$  is said to be a  \emph{linear provisional  ordering}   if $T$ is a
provisional ordering and \begin{displaymath}T\ms{1}{\cup}\ms{1}T^{\MPrev}\ms{4}{=}\ms{4}(T\ms{1}{\cap}\ms{1}T^{\MPrev}){\MPcomp}{\MPplattop}{\MPcomp}(T\ms{1}{\cap}\ms{1}T^{\MPrev})~~.\end{displaymath}\vspace{-9mm}
}
\MPendBox\end{Definition}

Definition \ref{p.ordering} weakens the equality in the standard notion of anti-symmetry to an inclusion.  The
standard definition of  a partial ordering ---an anti-symmetric preorder--- is weakened accordingly (as
mentioned earlier, in order to ``provide'' for our needs).

The following lemma anticipates the use of provisional preorders/orderings in examples presented later. 
\begin{Lemma}\label{p.ordering.dom}{\rm \ \ \ Suppose $T$ is a provisional ordering.  Then\begin{displaymath}T{\MPldom{}}\ms{5}{=}\ms{5}T\ms{1}{\cap}\ms{1}T^{\MPrev}\ms{5}{=}\ms{5}T{\MPrdom{}}~~.\end{displaymath}\vspace{-9mm}
}%
\end{Lemma}%
\MPendBox

\subsection{Squares and  Rectangles}\label{Difun:Rectangles}

 We now introduce the notions  of a ``rectangle'' and a ``square''; rectangles  are typically heterogeneous
whilst squares are, by definition, homogeneous relations.  Squares are rectangles;  properties of squares
are typically  obtained by specialising properties of rectangles.  
\begin{Definition}[Rectangle and Square]\label{rectangle}{\rm \ \ \ A relation $R$ is a \emph{rectangle} iff  $R\ms{1}{=}\ms{1}R{\MPcomp}{\MPplattop}{\MPcomp}R$.  A relation $R$
is a \emph{square} iff $R$ is a symmetric rectangle.
}
\MPendBox\end{Definition}

It is easily shown that a rectangle is a difunction and a square is a per.

\subsection{Isomorphic Relations}\label{Isomorphic Relations}

The (yet-to-be-defined) cores and indexes of a given relation are not unique; in common mathematical
jargon, they are unique ``up to isomorphism''.  In order to make this precise, we need to define the
notion of isomorphic relation and establish a number of properties.
\begin{Definition}\label{rel.iso}{\rm \ \ \ Suppose $R$ and $S$ are two relations (not necessarily of the same type).  Then we say
that $R$ and $S$ are \emph{isomorphic} and write $R\ms{1}{\cong}\ms{1}S$ iff
\begin{mpdisplay}{0.15em}{6.5mm}{0mm}{2}
	\push$\langle\exists\ms{1}\phi{,}\psi$\=	\>\+\\
	$:$	\>\+$\phi\ms{1}{\MPcomp}\ms{1}\phi^{\MPrev}\ms{2}{=}\ms{2}R{\MPldom{}}\ms{5}{\wedge}\ms{5}\phi^{\MPrev}\ms{1}{\MPcomp}\ms{1}\phi\ms{2}{=}\ms{2}S{\MPldom{}}\ms{5}{\wedge}\ms{5}\psi\ms{1}{\MPcomp}\ms{1}\psi^{\MPrev}\ms{2}{=}\ms{2}R{\MPrdom{}}\ms{5}{\wedge}\ms{5}\psi^{\MPrev}\ms{1}{\MPcomp}\ms{1}\psi\ms{2}{=}\ms{2}S{\MPrdom{}}$\-\\
	$:$	\>\+$R\ms{2}{=}\ms{2}\phi\ms{1}{\MPcomp}\ms{1}S\ms{1}{\MPcomp}\ms{1}\psi^{\MPrev}$\-\-\\
	$\rangle$\pop$~~.$
\end{mpdisplay} 
\vspace{-9mm}
}
\MPendBox\end{Definition}

\section{Indexes and Core Relations}\label{Indices General}

This section introduces the notions of ``index'' and ``core'' of a relation and records some of their
properties. An ``index'' is a special case of a ``core'' of a relation but, in general, it is more useful.  For a
detailed account of their properties (including proofs)  see \cite{VB2022,VB2023a}.

\subsection{Definitions}\label{Indexes General}

The definition of an ``index'' of a relation is as follows.  
\begin{Definition}[Index]\label{gen.index}{\rm \ \ \ An \emph{index} of a relation $R$ is a relation $J$ that  has the following properties:
\begin{description}
\item[(a)]$J\ms{1}{\subseteq}\ms{1}R~~,$ 
\item[(b)]$R{\MPperldom{}}\ms{1}{\MPcomp}\ms{1}J\ms{1}{\MPcomp}\ms{1}R{\MPperrdom{}}\ms{3}{=}\ms{3}R~~,$ 
\item[(c)]$J{\MPldom{}}\ms{1}{\MPcomp}\ms{1}R{\MPperldom{}}\ms{1}{\MPcomp}\ms{1}J{\MPldom{}}\ms{3}{=}\ms{3}J{\MPldom{}}~~,$ 
\item[(d)]$J{\MPrdom{}}\ms{1}{\MPcomp}\ms{1}R{\MPperrdom{}}\ms{1}{\MPcomp}\ms{1}J{\MPrdom{}}\ms{3}{=}\ms{3}J{\MPrdom{}}~~.$ 
 
\end{description}
\vspace{-7mm}
}
\MPendBox\end{Definition}

Indexes are a special case of what we call ``core'' relations.   
\begin{Definition}[Core]\label{core}{\rm \ \ \ Suppose $R$ is an arbitrary relation and suppose $C$ is a relation
such that \begin{displaymath}C\ms{4}{=}\ms{4}\lambda\ms{1}{\MPcomp}\ms{1}R\ms{1}{\MPcomp}\ms{1}\rho^{\MPrev}\end{displaymath}for some relations $\lambda$ and $\rho$ satisfying  \begin{displaymath}R{\MPperldom{}}\ms{3}{=}\ms{3}\lambda^{\MPrev}\ms{1}{\MPcomp}\ms{1}\lambda\ms{7}{\wedge}\ms{7}\lambda{\MPldom{}}\ms{2}{=}\ms{2}\lambda\ms{1}{\MPcomp}\ms{1}\lambda^{\MPrev}\ms{7}{\wedge}\ms{7}R{\MPperrdom{}}\ms{3}{=}\ms{3}\rho^{\MPrev}\ms{1}{\MPcomp}\ms{1}\rho\ms{7}{\wedge}\ms{7}\rho{\MPldom{}}\ms{2}{=}\ms{2}\rho\ms{1}{\MPcomp}\ms{1}\rho^{\MPrev}~~.\end{displaymath}Then $C$ is said to be a \emph{core of}  $R$ \emph{as witnessed by} $\lambda$ \emph{and} $\rho$.
%\vspace{-9mm}
}
\MPendBox\end{Definition}

Note particularly requirement \ref{gen.index}(a).  A consequence of this requirement is that an index of a
relation has the same type as the relation, which is not necessarily the case for cores.  

An index $J$ of a  relation $R$  is a core of the relation as witnessed by $J{\MPldom{}}\ms{1}{\MPcomp}\ms{1}R{\MPperldom{}}$ and $J{\MPrdom{}}\ms{1}{\MPcomp}\ms{1}R{\MPperrdom{}}$.   

 The property that is common to cores and indexes  is captured by the following definition. 
\begin{Definition}[Core Relation]\label{Core.gen}{\rm \ \ \ A relation $R$ is a \emph{core relation}  iff $R{\MPldom{}}\ms{1}{=}\ms{1}R{\MPperldom{}}$ and $R{\MPrdom{}}\ms{1}{=}\ms{1}R{\MPperrdom{}}$.
\vspace{-5mm}
}
\MPendBox\end{Definition}

We exploit the fact that both indexes and cores satisfy definition \ref{Core.gen} later.  (The proof of this fact in
\cite{VB2022} assumes that $R$ has an index.  The ---much less straightforward--- proof without this
assumption is given in \cite{RCB2020}.)

%**** Lemma core.char removed and put in file corechar.mpd.  Not sure that it is useful.

A number of properties of cores and indexes are needed below.  Suppose $C$ is a core of $R$ as witnessed by
 $\lambda$ and $\rho$.   Then
\begin{equation}\label{core.equn}
R\ms{4}{=}\ms{4}\lambda^{\MPrev}\ms{1}{\MPcomp}\ms{1}C\ms{1}{\MPcomp}\ms{1}\rho~~.
\end{equation}\begin{equation}\label{core.domains}
R{\MPldom{}}\ms{2}{=}\ms{2}\lambda{\MPrdom{}}\ms{5}{\wedge}\ms{5}C{\MPldom{}}\ms{2}{=}\ms{2}\lambda{\MPldom{}}\ms{5}{\wedge}\ms{5}R{\MPrdom{}}\ms{2}{=}\ms{2}\rho{\MPrdom{}}\ms{5}{\wedge}\ms{5}C{\MPrdom{}}\ms{2}{=}\ms{2}\rho{\MPldom{}}~~.
\end{equation}Suppose $J$ is an index of $R$.  Then
\begin{equation}\label{R-perleft}
R{\MPperldom{}}\ms{1}{\MPcomp}\ms{1}J{\MPldom{}}\ms{1}{\MPcomp}\ms{1}R{\MPperldom{}}\ms{3}{=}\ms{3}R{\MPperldom{}}\ms{7}{\wedge}\ms{7}R{\MPperrdom{}}\ms{1}{\MPcomp}\ms{1}J{\MPrdom{}}\ms{1}{\MPcomp}\ms{1}R{\MPperrdom{}}\ms{3}{=}\ms{3}R{\MPperrdom{}}~~.
\end{equation}

\subsection{Indexes of Pers}\label{Indices and Pers}

A relation $R$ is a per iff $R\ms{1}{=}\ms{1}R{\MPperldom{}}\ms{1}{=}\ms{1}R{\MPperrdom{}}$.  Using this property, the definition of index can be simplified for pers.

\begin{Definition}[Index of a Per]\label{per.index}{\rm \ \ \ Suppose $P$ is a per.  Then a (\emph{coreflexive}) 
 \emph{index} of $P$ is a relation $J$ such  that 
\begin{description}
\item[(a)]$J\ms{1}{\subseteq}\ms{1}P{\MPldom{}}~~,$ 
\item[(b)]$J{\MPcomp}P{\MPcomp}J\ms{2}{=}\ms{2}J~~,$ 
\item[(c)]$P{\MPcomp}J{\MPcomp}P\ms{2}{=}\ms{2}P~~.$ 
 
\end{description}
\vspace{-7mm}
}
\MPendBox\end{Definition}

To our axiom system we add the  postulate that every per has a coreflexive index.  
We call this the \emph{axiom of choice}.
\begin{Axiom}[Axiom of Choice]\label{Axiom of Choice}{\rm \ \ \ Every per has a coreflexive  index.
}
\MPendBox\end{Axiom}

Assuming our axiom of choice, it follows that every relation has an index.  Specficially, we have:
\begin{Theorem}\label{per.to.relation}{\rm \ \ \ Suppose  $J$ and $K$ are (coreflexive)  indices of $R{\MPperldom{}}$ and
$R{\MPperrdom{}}$, respectively.  Then $J{\MPcomp}R{\MPcomp}K$ is an index of $R$.
}
\MPendBox\end{Theorem}

It is also the case that Freyd and \v{S}\v{c}edrov's  \cite{FRSC90} so-called ``splittings'' of pers always exist.
\begin{Theorem}\label{split.existence}{\rm \ \ \ If per $P$  has a coreflexive index $J$ , then\begin{displaymath}P\ms{4}{=}\ms{4}(J{\MPcomp}P)^{\MPrev}\ms{1}{\MPcomp}\ms{1}(J{\MPcomp}P)\ms{8}{\wedge}\ms{8}J\ms{4}{=}\ms{4}(J{\MPcomp}P)\ms{1}{\MPcomp}\ms{1}(J{\MPcomp}P)^{\MPrev}~~.\end{displaymath}Thus, assuming the  axiom of choice,   for all relations $R$,\begin{displaymath}\mathsf{per}{.}R\ms{5}{\equiv}\ms{5}{\left\langle\exists{}f\ms{3}{:}\ms{3}f\ms{1}{\MPcomp}\ms{1}f^{\MPrev}\ms{3}{=}\ms{3}f{\MPldom{}}\ms{3}{:}\ms{3}R\ms{2}{=}\ms{2}f^{\MPrev}\ms{1}{\MPcomp}\ms{1}f\right\rangle}~~.\end{displaymath}\vspace{-7mm}
}
\MPendBox\end{Theorem}

The property that $R$ is a difunction is equivalent to  $R{\MPperldom{}}\ms{2}{=}\ms{2}R\ms{1}{\MPcomp}\ms{1}R^{\MPrev}$  (and symmetrically to  $R{\MPperrdom{}}\ms{2}{=}\ms{2}R^{\MPrev}\ms{1}{\MPcomp}\ms{1}R$).
Also, since $R\ms{2}{=}\ms{2}R\ms{1}{\MPcomp}\ms{1}R^{\MPrev}\ms{1}{\MPcomp}\ms{1}R$,  the
definition of an index of a difunction can be restated as follows.
\begin{Definition}[Difunction Index]\label{difunction.index}{\rm \ \ \ An index  of a difunction $R$ is a relation $J$ that  has the following properties:
\begin{description}
\item[(a)]$J\ms{1}{\subseteq}\ms{1}R~~,$ 
\item[(b)]$R\ms{1}{\MPcomp}\ms{1}J^{\MPrev}\ms{1}{\MPcomp}\ms{1}R\ms{3}{=}\ms{3}R~~.$ 
\item[(c)]$J{\MPldom{}}\ms{1}{\MPcomp}\ms{1}R\ms{1}{\MPcomp}\ms{1}R^{\MPrev}\ms{1}{\MPcomp}\ms{1}J{\MPldom{}}\ms{3}{=}\ms{3}J{\MPldom{}}~~,$ 
\item[(d)]$J{\MPrdom{}}\ms{1}{\MPcomp}\ms{1}R^{\MPrev}\ms{1}{\MPcomp}\ms{1}R\ms{1}{\MPcomp}\ms{1}J{\MPrdom{}}\ms{3}{=}\ms{3}J{\MPrdom{}}~~.$ 
 
\end{description}
\vspace{-7mm}
}
\MPendBox\end{Definition}

In the same way that pers are characterised by a single function $f$ ---see theorem \ref{split.existence}---
difunctions are characterised by a pair of functions $f$ and $g$: 
\begin{Theorem}\label{fwokg}{\rm \ \ \ Assuming the axiom of choice (axiom \ref{Axiom of Choice}),  for all relations $R$,\begin{displaymath}\mathsf{difunction}{.}R\ms{5}{\equiv}\ms{5}{\left\langle\exists\ms{1}{}f{,}g\ms{3}{:}\ms{3}f\ms{1}{\MPcomp}\ms{1}f^{\MPrev}\ms{3}{=}\ms{3}f{\MPldom{}}\ms{3}{=}\ms{3}g\ms{1}{\MPcomp}\ms{1}g^{\MPrev}\ms{3}{=}\ms{3}g{\MPldom{}}\ms{3}{:}\ms{3}R\ms{2}{=}\ms{2}f^{\MPrev}\ms{1}{\MPcomp}\ms{1}g\right\rangle}~~.\end{displaymath}\vspace{-7mm}
}
\MPendBox\end{Theorem}

\section{The Diagonal}\label{difunctional.Diagonal}

This section  anticipates the study of block-ordered relations 
%in section  \ref{Block-Ordered Relations}.    
We
introduce the notion of  the ``diagonal'' of a relation in section  \ref{difunctional.Diagonal.defn} and formulate
some basic properties in section \ref{difunctional.Diagonal.props}.   

%In section \ref{Polar Coverings}, we introduced the notion of a  polar covering of a relation.  Theorem \ref{polar.all}
%shows how to construct a polar covering for any given relation but example \ref{lt.rf} demonstrates  that the
%construction does not always produce a non-redundant covering.   In section \ref{Non-Redundant Coverings},  
%we explore conditions under which  the diagonal of the relation guarantees the non-redundancy of the  
%covering. 

\subsection{Definition and Examples}\label{difunctional.Diagonal.defn}

Straightforwardly from the definition of factors, properties of converse  and set intersection,\begin{equation}\label{difun.delta}
R\mbox{ is difunctional\ms{8}}{\equiv}\ms{8}R\ms{3}{=}\ms{3}R\ms{1}{\cap}\ms{1}(R{\setminus}R{/}R)^{\MPrev}~~.
\end{equation}More generally, we have:
 \begin{Lemma}\label{BD.difun}{\rm \ \ \ For all $R$,   $R\ms{1}{\cap}\ms{1}(R{\setminus}R{/}R)^{\MPrev}$ is difunctional.
}%
\end{Lemma}%
{\bf Proof}~~~Let $S$ denote $R\ms{1}{\cap}\ms{1}(R{\setminus}R{/}R)^{\MPrev}$.  We have to prove that $S$ is difunctional.  That is, by definition,\begin{displaymath}S\ms{1}{\MPcomp}\ms{1}S^{\MPrev}\ms{1}{\MPcomp}\ms{1}S\ms{3}{\subseteq}\ms{3}S~~.\end{displaymath}Since the right side is an intersection, this is equivalent to \begin{displaymath}S\ms{1}{\MPcomp}\ms{1}S^{\MPrev}\ms{1}{\MPcomp}\ms{1}S\ms{3}{\subseteq}\ms{3}R\ms{6}{\wedge}\ms{6}S\ms{1}{\MPcomp}\ms{1}S^{\MPrev}\ms{1}{\MPcomp}\ms{1}S\ms{3}{\subseteq}\ms{3}(R{\setminus}R{/}R)^{\MPrev}~~.\end{displaymath}The first is (almost) trivial:
\begin{mpdisplay}{0.15em}{6.5mm}{0mm}{2}
	$S\ms{1}{\MPcomp}\ms{1}S^{\MPrev}\ms{1}{\MPcomp}\ms{1}S$\push\-\\
	$\subseteq$	\>	\>$\{$	\>\+\+\+$S\ms{1}{\subseteq}\ms{1}R$, $S\ms{1}{\subseteq}\ms{1}(R{\setminus}R{/}R)^{\MPrev}$, \\
	converse,  monotonicity\-\-$~~~ \}$\pop\\
	$R\ms{1}{\MPcomp}\ms{1}R{\setminus}R{/}R\ms{1}{\MPcomp}\ms{1}R$\push\-\\
	$\subseteq$	\>	\>$\{$	\>\+\+\+cancellation\-\-$~~~ \}$\pop\\
	$R~~.$
\end{mpdisplay}
In the above calculation, the trick was to replace the outer occurrences of $S$ on the left side by $R$ and the
middle occurrence by $(R{\setminus}R{/}R)^{\MPrev}$.  The replacement is done the opposite way around in the second
calculation.
\begin{mpdisplay}{0.15em}{6.5mm}{0mm}{2}
	$S\ms{1}{\MPcomp}\ms{1}S^{\MPrev}\ms{1}{\MPcomp}\ms{1}S\ms{3}{\subseteq}\ms{3}(R{\setminus}R{/}R)^{\MPrev}$\push\-\\
	$\Leftarrow$	\>	\>$\{$	\>\+\+\+$S\ms{1}{\subseteq}\ms{1}(R{\setminus}R{/}R)^{\MPrev}$, $S\ms{1}{\subseteq}\ms{1}R$,  monotonicity and transitivity\-\-$~~~ \}$\pop\\
	$(R{\setminus}R{/}R)^{\MPrev}\ms{1}{\MPcomp}\ms{1}R^{\MPrev}\ms{1}{\MPcomp}\ms{1}(R{\setminus}R{/}R)^{\MPrev}\ms{3}{\subseteq}\ms{3}(R{\setminus}R{/}R)^{\MPrev}$\push\-\\
	$=$	\>	\>$\{$	\>\+\+\+converse\-\-$~~~ \}$\pop\\
	$R{\setminus}R{/}R\ms{1}{\MPcomp}\ms{1}R\ms{1}{\MPcomp}\ms{1}R{\setminus}R{/}R\ms{3}{\subseteq}\ms{3}R{\setminus}R{/}R$\push\-\\
	$=$	\>	\>$\{$	\>\+\+\+Galois connection\-\-$~~~ \}$\pop\\
	$R\ms{1}{\MPcomp}\ms{1}R{\setminus}R{/}R\ms{1}{\MPcomp}\ms{1}R\ms{1}{\MPcomp}\ms{1}R{\setminus}R{/}R\ms{1}{\MPcomp}\ms{1}R\ms{3}{\subseteq}\ms{3}R$\push\-\\
	$=$	\>	\>$\{$	\>\+\+\+cancellation,  monotonicity and transitivity\-\-$~~~ \}$\pop\\
	$\mathsf{true}~~.$
\end{mpdisplay}
\vspace{-7mm}
\MPendBox

We call the relation $R\ms{1}{\cap}\ms{1}(R{\setminus}R{/}R)^{\MPrev}$ the
\emph{diagonal} of $R$;  Riguet \cite{Riguet51} calls it the ``diff\'{e}rence''.  (Riguet's definition does
not use factors but  is equivalent.; indeed, rewriting the definition using 
(\ref{underover.to.not}) the diagonal of $R$ is the ``diff\'{e}rence'' of $R$ and $R^{\MPrev}\ms{1}{\MPcomp}\ms{1}{\neg}R\ms{1}{\MPcomp}\ms{1}R^{\MPrev}$.)  
\begin{Definition}[Diagonal]\label{BD:diagonal}{\rm \ \ \ The \emph{diagonal} of relation $R$ is the relation $R\ms{1}{\cap}\ms{1}(R{\setminus}R{/}R)^{\MPrev}$. 
For brevity,  $R\ms{1}{\cap}\ms{1}(R{\setminus}R{/}R)^{\MPrev}$ will  be denoted by $\Delta{}R$.
}
\MPendBox\end{Definition}

Many readers will be familiar with the decomposition of a preorder into a partial ordering on a set of
equivalence classes.  The diagonal of a preorder $T$  is the  equivalence relation $T\ms{1}{\cap}\ms{1}T^{\MPrev}$.  More 
generally:

\begin{Example}\label{example.ppreorder}{\rm \ \ \ The diagonal of a provisional preorder $T$ is $T\ms{1}{\cap}\ms{1}T^{\MPrev}$.  This is because,
for an arbitrary relation $T$,  \begin{displaymath}T\ms{1}{\cap}\ms{1}(T{\setminus}T{/}T)^{\MPrev}\ms{7}{=}\ms{7}T\ms{3}{\cap}\ms{3}T{\MPldom{}}\ms{1}{\MPcomp}\ms{1}(T{\setminus}T{/}T)^{\MPrev}\ms{1}{\MPcomp}\ms{1}T{\MPrdom{}}~~.\end{displaymath}But,  if $T$ is a provisional preorder,  \begin{displaymath}T{\MPldom{}}\ms{1}{\MPcomp}\ms{1}(T{\setminus}T{/}T)^{\MPrev}\ms{1}{\MPcomp}\ms{1}T{\MPrdom{}}\ms{5}{=}\ms{5}T^{\MPrev}~~.\end{displaymath}(See lemmas \ref{provisionalpreorder.domains} and  \ref{ppreorder.left}.)
}%
\MPendBox\end{Example}
\begin{Example}\label{difun.graphs0}{\rm \ \ \ A particular instance of example \ref{example.ppreorder} is if   $G$ is the edge
relation of a finite  graph.  Then $\Delta(G^{*})$ is $G^{*}\ms{1}{\cap}\ms{1}(G^{\MPrev})^{*}$, the relation that holds between nodes $a$ and $b$ if there
is a path from $a$ to $b$ and a path from $b$ to $a$ in the graph.     Thus $\Delta(G^{*})$  is the equivalence relation that
holds between nodes that are in the same strongly  connected  component of $G.$
}%
\MPendBox\end{Example}

 \begin{Example}\label{diagonal.lessthan}{\rm \ \ \ In this example, we consider three versions of the less-than relation: the
homogeneous less-than relation on integers, which we denote by ${<}_{\MPInt}$,  the homogeneous less-than
relation  on real numbers, which we denote by ${<}_{\MPReal}$, and the heterogeneous less-than relation on integers
and real numbers, which we denote by ${}_{\MPInt}{<}_{\MPReal}$.    Specifically, the relation ${}_{\MPInt}{<}_{\MPReal}$ relates integer $m$ to real
number $x$ when  $m\ms{1}{<}\ms{1}x$  (using the conventional over-loaded notation).  We also subscript the at-most
symbol ${\leq}$ in the same way in order to indicate the type of the relation in question.

The diagonal of the less-than relation on integers is the predecessor
relation (i.e. it relates integer $m$ to integer $n$ exactly when $n\ms{1}{=}\ms{1}m{+}1$).   
This is because ${<}_{\MPInt}{\setminus}{<}_{\MPInt}\ms{2}{=}\ms{2}{\leq}_{\MPInt}$,   and ${\leq}_{\MPInt}{/}{<}_{\MPInt}$ relates integer $m$ to integer $n$ exactly when $m\ms{1}{\leq}_{\MPInt}\ms{1}n{+}1$ (where 
the subscript $\MPInt$ indicates the type of the ordering).  The diagonal is thus functional and injective.

The diagonal of the less-than  relation on real numbers is the empty relation.  This is because 
 ${<}_{\MPReal}{\setminus}{<}_{\MPReal}\ms{2}{=}\ms{2}{\leq}_{\MPReal}$,   ${\leq}_{\MPReal}{/}{<}_{\MPReal}\ms{2}{=}\ms{2}{\leq}_{\MPReal}$  and ${<}_{\MPReal}\ms{1}{\cap}\ms{1}{\geq}_{\MPReal}\ms{1}{=}\ms{1}{\MPplatbottom}_{\MPReal}$.  (Again, the subscript indicates the type of the ordering.)

The diagonal of the heterogeneous less-than relation ${}_{\MPInt}{<}_{\MPReal}$ relates integer $m$ to real number $x$ when
 $m\ms{1}{<}\ms{1}x\ms{1}{\leq}\ms{1}m{+}1$.  This is equivalent to $\lceil{}x\rceil\ms{1}{=}\ms{1}m{+}1$.   The diagonal  is thus a difunctional relation characterised 
by   ---in the sense of theorem \ref{fwokg}--- the  ceiling function ${\left\langle{}x\ms{1}{:}{:}\ms{1}\lceil{}x\rceil\right\rangle}$  and the successor function 
${\left\langle{}m\ms{2}{:}{:}\ms{2}m{+}1\right\rangle}$.  
%We leave the reader to check the details of this example.  
%See also  examples \ref{lt.rf} and \ref{heterogeneous.atmost}.
%and \ref{less-than},  and theorem \ref{staircase.neq.blockordered}.
}%
\MPendBox\end{Example}

The following example introduces a general mechanism for constructing illustrative examples of the
concepts introduced later.  The example exploits the observation that $\Delta{}R$ is injective if the preorder $R{\setminus}R$
is anti-symmetric; that is, $\Delta{}R$ is injective if $R{\setminus}R$ is a partial ordering.  (Equivalently, $\Delta{}R$ is functional if 
 $R{/}R$ is a partial ordering.)  We leave the straightforward proof to the reader.  
\begin{Example}\label{membership.diagonal}{\rm \ \ \ Suppose $\mathcal{X}$ is a finite type.  We use dummy $x$ to range over
elements of type $\mathcal{X}$.   Let $\mathcal{S}$ denote a subset of $2^{\mathcal{X}}$.   
Let $\mathsf{in}$ denote the membership relation of type $\mathcal{X}{\sim}\mathcal{S}$.  That is, if   $S$ is a subset of  $\mathcal{S}$,  $x{\MPcomp}{\MPplattop}{\MPcomp}S\ms{1}{\subseteq}\ms{1}\mathsf{in}$ exactly 
when $x$ is an element of the set $S$.   The relation $\mathsf{in}{\setminus}\mathsf{in}$ is the subset relation of type $\mathcal{S}{\sim}\mathcal{S}$.  

 (Conventionally,  $\mathsf{in}$ is denoted by the
symbol ``${\in}$'' and one writes $x{\in}S$ instead of $x{\MPcomp}{\MPplattop}{\MPcomp}S\ms{1}{\subseteq}\ms{1}\mathsf{in}$.  Also, the relation $\mathsf{in}{\setminus}\mathsf{in}$ is conventionally denoted by
the symbol ``${\subseteq}$''.  That is,  if $S$ and $S'$ are both elements of $\mathcal{S}$,   $S{\MPcomp}{\MPplattop}{\MPcomp}S'\ms{1}{\subseteq}\ms{1}\mathsf{in}{\setminus}\mathsf{in}$  exactly when $S\ms{1}{\subseteq}\ms{1}S'$.  
Were we to adopt conventional practice, the 
overloading  of the notation that occurs is  likely to cause confusion, so we choose to avoid it.)  

The relation $\mathsf{in}{\setminus}\mathsf{in}$ is anti-symmetric.  As a consequence, $\Delta\mathsf{in}$ is injective.  (Equivalently,  $(\Delta\mathsf{in})^{\MPrev}$ is
functional.)  Specifically, for all $x$ of type $\mathcal{X}$ and $S$ of type $\mathcal{S}$, \begin{displaymath}x{\MPcomp}{\MPplattop}{\MPcomp}S\ms{2}{\subseteq}\ms{2}\Delta\mathsf{in}\ms{6}{\equiv}\ms{6}x{\MPcomp}{\MPplattop}{\MPcomp}S\ms{1}{\subseteq}\ms{1}\mathsf{in}\ms{3}{\wedge}\ms{3}{\left\langle\forall{}S'\ms{2}{:}\ms{2}x{\MPcomp}{\MPplattop}{\MPcomp}S'\ms{1}{\subseteq}\ms{1}\mathsf{in}\ms{2}{:}\ms{2}S{\MPcomp}{\MPplattop}{\MPcomp}S'\ms{1}{\subseteq}\ms{1}\mathsf{in}{\setminus}\mathsf{in}\right\rangle}~~,\end{displaymath}where dummy $S'$ ranges over elements of $\mathcal{S}$.   Using conventional notation, the right side of this equation is 
recognised as the definition of a minimum, and  one might write\begin{displaymath}x\ms{2}\MPdopen\Delta\mathsf{in}\MPdclose\ms{2}{}S\ms{7}{\equiv}\ms{7}S\ms{2}\tuberight\ms{2}{\left\langle\mathsf{MIN}\ms{1}S'\ms{1}{:}\ms{1}x{\in}S'\ms{1}{:}\ms{1}S'\right\rangle}\end{displaymath}where  the venturi tube ``$\tuberight$'' indicates an equality assuming the well-definedness of the expression on its 
right side.

\begin{figure}[h]
\centering \includegraphics{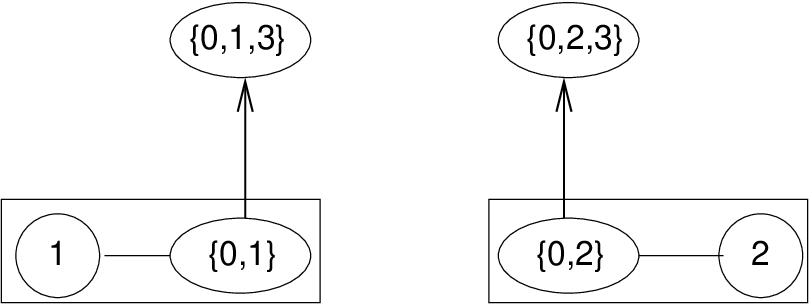}
 
\caption{Diagonal of an Instance of the Membership Relation}\label{membershipdiagFig}
\end{figure}

Fig.\ \ref{membershipdiagFig} shows a particular instance.  The set $\mathcal{X}$ is the set of numbers from $0$ to $3$.  The
set $\mathcal{S}$ is a subset of $2^{\{0{,}1{,}2{,}3\}}$; the chosen subset and the  relation $\mathsf{in}{\setminus}\mathsf{in}$ for this choice are depicted by  the
directed graph  forming the central portion of fig.\ \ref{membershipdiagFig}.  The relation $\Delta\mathsf{in}$ of type $\mathcal{X}\ms{1}{\sim}\ms{1}\mathcal{S}$ is
depicted by the undirected graph whereby  the atoms of the relation are depicted as rectangles.  Note
that the numbers  $0$ and $3$  are  not related by $\Delta\mathsf{in}$ to any of the elements of $\mathcal{S}$.  
%***** See example \ref{membership.analogie}    for  further discussion of this example.
}%
\MPendBox\end{Example}

\subsection{Basic Properties}\label{difunctional.Diagonal.props}

Primarily for notational convenience, we note a simple property of the diagonal:
\begin{Lemma}\label{diag.conv}{\rm \ \ \ \begin{displaymath}(\Delta{}R)^{\MPrev}\ms{4}{=}\ms{4}\Delta(R^{\MPrev})~~.\end{displaymath}
}%
\end{Lemma}%
{\bf Proof}~~~
\begin{mpdisplay}{0.15em}{6.5mm}{0mm}{2}
	$(\Delta{}R)^{\MPrev}$\push\-\\
	$=$	\>	\>$\{$	\>\+\+\+definition and distributivity\-\-$~~~ \}$\pop\\
	$R^{\MPrev}\ms{1}{\cap}\ms{1}R{\setminus}R{/}R$\push\-\\
	$=$	\>	\>$\{$	\>\+\+\+factors\-\-$~~~ \}$\pop\\
	$R^{\MPrev}\ms{1}{\cap}\ms{1}(R^{\MPrev}\ms{1}{\setminus}\ms{1}R^{\MPrev}\ms{1}{/}\ms{1}R^{\MPrev})^{\MPrev}$\push\-\\
	$=$	\>	\>$\{$	\>\+\+\+definition\-\-$~~~ \}$\pop\\
	$\Delta(R^{\MPrev})~~.$
\end{mpdisplay}
\vspace{-7mm}
\MPendBox

A consequence of lemma \ref{diag.conv} is that we can write $\Delta{}R^{\MPrev}$ without ambiguity.  This we do from now
on.

Very straightforwardly,  the relation  $R\ms{1}{\MPcomp}\ms{1}R^{\MPrev}$ is a per if $R$ is difunctional.  For a difunctional relation $R$,  
the relation $R\ms{1}{\MPcomp}\ms{1}R^{\MPrev}$ is the left per domain of $R$.  (Symmetrically, $R^{\MPrev}\ms{1}{\MPcomp}\ms{1}R$ is the right per
domain of $R$.  See theorem \ref{difunctional.strongdef}, parts (iv) and (vi).)
Thus $\Delta{}R\ms{1}{\MPcomp}\ms{1}(\Delta{}R)^{\MPrev}$ is the left per domain of  the diagonal of $R$.  
The following lemma is the basis of the construction,  in certain cases,  of an economic representation 
of the diagonal of $R$ and, hence,  of $R$ itself.  

%****See definition \ref{polar.covering} and  theorems \ref{difun.cover.refined}  and \ref{polar.nonred}. ****
\begin{Lemma}\label{per.difun.div}{\rm \ \ \ For all relations $R$, \begin{displaymath}(\Delta{}R){\MPperldom{}}\ms{5}{=}\ms{5}(\Delta{}R){\MPldom{}}\ms{1}{\MPcomp}\ms{1}R{\MPperldom{}}\ms{5}{=}\ms{5}R{\MPperldom{}}\ms{1}{\MPcomp}\ms{1}(\Delta{}R){\MPldom{}}~~.\end{displaymath}Dually,\begin{displaymath}(\Delta{}R){\MPperrdom{}}\ms{5}{=}\ms{5}R{\MPperrdom{}}\ms{1}{\MPcomp}\ms{1}(\Delta{}R){\MPrdom{}}\ms{5}{=}\ms{5}(\Delta{}R){\MPrdom{}}\ms{1}{\MPcomp}\ms{1}R{\MPperrdom{}}~~.\end{displaymath}
}%
\end{Lemma}%
{\bf Proof}~~~We prove the first equation by mutual inclusion.  First,  
\begin{mpdisplay}{0.15em}{6.5mm}{0mm}{2}
	$(\Delta{}R){\MPperldom{}}\ms{3}{\subseteq}\ms{3}(\Delta{}R){\MPldom{}}\ms{1}{\MPcomp}\ms{1}R{\MPperldom{}}$\push\-\\
	$=$	\>	\>$\{$	\>\+\+\+$\Delta{}R$ is difunctional, theorem \ref{difunctional.strongdef};  definition:  (\ref{per.rightdomain})\-\-$~~~ \}$\pop\\
	$\Delta{}R\ms{1}{\MPcomp}\ms{1}\Delta{}R^{\MPrev}\ms{3}{\subseteq}\ms{3}(\Delta{}R){\MPldom{}}\ms{1}{\MPcomp}\ms{1}R\leftsymdivision{}R$\push\-\\
	$\Leftarrow$	\>	\>$\{$	\>\+\+\+domains and monotonicity\-\-$~~~ \}$\pop\\
	$\Delta{}R\ms{1}{\MPcomp}\ms{1}\Delta{}R^{\MPrev}\ms{3}{\subseteq}\ms{3}R\leftsymdivision{}R$\push\-\\
	$=$	\>	\>$\{$	\>\+\+\+definition of $R\leftsymdivision{}R$,  converse and factors\-\-$~~~ \}$\pop\\
	$\Delta{}R\ms{1}{\MPcomp}\ms{1}\Delta{}R^{\MPrev}\ms{1}{\MPcomp}\ms{1}R\ms{3}{\subseteq}\ms{3}R$\push\-\\
	$=$	\>	\>$\{$	\>\+\+\+$\Delta{}R\ms{1}{\subseteq}\ms{1}R$;  $\Delta{}R^{\MPrev}\ms{1}{\subseteq}\ms{1}R{\setminus}R{/}R$ and cancellation\-\-$~~~ \}$\pop\\
	$\mathsf{true}~~.$
\end{mpdisplay}
Second,
\begin{mpdisplay}{0.15em}{6.5mm}{0mm}{2}
	$(\Delta{}R){\MPldom{}}\ms{1}{\MPcomp}\ms{1}R{\MPperldom{}}\ms{3}{\subseteq}\ms{3}(\Delta{}R){\MPperldom{}}$\push\-\\
	$=$	\>	\>$\{$	\>\+\+\+$\Delta{}R$ is difunctional,  theorem \ref{difunctional.strongdef}\-\-$~~~ \}$\pop\\
	$(\Delta{}R){\MPldom{}}\ms{1}{\MPcomp}\ms{1}R{\MPperldom{}}\ms{3}{\subseteq}\ms{3}\Delta{}R\ms{1}{\MPcomp}\ms{1}\Delta{}R^{\MPrev}$\push\-\\
	$\Leftarrow$	\>	\>$\{$	\>\+\+\+domains and definition:  (\ref{per.rightdomain})\-\-$~~~ \}$\pop\\
	$\Delta{}R\ms{1}{\MPcomp}\ms{1}\Delta{}R^{\MPrev}\ms{1}{\MPcomp}\ms{1}R\leftsymdivision{}R\ms{3}{\subseteq}\ms{3}\Delta{}R\ms{1}{\MPcomp}\ms{1}\Delta{}R^{\MPrev}$\push\-\\
	$\Leftarrow$	\>	\>$\{$	\>\+\+\+monotonicity and converse\-\-$~~~ \}$\pop\\
	$R\leftsymdivision{}R\ms{1}{\MPcomp}\ms{1}\Delta{}R\ms{3}{\subseteq}\ms{3}\Delta{}R$\push\-\\
	$=$	\>	\>$\{$	\>\+\+\+definition of diagonal\-\-$~~~ \}$\pop\\
	$R\leftsymdivision{}R\ms{1}{\MPcomp}\ms{1}\Delta{}R\ms{3}{\subseteq}\ms{3}R\ms{6}{\wedge}\ms{6}R\leftsymdivision{}R\ms{1}{\MPcomp}\ms{1}\Delta{}R\ms{3}{\subseteq}\ms{3}(R{\setminus}R{/}R)^{\MPrev}$\push\-\\
	$\Leftarrow$	\>	\>$\{$	\>\+\+\+$\Delta{}R\ms{1}{\subseteq}\ms{1}R$~;  converse \-\-$~~~ \}$\pop\\
	$R\leftsymdivision{}R\ms{1}{\MPcomp}\ms{1}R\ms{3}{\subseteq}\ms{3}R\ms{6}{\wedge}\ms{6}\Delta{}R^{\MPrev}\ms{1}{\MPcomp}\ms{1}R\leftsymdivision{}R\ms{3}{\subseteq}\ms{3}R{\setminus}R{/}R$\push\-\\
	$=$	\>	\>$\{$	\>\+\+\+cancellation; factors\-\-$~~~ \}$\pop\\
	$\mathsf{true}\ms{5}{\wedge}\ms{5}R\ms{1}{\MPcomp}\ms{1}\Delta{}R^{\MPrev}\ms{1}{\MPcomp}\ms{1}R\leftsymdivision{}R\ms{1}{\MPcomp}\ms{1}R\ms{3}{\subseteq}\ms{3}R$\push\-\\
	$\Leftarrow$	\>	\>$\{$	\>\+\+\+cancellation  and  $\Delta{}R^{\MPrev}\ms{1}{\subseteq}\ms{1}R{\setminus}R{/}R$\-\-$~~~ \}$\pop\\
	$R\ms{1}{\MPcomp}\ms{1}R{\setminus}R{/}R\ms{1}{\MPcomp}\ms{1}R\ms{3}{\subseteq}\ms{3}R$\push\-\\
	$=$	\>	\>$\{$	\>\+\+\+cancellation\-\-$~~~ \}$\pop\\
	$\mathsf{true}~~.$
\end{mpdisplay}
The remaining three  equalities are  simple consequences of the properties of converse, pers and
coreflexives.
%\vspace{-7mm}
\MPendBox

The following  corollary  of lemma \ref{per.difun.div} proves to be crucial later: 
%****see the discussion following  lemma \ref{BD.diagdom.if}.***
\begin{Lemma}\label{diag.dom.perdom}{\rm \ \ \ For all relations $R$,   \begin{displaymath}(\Delta{}R){\MPperldom{}}\ms{3}{=}\ms{3}R{\MPperldom{}}\ms{6}{\equiv}\ms{6}(\Delta{}R){\MPldom{}}\ms{2}{=}\ms{2}R{\MPldom{}}~~.\end{displaymath}Dually,\begin{displaymath}(\Delta{}R){\MPperrdom{}}\ms{3}{=}\ms{3}R{\MPperrdom{}}\ms{6}{\equiv}\ms{6}(\Delta{}R){\MPrdom{}}\ms{2}{=}\ms{2}R{\MPrdom{}}~~.\end{displaymath}
}%
\end{Lemma}%
{\bf Proof}~~~The proof is by mutual implication:
\begin{mpdisplay}{0.15em}{6.5mm}{0mm}{2}
	$(\Delta{}R){\MPldom{}}\ms{2}{=}\ms{2}R{\MPldom{}}$\push\-\\
	$\Rightarrow$	\>	\>$\{$	\>\+\+\+lemma \ref{per.difun.div} and Leibniz\-\-$~~~ \}$\pop\\
	$(\Delta{}R){\MPperldom{}}\ms{4}{=}\ms{4}R{\MPldom{}}\ms{1}{\MPcomp}\ms{1}R{\MPperldom{}}$\push\-\\
	$=$	\>	\>$\{$	\>\+\+\+dual of (\ref{per.rightdomain.doms})\-\-$~~~ \}$\pop\\
	$(\Delta{}R){\MPperldom{}}\ms{3}{=}\ms{3}R{\MPperldom{}}$\push\-\\
	$\Rightarrow$	\>	\>$\{$	\>\+\+\+Leibniz\-\-$~~~ \}$\pop\\
	$((\Delta{}R){\MPperldom{}}){\MPldom{}}\ms{3}{=}\ms{3}(R{\MPperldom{}}){\MPldom{}}$\push\-\\
	$=$	\>	\>$\{$	\>\+\+\+dual of (\ref{per.rightdomain.doms}) with $R\ms{1}{:=}\ms{1}\Delta{}R$ and $R\ms{1}{:=}\ms{1}R$\-\-$~~~ \}$\pop\\
	$(\Delta{}R){\MPldom{}}\ms{3}{=}\ms{3}R{\MPldom{}}~~.$
\end{mpdisplay}
\vspace{-7mm}
\MPendBox

\subsection{Reduction to the Core}\label{sec:Kernel}

In this section our goal is to prove that if $J$ is an index of relation $R$ then $\Delta{}J$ is an index of $\Delta{}R$. 
Instantiating definition \ref{difunction.index} with $J{,}R\ms{1}{:=}\ms{1}\Delta{}J{,}\Delta{}R$the properties we have to prove are as follows.

\begin{description}
\item[(a)]$\Delta{}J\ms{1}{\subseteq}\ms{1}\Delta{}R~~,$ 
\item[(b)]$\Delta{}R\ms{1}{\MPcomp}\ms{1}\Delta{}J^{\MPrev}\ms{1}{\MPcomp}\ms{1}\Delta{}R\ms{3}{=}\ms{3}\Delta{}R~~.$ 
\item[(c)]$(\Delta{}J){\MPldom{}}\ms{1}{\MPcomp}\ms{1}\Delta{}R\ms{1}{\MPcomp}\ms{1}\Delta{}R^{\MPrev}\ms{1}{\MPcomp}\ms{1}(\Delta{}J){\MPldom{}}\ms{3}{=}\ms{3}(\Delta{}J){\MPldom{}}~~,$ 
\item[(d)]$(\Delta{}J){\MPrdom{}}\ms{1}{\MPcomp}\ms{1}\Delta{}R^{\MPrev}\ms{1}{\MPcomp}\ms{1}\Delta{}R\ms{1}{\MPcomp}\ms{1}(\Delta{}J){\MPrdom{}}\ms{3}{=}\ms{3}(\Delta{}J){\MPrdom{}}~~.$ 
 
\end{description}
 
Of these, the hardest to prove is (b).  For properties (a), (c) and (d), all we need is that $J$ is an arbitrary
index of $R$.  For property (b), we use the fact that
an index of an arbitrary relation $R$ is defined to be $J{\MPcomp}R{\MPcomp}K$ where $J$ is an index of $R{\MPperldom{}}$ and $K$ is an index of 
$R{\MPperrdom{}}$.  

We begin with the easier properties.
\begin{Lemma}\label{diag.index(a)}{\rm \ \ \ Suppose $J$ is an index of $R$. Then \begin{displaymath}\Delta{}J\ms{1}{\subseteq}\ms{1}\Delta{}R~~.\end{displaymath}
}%
\end{Lemma}%
{\bf Proof}~~~
\begin{mpdisplay}{0.15em}{6.5mm}{0mm}{2}
	$\Delta{}J\ms{1}{\subseteq}\ms{1}\Delta{}R$\push\-\\
	$=$	\>	\>$\{$	\>\+\+\+definition \ref{BD:diagonal}\-\-$~~~ \}$\pop\\
	$J\ms{1}{\cap}\ms{1}(J{\setminus}J{/}J)^{\MPrev}\ms{4}{\subseteq}\ms{4}R\ms{1}{\cap}\ms{1}(R{\setminus}R{/}R)^{\MPrev}$\push\-\\
	$=$	\>	\>$\{$	\>\+\+\+domains\-\-$~~~ \}$\pop\\
	$J\ms{2}{\cap}\ms{2}J{\MPldom{}}\ms{1}{\MPcomp}\ms{1}(J{\setminus}J{/}J)^{\MPrev}\ms{1}{\MPcomp}\ms{1}J{\MPrdom{}}\ms{5}{\subseteq}\ms{5}R\ms{1}{\cap}\ms{1}(R{\setminus}R{/}R)^{\MPrev}$\push\-\\
	$\Leftarrow$	\>	\>$\{$	\>\+\+\+$J$ is an index of $R$, so $J\ms{1}{\subseteq}\ms{1}R$; monotonicity\-\-$~~~ \}$\pop\\
	$J{\MPldom{}}\ms{1}{\MPcomp}\ms{1}(J{\setminus}J{/}J)^{\MPrev}\ms{1}{\MPcomp}\ms{1}J{\MPrdom{}}\ms{4}{\subseteq}\ms{4}(R{\setminus}R{/}R)^{\MPrev}$\push\-\\
	$=$	\>	\>$\{$	\>\+\+\+converse\-\-$~~~ \}$\pop\\
	$J{\MPrdom{}}\ms{1}{\MPcomp}\ms{1}J{\setminus}J{/}J\ms{1}{\MPcomp}\ms{1}J{\MPldom{}}\ms{4}{\subseteq}\ms{4}R{\setminus}R{/}R$\push\-\\
	$=$	\>	\>$\{$	\>\+\+\+factors\-\-$~~~ \}$\pop\\
	$R\ms{1}{\MPcomp}\ms{1}J{\MPrdom{}}\ms{1}{\MPcomp}\ms{1}J{\setminus}J{/}J\ms{1}{\MPcomp}\ms{1}J{\MPldom{}}\ms{1}{\MPcomp}\ms{1}R\ms{4}{\subseteq}\ms{4}R$\push\-\\
	$=$	\>	\>$\{$	\>\+\+\+$J$ is an index of $R$, definition \ref{gen.index}(b); per domains\-\-$~~~ \}$\pop\\
	$R{\MPperldom{}}\ms{1}{\MPcomp}\ms{1}J\ms{1}{\MPcomp}\ms{1}R{\MPperrdom{}}\ms{1}{\MPcomp}\ms{1}J{\MPrdom{}}\ms{1}{\MPcomp}\ms{1}J{\setminus}J{/}J\ms{1}{\MPcomp}\ms{1}J{\MPldom{}}\ms{1}{\MPcomp}\ms{1}R{\MPperldom{}}\ms{1}{\MPcomp}\ms{1}J\ms{1}{\MPcomp}\ms{1}R{\MPperrdom{}}\ms{4}{\subseteq}\ms{4}R{\MPperldom{}}\ms{1}{\MPcomp}\ms{1}R\ms{1}{\MPcomp}\ms{1}R{\MPperrdom{}}$\push\-\\
	$\Leftarrow$	\>	\>$\{$	\>\+\+\+monotonicity\-\-$~~~ \}$\pop\\
	$J\ms{1}{\MPcomp}\ms{1}R{\MPperrdom{}}\ms{1}{\MPcomp}\ms{1}J{\MPrdom{}}\ms{1}{\MPcomp}\ms{1}J{\setminus}J{/}J\ms{1}{\MPcomp}\ms{1}J{\MPldom{}}\ms{1}{\MPcomp}\ms{1}R{\MPperldom{}}\ms{1}{\MPcomp}\ms{1}J\ms{4}{\subseteq}\ms{4}R~~.$
\end{mpdisplay}
Continuing with the left side of the inclusion:
\begin{mpdisplay}{0.15em}{6.5mm}{0mm}{2}
	$J\ms{1}{\MPcomp}\ms{1}R{\MPperrdom{}}\ms{1}{\MPcomp}\ms{1}J{\MPrdom{}}\ms{1}{\MPcomp}\ms{1}J{\setminus}J{/}J\ms{1}{\MPcomp}\ms{1}J{\MPldom{}}\ms{1}{\MPcomp}\ms{1}R{\MPperldom{}}\ms{1}{\MPcomp}\ms{1}J$\push\-\\
	$=$	\>	\>$\{$	\>\+\+\+domains\-\-$~~~ \}$\pop\\
	$J\ms{1}{\MPcomp}\ms{1}J{\MPrdom{}}\ms{1}{\MPcomp}\ms{1}R{\MPperrdom{}}\ms{1}{\MPcomp}\ms{1}J{\MPrdom{}}\ms{1}{\MPcomp}\ms{1}J{\setminus}J{/}J\ms{1}{\MPcomp}\ms{1}J{\MPldom{}}\ms{1}{\MPcomp}\ms{1}R{\MPperldom{}}\ms{1}{\MPcomp}\ms{1}J{\MPldom{}}\ms{1}{\MPcomp}\ms{1}J$\push\-\\
	$=$	\>	\>$\{$	\>\+\+\+$J$ is an index of $R$; definition \ref{gen.index}(c) and (d)\-\-$~~~ \}$\pop\\
	$J\ms{1}{\MPcomp}\ms{1}J{\MPrdom{}}\ms{1}{\MPcomp}\ms{1}J{\setminus}J{/}J\ms{1}{\MPcomp}\ms{1}J{\MPldom{}}\ms{1}{\MPcomp}\ms{1}J$\push\-\\
	$\subseteq$	\>	\>$\{$	\>\+\+\+domains and cancellation\-\-$~~~ \}$\pop\\
	$J$\push\-\\
	$\subseteq$	\>	\>$\{$	\>\+\+\+$J$ is an index of $R$; definition \ref{gen.index}(a)\-\-$~~~ \}$\pop\\
	$R~~.$
\end{mpdisplay}
\vspace{-7mm}
\MPendBox

\begin{Lemma}\label{diag.index(c)}{\rm \ \ \ Suppose $J$ is an index of $R$.   Then \begin{displaymath}(\Delta{}J){\MPldom{}}\ms{2}{\MPcomp}\ms{2}\Delta{}R\ms{2}{\MPcomp}\ms{2}\Delta{}R^{\MPrev}\ms{2}{\MPcomp}\ms{2}(\Delta{}J){\MPldom{}}\ms{6}{=}\ms{6}(\Delta{}J){\MPldom{}}~~.\end{displaymath}Dually, \begin{displaymath}(\Delta{}J){\MPrdom{}}\ms{2}{\MPcomp}\ms{2}\Delta{}R^{\MPrev}\ms{2}{\MPcomp}\ms{2}\Delta{}R\ms{2}{\MPcomp}\ms{2}(\Delta{}J){\MPrdom{}}\ms{6}{=}\ms{6}(\Delta{}J){\MPrdom{}}~~.\end{displaymath}
}%
\end{Lemma}%
{\bf Proof}~~~ 
\begin{mpdisplay}{0.15em}{6.5mm}{0mm}{2}
	$(\Delta{}J){\MPldom{}}\ms{2}{\MPcomp}\ms{2}\Delta{}R\ms{2}{\MPcomp}\ms{2}\Delta{}R^{\MPrev}\ms{2}{\MPcomp}\ms{2}(\Delta{}J){\MPldom{}}$\push\-\\
	$=$	\>	\>$\{$	\>\+\+\+$\Delta{}R$ is a difunction, theorem \ref{difunctional.strongdef}\-\-$~~~ \}$\pop\\
	$(\Delta{}J){\MPldom{}}\ms{2}{\MPcomp}\ms{2}(\Delta{}R){\MPperldom{}}\ms{2}{\MPcomp}\ms{2}(\Delta{}J){\MPldom{}}$\push\-\\
	$=$	\>	\>$\{$	\>\+\+\+lemma \ref{per.difun.div} (and symmetry)\-\-$~~~ \}$\pop\\
	$(\Delta{}J){\MPldom{}}\ms{2}{\MPcomp}\ms{2}(\Delta{}R){\MPldom{}}\ms{1}{\MPcomp}\ms{1}R{\MPperldom{}}\ms{1}{\MPcomp}\ms{1}(\Delta{}R){\MPldom{}}\ms{2}{\MPcomp}\ms{2}(\Delta{}J){\MPldom{}}$\push\-\\
	$=$	\>	\>$\{$	\>\+\+\+by lemma \ref{diag.index(a)} and monotonicity,  $(\Delta{}J){\MPldom{}}\ms{1}{\subseteq}\ms{1}(\Delta{}R){\MPldom{}}$\-\-$~~~ \}$\pop\\
	$(\Delta{}J){\MPldom{}}\ms{1}{\MPcomp}\ms{1}R{\MPperldom{}}\ms{1}{\MPcomp}\ms{1}(\Delta{}J){\MPldom{}}$\push\-\\
	$=$	\>	\>$\{$	\>\+\+\+$(\Delta{}J){\MPldom{}}\ms{1}{\subseteq}\ms{1}J{\MPldom{}}$ (since $\Delta{}J\ms{1}{\subseteq}\ms{1}J$)\-\-$~~~ \}$\pop\\
	$(\Delta{}J){\MPldom{}}\ms{1}{\MPcomp}\ms{1}J{\MPldom{}}\ms{1}{\MPcomp}\ms{1}R{\MPperldom{}}\ms{1}{\MPcomp}\ms{1}J{\MPldom{}}\ms{1}{\MPcomp}\ms{1}(\Delta{}J){\MPldom{}}$\push\-\\
	$=$	\>	\>$\{$	\>\+\+\+$J$ is an index of $R$,  definition \ref{gen.index}(c)\-\-$~~~ \}$\pop\\
	$(\Delta{}J){\MPldom{}}\ms{1}{\MPcomp}\ms{1}J{\MPldom{}}\ms{1}{\MPcomp}\ms{1}(\Delta{}J){\MPldom{}}$\push\-\\
	$=$	\>	\>$\{$	\>\+\+\+$(\Delta{}J){\MPldom{}}\ms{1}{\subseteq}\ms{1}J{\MPldom{}}$ (since $\Delta{}J\ms{1}{\subseteq}\ms{1}J$)\-\-$~~~ \}$\pop\\
	$(\Delta{}J){\MPldom{}}~~.$
\end{mpdisplay}
\vspace{-7mm}
\MPendBox

In order to prove (b), we prove a more general theorem on cores.  First, a lemma: 
\begin{Lemma}\label{core.potential}{\rm \ \ \ Suppose $R$,  $C$,   $\lambda$ and $\rho$ are as in definition  \ref{core}.  Then  \begin{displaymath}R{\MPrdom{}}\ms{1}{\MPcomp}\ms{1}R{\setminus}R{/}R\ms{1}{\MPcomp}\ms{1}R{\MPldom{}}\ms{5}{=}\ms{5}\rho^{\MPrev}\ms{1}{\MPcomp}\ms{1}C{\setminus}C{/}C\ms{1}{\MPcomp}\ms{1}\lambda~~.\end{displaymath}
}%
\end{Lemma}%
{\bf Proof}~~~ 
\begin{mpdisplay}{0.15em}{6.5mm}{0mm}{2}
	$R{\MPrdom{}}\ms{1}{\MPcomp}\ms{1}R{\setminus}R{/}R\ms{1}{\MPcomp}\ms{1}R{\MPldom{}}$\push\-\\
	$=$	\>	\>$\{$	\>\+\+\+(\ref{per.rightdomain.doms})\-\-$~~~ \}$\pop\\
	$(R{\MPperrdom{}}){\MPrdom{}}\ms{1}{\MPcomp}\ms{1}R{\setminus}R{/}R\ms{1}{\MPcomp}\ms{1}(R{\MPperldom{}}){\MPldom{}}$\push\-\\
	$=$	\>	\>$\{$	\>\+\+\+$R{\MPperldom{}}\ms{3}{=}\ms{3}\lambda^{\MPrev}\ms{1}{\MPcomp}\ms{1}\lambda$,   $R{\MPperrdom{}}\ms{3}{=}\ms{3}\rho^{\MPrev}\ms{1}{\MPcomp}\ms{1}\rho$, and domains\-\-$~~~ \}$\pop\\
	$\rho{\MPrdom{}}\ms{1}{\MPcomp}\ms{1}R{\setminus}R{/}R\ms{1}{\MPcomp}\ms{1}\lambda{\MPrdom{}}$\push\-\\
	$=$	\>	\>$\{$	\>\+\+\+(\ref{core.equn})\-\-$~~~ \}$\pop\\
	$\rho{\MPrdom{}}\ms{1}{\MPcomp}\ms{1}(\lambda^{\MPrev}\ms{1}{\MPcomp}\ms{1}C\ms{1}{\MPcomp}\ms{1}\rho){\setminus}(\lambda^{\MPrev}\ms{1}{\MPcomp}\ms{1}C\ms{1}{\MPcomp}\ms{1}\rho){/}(\lambda^{\MPrev}\ms{1}{\MPcomp}\ms{1}C\ms{1}{\MPcomp}\ms{1}\rho)\ms{1}{\MPcomp}\ms{1}\lambda{\MPrdom{}}$\push\-\\
	$=$	\>	\>$\{$	\>\+\+\+lemma \ref{f.under.over.g} with $f{,}g{,}U{,}V{,}W\ms{1}{:=}\ms{1}\rho{,}\lambda{,}C{,}C{,}C$\-\-$~~~ \}$\pop\\
	$\rho^{\MPrev}\ms{1}{\MPcomp}\ms{1}(\lambda{\MPldom{}}\ms{1}{\MPcomp}\ms{1}C){\setminus}C{/}(C\ms{1}{\MPcomp}\ms{1}\rho{\MPldom{}})\ms{1}{\MPcomp}\ms{1}\lambda$\push\-\\
	$=$	\>	\>$\{$	\>\+\+\+$C\ms{2}{=}\ms{2}\lambda\ms{1}{\MPcomp}\ms{1}R\ms{1}{\MPcomp}\ms{1}\rho^{\MPrev}$; so $\lambda{\MPldom{}}\ms{1}{\MPcomp}\ms{1}C\ms{2}{=}\ms{2}C\ms{2}{=}\ms{2}C\ms{1}{\MPcomp}\ms{1}\rho{\MPldom{}}$\-\-$~~~ \}$\pop\\
	$\rho^{\MPrev}\ms{1}{\MPcomp}\ms{1}C{\setminus}C{/}C\ms{1}{\MPcomp}\ms{1}\lambda~~.$
\end{mpdisplay}
\vspace{-7mm}
\MPendBox

\begin{Theorem}\label{diagonal.core}{\rm \ \ \ Suppose $R$, $C$,   $\lambda$ and $\rho$ are as in definition \ref{core}.  Then \begin{displaymath}\Delta{}R\ms{4}{=}\ms{4}\lambda^{\MPrev}\ms{1}{\MPcomp}\ms{1}\Delta{}C\ms{1}{\MPcomp}\ms{1}\rho\ms{8}{\wedge}\ms{8}\Delta{}C\ms{4}{=}\ms{4}\lambda\ms{1}{\MPcomp}\ms{1}\Delta{}R\ms{1}{\MPcomp}\ms{1}\rho^{\MPrev}~~.\end{displaymath}In  words, if $\lambda$ and $\rho$ witness that $C$ is a core of $R$, then $\lambda$ and $\rho$ witness that $\Delta{}C$ is a core of $\Delta{}R$.
}
\end{Theorem}
{\bf Proof}~~~    
\begin{mpdisplay}{0.15em}{6.5mm}{0mm}{2}
	$\Delta{}R$\push\-\\
	$=$	\>	\>$\{$	\>\+\+\+definition \-\-$~~~ \}$\pop\\
	$R\ms{1}{\cap}\ms{1}(R{\setminus}R{/}R)^{\MPrev}$\push\-\\
	$=$	\>	\>$\{$	\>\+\+\+domains and converse\-\-$~~~ \}$\pop\\
	$R\ms{3}{\cap}\ms{3}(R{\MPrdom{}}\ms{1}{\MPcomp}\ms{1}R{\setminus}R{/}R\ms{1}{\MPcomp}\ms{1}R{\MPldom{}})^{\MPrev}$\push\-\\
	$=$	\>	\>$\{$	\>\+\+\+lemma \ref{core.potential}\-\-$~~~ \}$\pop\\
	$R\ms{3}{\cap}\ms{3}(\rho^{\MPrev}\ms{1}{\MPcomp}\ms{1}C{\setminus}C{/}C\ms{1}{\MPcomp}\ms{1}\lambda)^{\MPrev}$\push\-\\
	$=$	\>	\>$\{$	\>\+\+\+straightforwardly)  $R\ms{4}{=}\ms{4}\lambda^{\MPrev}\ms{1}{\MPcomp}\ms{1}C\ms{1}{\MPcomp}\ms{1}\rho$\-\-$~~~ \}$\pop\\
	$\lambda^{\MPrev}\ms{1}{\MPcomp}\ms{1}C\ms{1}{\MPcomp}\ms{1}\rho\ms{4}{\cap}\ms{4}(\rho^{\MPrev}\ms{1}{\MPcomp}\ms{1}C{\setminus}C{/}C\ms{1}{\MPcomp}\ms{1}\lambda)^{\MPrev}$\push\-\\
	$=$	\>	\>$\{$	\>\+\+\+distributivity of converse and  functional relations\-\-$~~~ \}$\pop\\
	$\lambda^{\MPrev}\ms{1}{\MPcomp}\ms{1}(C\ms{1}{\cap}\ms{1}(C{\setminus}C{/}C)^{\MPrev})\ms{1}{\MPcomp}\ms{1}\rho$\push\-\\
	$=$	\>	\>$\{$	\>\+\+\+definition \ref{BD:diagonal}\-\-$~~~ \}$\pop\\
	$\lambda^{\MPrev}\ms{1}{\MPcomp}\ms{1}\Delta{}C\ms{1}{\MPcomp}\ms{1}\rho~~.$
\end{mpdisplay}
Hence
\begin{mpdisplay}{0.15em}{6.5mm}{0mm}{2}
	$\lambda\ms{1}{\MPcomp}\ms{1}\Delta{}R\ms{1}{\MPcomp}\ms{1}\rho^{\MPrev}$\push\-\\
	$=$	\>	\>$\{$	\>\+\+\+above\-\-$~~~ \}$\pop\\
	$\lambda\ms{1}{\MPcomp}\ms{1}\lambda^{\MPrev}\ms{1}{\MPcomp}\ms{1}\Delta{}C\ms{1}{\MPcomp}\ms{1}\rho\ms{1}{\MPcomp}\ms{1}\rho^{\MPrev}$\push\-\\
	$=$	\>	\>$\{$	\>\+\+\+$\lambda$ and $\rho$ are functional\-\-$~~~ \}$\pop\\
	$\lambda{\MPldom{}}\ms{1}{\MPcomp}\ms{1}\Delta{}C\ms{1}{\MPcomp}\ms{1}\rho{\MPldom{}}$\push\-\\
	$=$	\>	\>$\{$	\>\+\+\+$\Delta{}C\ms{1}{\subseteq}\ms{1}C$;  so  $(\Delta{}C){\MPldom{}}\ms{2}{\subseteq}\ms{2}C{\MPldom{}}$ and $(\Delta{}C){\MPrdom{}}\ms{2}{\subseteq}\ms{2}C{\MPrdom{}}$\\
	 (\ref{core.domains}) and domains\-\-$~~~ \}$\pop\\
	$\Delta{}C~~.$
\end{mpdisplay}
\vspace{-9mm}
\MPendBox

We are now in a position to prove the final property (b) above.
\begin{Lemma}\label{diag.index(b)}{\rm \ \ \ Suppose $J$ is an index of $R$.   Then \begin{displaymath}\Delta{}R\ms{1}{\MPcomp}\ms{1}\Delta{}J^{\MPrev}\ms{1}{\MPcomp}\ms{1}\Delta{}R\ms{3}{=}\ms{3}\Delta{}R~~.\end{displaymath}
}%
\end{Lemma}%
{\bf Proof}~~~We begin by noting that theorem  \ref{diagonal.core} applies with $C$ instantiated to $J$ and 
 $\lambda$ and $\rho$ defined by $\lambda\ms{2}{=}\ms{2}J{\MPldom{}}\ms{1}{\MPcomp}\ms{1}R{\MPperldom{}}$ and $\rho\ms{2}{=}\ms{2}J{\MPrdom{}}\ms{1}{\MPcomp}\ms{1}R{\MPperrdom{}}$.  This is because $J$ is a core of $R$.  So 
\begin{mpdisplay}{0.15em}{6.5mm}{0mm}{2}
	$\Delta{}R\ms{1}{\MPcomp}\ms{1}\Delta{}J^{\MPrev}\ms{1}{\MPcomp}\ms{1}\Delta{}R$\push\-\\
	$=$	\>	\>$\{$	\>\+\+\+theorem  \ref{diagonal.core} with $C{,}\lambda{,}\rho\ms{3}{:=}\ms{3}J\ms{2}{,}\ms{2}J{\MPldom{}}\ms{1}{\MPcomp}\ms{1}R{\MPperldom{}}\ms{2}{,}\ms{2}J{\MPrdom{}}\ms{1}{\MPcomp}\ms{1}R{\MPperrdom{}}$\-\-$~~~ \}$\pop\\
	$\Delta{}R\ms{2}{\MPcomp}\ms{2}(\lambda\ms{2}{\MPcomp}\ms{2}\Delta{}R\ms{2}{\MPcomp}\ms{2}\rho^{\MPrev})^{\MPrev}\ms{2}{\MPcomp}\ms{2}\Delta{}R$\push\-\\
	$=$	\>	\>$\{$	\>\+\+\+converse\-\-$~~~ \}$\pop\\
	$\Delta{}R\ms{2}{\MPcomp}\ms{2}\rho\ms{2}{\MPcomp}\ms{2}\Delta{}R^{\MPrev}\ms{2}{\MPcomp}\ms{2}\lambda^{\MPrev}\ms{2}{\MPcomp}\ms{2}\Delta{}R$\push\-\\
	$=$	\>	\>$\{$	\>\+\+\+definition of $\rho$ and $\lambda$,  $(J{\MPldom{}}\ms{1}{\MPcomp}\ms{1}R{\MPperldom{}})^{\MPrev}\ms{2}{=}\ms{2}R{\MPperldom{}}\ms{1}{\MPcomp}\ms{1}J{\MPldom{}}$\-\-$~~~ \}$\pop\\
	$\Delta{}R\ms{2}{\MPcomp}\ms{2}J{\MPrdom{}}\ms{1}{\MPcomp}\ms{1}R{\MPperrdom{}}\ms{2}{\MPcomp}\ms{2}\Delta{}R^{\MPrev}\ms{2}{\MPcomp}\ms{2}R{\MPperldom{}}\ms{1}{\MPcomp}\ms{1}J{\MPldom{}}\ms{2}{\MPcomp}\ms{2}\Delta{}R$\push\-\\
	$=$	\>	\>$\{$	\>\+\+\+per domains\-\-$~~~ \}$\pop\\
	$\Delta{}R\ms{1}{\MPcomp}\ms{1}(\Delta{}R){\MPperrdom{}}\ms{2}{\MPcomp}\ms{2}J{\MPrdom{}}\ms{1}{\MPcomp}\ms{1}R{\MPperrdom{}}\ms{2}{\MPcomp}\ms{2}\Delta{}R^{\MPrev}\ms{2}{\MPcomp}\ms{2}R{\MPperldom{}}\ms{1}{\MPcomp}\ms{1}J{\MPldom{}}\ms{1}{\MPcomp}\ms{1}(\Delta{}R){\MPperldom{}}\ms{2}{\MPcomp}\ms{2}\Delta{}R$\push\-\\
	$=$	\>	\>$\{$	\>\+\+\+ lemma \ref{per.difun.div}\-\-$~~~ \}$\pop\\
	$\Delta{}R\ms{1}{\MPcomp}\ms{1}(\Delta{}R){\MPrdom{}}\ms{1}{\MPcomp}\ms{1}R{\MPperrdom{}}\ms{2}{\MPcomp}\ms{2}J{\MPrdom{}}\ms{1}{\MPcomp}\ms{1}R{\MPperrdom{}}\ms{2}{\MPcomp}\ms{2}\Delta{}R^{\MPrev}\ms{2}{\MPcomp}\ms{2}R{\MPperldom{}}\ms{1}{\MPcomp}\ms{1}J{\MPldom{}}\ms{1}{\MPcomp}\ms{1}R{\MPperldom{}}\ms{1}{\MPcomp}\ms{1}(\Delta{}R){\MPldom{}}\ms{2}{\MPcomp}\ms{2}\Delta{}R$\push\-\\
	$=$	\>	\>$\{$	\>\+\+\+lemma \ref{R-perleft}\-\-$~~~ \}$\pop\\
	$\Delta{}R\ms{1}{\MPcomp}\ms{1}(\Delta{}R){\MPrdom{}}\ms{1}{\MPcomp}\ms{1}R{\MPperrdom{}}\ms{2}{\MPcomp}\ms{2}\Delta{}R^{\MPrev}\ms{2}{\MPcomp}\ms{2}R{\MPperldom{}}\ms{1}{\MPcomp}\ms{1}(\Delta{}R){\MPldom{}}\ms{2}{\MPcomp}\ms{2}\Delta{}R$\push\-\\
	$=$	\>	\>$\{$	\>\+\+\+lemma \ref{per.difun.div}\-\-$~~~ \}$\pop\\
	$\Delta{}R\ms{1}{\MPcomp}\ms{1}(\Delta{}R){\MPperrdom{}}\ms{2}{\MPcomp}\ms{2}\Delta{}R^{\MPrev}\ms{2}{\MPcomp}\ms{2}(\Delta{}R){\MPperldom{}}\ms{2}{\MPcomp}\ms{2}\Delta{}R$\push\-\\
	$=$	\>	\>$\{$	\>\+\+\+per domains\-\-$~~~ \}$\pop\\
	$\Delta{}R\ms{2}{\MPcomp}\ms{2}\Delta{}R^{\MPrev}\ms{2}{\MPcomp}\ms{2}\Delta{}R$\push\-\\
	$=$	\>	\>$\{$	\>\+\+\+$\Delta{}R$ is difunctional, theorem \ref{difunctional.strongdef}\-\-$~~~ \}$\pop\\
	$\Delta{}R~~.$
\end{mpdisplay}
\vspace{-7mm}
\MPendBox

Putting all the lemmas together, we have:
\begin{Theorem}\label{diag.index.is.index}{\rm \ \ \ Suppose $J$ is an index of $R$.   Then 
 $\Delta{}J$ is an index of $\Delta{}R$.
}
\end{Theorem}
{\bf Proof}~~~Lemmas \ref{diag.index(a)},  \ref{diag.index(c)} and 
\ref{diag.index(b)}  combined with definition \ref{difunction.index} (instantiated with $J{,}R\ms{1}{:=}\ms{1}\Delta{}J{,}\Delta{}R$).
\MPendBox

We conclude with a beautiful theorem.  

\begin{Theorem}\label{deltaJdeltaR}{\rm \ \ \ Suppose $J$ is an index of $R$.   Then\begin{displaymath}\Delta{}J\ms{3}{=}\ms{3}J{\MPldom{}}\ms{1}{\MPcomp}\ms{1}\Delta{}R\ms{1}{\MPcomp}\ms{1}J{\MPrdom{}}\ms{9}{\wedge}\ms{9}\Delta{}R\ms{3}{=}\ms{3}R{\MPperldom{}}\ms{1}{\MPcomp}\ms{1}\Delta{}J\ms{1}{\MPcomp}\ms{1}R{\MPperrdom{}}~~.\end{displaymath}
}
\end{Theorem}
{\bf Proof}~~~We first prove, by mutual implication, that the two equations are equivalent.  Assume that\begin{displaymath}\Delta{}R\ms{4}{=}\ms{4}R{\MPperldom{}}\ms{1}{\MPcomp}\ms{1}\Delta{}J\ms{1}{\MPcomp}\ms{1}R{\MPperrdom{}}~~.\end{displaymath}Then,
\begin{mpdisplay}{0.15em}{6.5mm}{0mm}{2}
	$J{\MPldom{}}\ms{1}{\MPcomp}\ms{1}\Delta{}R\ms{1}{\MPcomp}\ms{1}J{\MPrdom{}}$\push\-\\
	$=$	\>	\>$\{$	\>\+\+\+assumption\-\-$~~~ \}$\pop\\
	$J{\MPldom{}}\ms{1}{\MPcomp}\ms{1}R{\MPperldom{}}\ms{1}{\MPcomp}\ms{1}\Delta{}J\ms{1}{\MPcomp}\ms{1}R{\MPperrdom{}}\ms{1}{\MPcomp}\ms{1}J{\MPrdom{}}$\push\-\\
	$=$	\>	\>$\{$	\>\+\+\+$\Delta{}J\ms{1}{\subseteq}\ms{1}J$, so $(\Delta{}J){\MPldom{}}\ms{1}{\subseteq}\ms{1}J{\MPldom{}}$ and $(\Delta{}J){\MPrdom{}}\ms{1}{\subseteq}\ms{1}J{\MPrdom{}}$; domains\-\-$~~~ \}$\pop\\
	$J{\MPldom{}}\ms{1}{\MPcomp}\ms{1}R{\MPperldom{}}\ms{1}{\MPcomp}\ms{1}J{\MPldom{}}\ms{1}{\MPcomp}\ms{1}\Delta{}J\ms{1}{\MPcomp}\ms{1}J{\MPrdom{}}\ms{1}{\MPcomp}\ms{1}R{\MPperrdom{}}\ms{1}{\MPcomp}\ms{1}J{\MPrdom{}}$\push\-\\
	$=$	\>	\>$\{$	\>\+\+\+$J$ is an index of $R$,  definition \ref{gen.index}(c) and (d)\-\-$~~~ \}$\pop\\
	$J{\MPldom{}}\ms{1}{\MPcomp}\ms{1}\Delta{}J\ms{1}{\MPcomp}\ms{1}J{\MPrdom{}}$\push\-\\
	$=$	\>	\>$\{$	\>\+\+\+reverse of middle step\-\-$~~~ \}$\pop\\
	$\Delta{}J~~.$
\end{mpdisplay}
Conversely, assume\begin{displaymath}\Delta{}J\ms{2}{=}\ms{2}J{\MPldom{}}\ms{1}{\MPcomp}\ms{1}\Delta{}R\ms{1}{\MPcomp}\ms{1}J{\MPrdom{}}~~.\end{displaymath}Then,
\begin{mpdisplay}{0.15em}{6.5mm}{0mm}{2}
	$R{\MPperldom{}}\ms{1}{\MPcomp}\ms{1}\Delta{}J\ms{1}{\MPcomp}\ms{1}R{\MPperrdom{}}$\push\-\\
	$=$	\>	\>$\{$	\>\+\+\+assumption\-\-$~~~ \}$\pop\\
	$R{\MPperldom{}}\ms{1}{\MPcomp}\ms{1}J{\MPldom{}}\ms{1}{\MPcomp}\ms{1}\Delta{}R\ms{1}{\MPcomp}\ms{1}J{\MPrdom{}}\ms{1}{\MPcomp}\ms{1}R{\MPperrdom{}}$\push\-\\
	$=$	\>	\>$\{$	\>\+\+\+lemma \ref{per.difun.div}\-\-$~~~ \}$\pop\\
	$R{\MPperldom{}}\ms{1}{\MPcomp}\ms{1}J{\MPldom{}}\ms{1}{\MPcomp}\ms{1}(\Delta{}R){\MPldom{}}\ms{1}{\MPcomp}\ms{1}R{\MPperldom{}}\ms{1}{\MPcomp}\ms{1}\Delta{}R\ms{1}{\MPcomp}\ms{1}R{\MPperrdom{}}\ms{1}{\MPcomp}\ms{1}(\Delta{}R){\MPrdom{}}\ms{1}{\MPcomp}\ms{1}J{\MPrdom{}}\ms{1}{\MPcomp}\ms{1}R{\MPperrdom{}}$\push\-\\
	$=$	\>	\>$\{$	\>\+\+\+lemma \ref{diag.index(a)} and domains\-\-$~~~ \}$\pop\\
	$R{\MPperldom{}}\ms{1}{\MPcomp}\ms{1}J{\MPldom{}}\ms{1}{\MPcomp}\ms{1}R{\MPperldom{}}\ms{1}{\MPcomp}\ms{1}\Delta{}R\ms{1}{\MPcomp}\ms{1}R{\MPperrdom{}}\ms{1}{\MPcomp}\ms{1}J{\MPrdom{}}\ms{1}{\MPcomp}\ms{1}R{\MPperrdom{}}$\push\-\\
	$=$	\>	\>$\{$	\>\+\+\+definition \ref{gen.index}(c) and \ref{gen.index}(d)\-\-$~~~ \}$\pop\\
	$R{\MPperldom{}}\ms{1}{\MPcomp}\ms{1}\Delta{}R\ms{1}{\MPcomp}\ms{1}R{\MPperrdom{}}$\push\-\\
	$=$	\>	\>$\{$	\>\+\+\+lemma \ref{per.difun.div}  and domains\-\-$~~~ \}$\pop\\
	$\Delta{}R~~.$
\end{mpdisplay}
Combining the two calculations,  the two equations are equivalent
and, therefore, it  suffices to prove just one of them\footnote{It is not necessary to prove the
equivalence of the two statements in order to prove the theorem; we could have omitted the second
calculation.   But some redundancy in proofs
enhances their reliability.}.  We prove the second by mutual inclusion:
\begin{mpdisplay}{0.15em}{6.5mm}{0mm}{2}
	$\Delta{}R$\push\-\\
	$=$	\>	\>$\{$	\>\+\+\+$\Delta{}R$ is difunctional\-\-$~~~ \}$\pop\\
	$\Delta{}R\ms{1}{\MPcomp}\ms{1}\Delta{}R^{\MPrev}\ms{1}{\MPcomp}\ms{1}\Delta{}R$\push\-\\
	$=$	\>	\>$\{$	\>\+\+\+lemma \ref{diag.index(b)}, converse\-\-$~~~ \}$\pop\\
	$\Delta{}R\ms{1}{\MPcomp}\ms{1}\Delta{}R^{\MPrev}\ms{1}{\MPcomp}\ms{1}\Delta{}J\ms{1}{\MPcomp}\ms{1}\Delta{}R^{\MPrev}\ms{1}{\MPcomp}\ms{1}\Delta{}R$\push\-\\
	$=$	\>	\>$\{$	\>\+\+\+$\Delta{}R$ is difunctional, theorem \ref{difunctional.strongdef}(iv) and (vi)\-\-$~~~ \}$\pop\\
	$(\Delta{}R){\MPperldom{}}\ms{1}{\MPcomp}\ms{1}\Delta{}J\ms{1}{\MPcomp}\ms{1}(\Delta{}R){\MPperrdom{}}$\push\-\\
	$=$	\>	\>$\{$	\>\+\+\+lemma \ref{per.difun.div}\-\-$~~~ \}$\pop\\
	$(\Delta{}R){\MPldom{}}\ms{1}{\MPcomp}\ms{1}R{\MPperldom{}}\ms{1}{\MPcomp}\ms{1}\Delta{}J\ms{1}{\MPcomp}\ms{1}R{\MPperrdom{}}\ms{1}{\MPcomp}\ms{1}(\Delta{}R){\MPrdom{}}$\push\-\\
	$\subseteq$	\>	\>$\{$	\>\+\+\+domains are coreflexive\-\-$~~~ \}$\pop\\
	$R{\MPperldom{}}\ms{1}{\MPcomp}\ms{1}\Delta{}J\ms{1}{\MPcomp}\ms{1}R{\MPperrdom{}}$\push\-\\
	$\subseteq$	\>	\>$\{$	\>\+\+\+lemma \ref{diag.index(a)} and monotonicity\-\-$~~~ \}$\pop\\
	$R{\MPperldom{}}\ms{1}{\MPcomp}\ms{1}\Delta{}R\ms{1}{\MPcomp}\ms{1}R{\MPperrdom{}}$\push\-\\
	$=$	\>	\>$\{$	\>\+\+\+lemma \ref{per.difun.div}, domains\-\-$~~~ \}$\pop\\
	$\Delta{}R~~.$
\end{mpdisplay}
\vspace{-7mm}
\MPendBox

\section{Block-Ordered Relations}\label{Block-Ordered Relations}

In general, dividing a subset of a set $A$ into blocks  
is formulated by specifying a functional relation  $f$, say, with source\footnote{In
the terminology we use, a 
relation  of type $A{\sim}B$  has \emph{target} $A$ and \emph{source} $B$.} the set $A$;
elements $a0$ and $a1$ are in the same block equivales  $f{.}a0$ and $f{.}a1$ are both defined and  $f{.}a0\ms{1}{=}\ms{1}f{.}a1$.   
In mathematical terminology, a functional relation $f$ defines the  \emph{partial equivalence relation} $f^{\MPrev}\ms{1}{\MPcomp}\ms{1}f$ and 
the  ``blocks'' are the equivalence classes of $f^{\MPrev}\ms{1}{\MPcomp}\ms{1}f$.  (Partiality means that some elements may not be in an
equivalence class.)

Given functional relations  $f$ and $g$ with sources $A$ and $B$, respectively, and equal  left  domains,   relation
$R$ of type $A{\sim}B$ is said to be \emph{block-structured} by $f$ and $g$ if there is a relation $S$  such that $R\ms{2}{=}\ms{2}f^{\MPrev}\ms{1}{\MPcomp}\ms{1}S\ms{1}{\MPcomp}\ms{1}g$.  
Informally, whether or not $a$ and $b$ are related by $R$ depends entirely on the ``block'' $(f{.}a\ms{1},\ms{1}g{.}b)$ to which
they belong.  Note that it is not required that $f$ and $g$ be total functions: it suffices that $f{\MPrdom{}}\ms{1}{=}\ms{1}R{\MPldom{}}$ and $g{\MPrdom{}}\ms{1}{=}\ms{1}R{\MPrdom{}}$.  
The type of $S$ is  $C{\sim}C$ where $C$ includes   $\{a{:}\ms{3}a\ms{1}{\MPcomp}\ms{1}f{\MPrdom{}}\ms{2}{=}\ms{2}a{:}\ms{3}f{.}a\}$ (equally $\{b{:}\ms{3}b\ms{1}{\MPcomp}\ms{1}f{\MPrdom{}}\ms{2}{=}\ms{2}b{:}\ms{3}g{.}b\}$).  
\begin{Definition}[Block-Ordered Relation]\label{def:Block-Ordered Relation}{\rm \ \ \ 
Suppose $T$ is  a relation of type $C{\sim}C$,   $f$ is a   relation of  type $C{\sim}A$   and $g$ is a   relation 
of type   $C{\sim}B$.   Suppose further that  $T$ is  a provisional ordering and 
that $f$ and $g$ are functional and onto the domain of $T$.  That is, suppose \begin{equation}\label{def:Block-Ordered Relationfg}
f\ms{1}{\MPcomp}\ms{1}f^{\MPrev}\ms{4}{=}\ms{4}f{\MPldom{}}\ms{4}{=}\ms{4}T\ms{1}{\cap}\ms{1}T^{\MPrev}\ms{4}{=}\ms{4}g{\MPldom{}}\ms{4}{=}\ms{4}g\ms{1}{\MPcomp}\ms{1}g^{\MPrev}~~.
\end{equation}Then we say that the relation   $f^{\MPrev}\ms{1}{\MPcomp}\ms{1}T\ms{1}{\MPcomp}\ms{1}g$ is a \emph{block-ordered relation}.   A relation $R$ of
type $A{\sim}B$ is said to be \emph{block-ordered} by $f$, $g$ and $T$ if $R\ms{2}{=}\ms{2}f^{\MPrev}\ms{1}{\MPcomp}\ms{1}T\ms{1}{\MPcomp}\ms{1}g$  and 
$f^{\MPrev}\ms{1}{\MPcomp}\ms{1}T\ms{1}{\MPcomp}\ms{1}g$ is a block-ordered relation.
}
\MPendBox\end{Definition}

The archetypical   example of a block-ordered relation is
a   preorder.   Informally, if  $R$ is a preorder,  its  symmetric closure     $R\ms{1}{\cap}\ms{1}R^{\MPrev}$   is an equivalence relation,
and the relation $R$ defines a partial  ordering on the equivalence classes.   Equivalently,  if a representative element
is chosen for each equivalence class, the relation $R$ is a partial ordering on the representatives.  
Theorem \ref{provpreorder.decom}  makes this precise.

\begin{Theorem}\label{provpreorder.decom}{\rm \ \ \ Suppose $T$ is a provisional preorder  and suppose $J$ is a (coreflexive) index of 
$T{\MPperldom{}}$.  Then $J{\MPcomp}T{\MPcomp}J$ is an index of $T$ and is a provisional ordering.  Hence,  $T$ is a
block-ordered relation.
}
\end{Theorem}
{\bf Proof}~~~That $J{\MPcomp}T{\MPcomp}J$ is an index of $T$ is the combination of lemma \ref{ppreorder.perdomain} and 
theorem \ref{per.to.relation}.  It is a provisional preorder because $T$ is a preorder and $J$ is coreflexive.
So, it remains to show that $J{\MPcomp}T{\MPcomp}J$ is provisionally anti-symmetric.  That is, we must show that
 $J{\MPcomp}T{\MPcomp}J\ms{2}{\cap}\ms{2}(J{\MPcomp}T{\MPcomp}J)^{\MPrev}\ms{3}{\subseteq}\ms{3}I$. 
\begin{mpdisplay}{0.15em}{6.5mm}{0mm}{2}
	$J{\MPcomp}T{\MPcomp}J\ms{2}{\cap}\ms{2}(J{\MPcomp}T{\MPcomp}J)^{\MPrev}$\push\-\\
	$=$	\>	\>$\{$	\>\+\+\+$J$ is coreflexive, distributivity\-\-$~~~ \}$\pop\\
	$J{\MPcomp}(T\ms{1}{\cap}\ms{1}T^{\MPrev}){\MPcomp}J$\push\-\\
	$\subseteq$	\>	\>$\{$	\>\+\+\+\ref{ppreorder.perdomain}\-\-$~~~ \}$\pop\\
	$J\ms{1}{\MPcomp}\ms{1}T{\MPperldom{}}\ms{1}{\MPcomp}\ms{1}J$\push\-\\
	$=$	\>	\>$\{$	\>\+\+\+$J$ is an index of $T{\MPperldom{}}$, definition  \ref{per.index}(b) with $P\ms{1}{:=}\ms{1}T{\MPperldom{}}$\-\-$~~~ \}$\pop\\
	$J$\push\-\\
	$\subseteq$	\>	\>$\{$	\>\+\+\+$J$ is coreflexive\-\-$~~~ \}$\pop\\
	$I~~.$
\end{mpdisplay}
\vspace{-7mm}
\MPendBox

Identifying a block-ordering  of a  relation ---if it exists--- is important for efficiency.    Although a
relation is defined to be a set of pairs, relations ---even relations on finite sets--- are rarely stored as such;
instead some base set of pairs is stored and an algorithm used to generate, on demand, additional
information about the relation.  This is particularly so of ordering relations.  For example, a test $m\ms{1}{<}\ms{1}n$ on
integers $m$ and $n$ in a computer program is never implemented as a table lookup; instead an algorithm is
used to infer from the basic relations  $0\ms{1}{<}\ms{1}1$ together with the internal  representation of $m$ and $n$ what the
value of the test is.  In the case of block-structured relations,  functional relations  $f$ and $g$  define partial
equivalence relations $f^{\MPrev}\ms{1}{\MPcomp}\ms{1}f$ and $g^{\MPrev}\ms{1}{\MPcomp}\ms{1}g$ on their respective sources.  (The relations $f^{\MPrev}\ms{1}{\MPcomp}\ms{1}f$ and  $g^{\MPrev}\ms{1}{\MPcomp}\ms{1}g$ are
partial because $f$ and $g$ are not required to be total.)  Combining the functional relations  with an
ordering relation on their  (common) target is an effective way of implementing a  relation
(assuming the ordering relation is also implemented effectively).

\begin{Example}\label{bd:stronglyconnectecomponents}{\rm \ \ \ Suppose $G$ is the edge relation of a finite graph.   The
relation $G^{*}$ is, of course, a preorder and so is block-ordered.  The block-ordering of $G^{*}$  given by theorem  
\ref{provpreorder.decom}  is, however, not very useful.  For practical 
purposes a block-ordering constructed from  $G$ (rather than $G^{*}$) is preferable.   Here we outline how this is 
done.

% As aid to making changes, labels in this example are to Basic Graph Theory with prefix "bd:".

Recall from example \ref{difun.graphs0}, that the diagonal $\Delta(G^{*})$ is the relation $G^{*}\ms{1}{\cap}\ms{1}(G^{\MPrev})^{*}$
and that this  is an equivalence relation on the nodes of $G$, whereby the equivalence classes are
the \emph{strongly connected components} of $G$.  Let $N$ denote the nodes of $G$ and $C$ denote the set of strongly
connected components  of $G.$  By theorem \ref{split.existence},  there is a function $\mathsf{sc}$ of  type $C{\leftarrow}N$ such that \begin{equation}\label{bd:sc.partition}
G^{*}\ms{1}{\cap}\ms{1}(G^{\MPrev})^{*}\ms{4}{=}\ms{4}\mathsf{sc}^{\MPrev}\ms{1}{\MPcomp}\ms{1}\mathsf{sc}~~.
\end{equation}The relation $\mathcal{A}$ defined by \begin{displaymath}\mathsf{sc}\ms{1}{\MPcomp}\ms{1}G\ms{1}{\MPcomp}\ms{1}\mathsf{sc}^{\MPrev}\ms{4}{\cap}\ms{4}{\neg}I_{C}\end{displaymath}is a homogeneous relation on the strongly connected components of $G$, i.e. a relation of type $C{\sim}C$.  
Informally, it is a graph obtained from the graph $G$ by coalescing the nodes in a strongly connected
component of $G$ into a single node whilst retaining the edges of $G$ that connect nodes in distinct strongly
connected components.   A fundamental  theorem is  that \begin{equation}\label{bd:SCgraph1}
G^{*}\ms{6}{=}\ms{6}\mathsf{sc}^{\MPrev}\ms{1}{\MPcomp}\ms{1}\mathcal{A}^{*}\ms{1}{\MPcomp}\ms{1}\mathsf{sc}~~.
\end{equation}Moreover,  $\mathcal{A}$ is acyclic.   That is,\begin{equation}\label{bd:SCgraph2}
I_{C}\ms{2}{\cap}\ms{2}\mathcal{A}^{+}\ms{5}{=}\ms{5}{\MPplatbottom}~~.
\end{equation}(See \cite{BACKHOUSE2022100730,RCB2022} for the  details of the proof of (\ref{bd:SCgraph1}) and (\ref{bd:SCgraph2}). In fact
the theorem is valid for all relations $G$; finiteness is not required.)

The relation $\mathcal{A}^{*}$ is, of course,  transitive.  It is also reflexive;  combined with its acyclicity, it follows that \begin{equation}\label{bd:acyclic.equiv.defs}
\mathcal{A}^{*}\ms{1}{\cap}\ms{1}(\mathcal{A}^{*})^{\MPrev}\ms{4}{=}\ms{4}I_{C}~~.
\end{equation}That is, $\mathcal{A}^{*}$ is a (total) provisional ordering on $C.$    The conclusion is that $G^{*}$ is
block-ordered by $\mathsf{sc}$, $\mathsf{sc}$ and $\mathcal{A}^{*}$.  

Informally, a finite graph can always be decomposed into its strongly connected components together
with an acyclic graph connecting the components.

Although the informal interpretation of this theorem is well-known,  the  formal proof is non-trivial.  
Although not formulated in the same way, it is essentially the ``transitive reduction'' of an arbitrary (not
necessarily acyclic) graph formulated by Aho, Garey and Ullman  \cite[Theorem 2]{AGU1972}.  

The  decomposition (\ref{bd:SCgraph1})  is (implicitly)  exploited when computing
the inverse $\mathbf{A}^{{-}1}$ of a real matrix $\mathbf{A}$ in order to minimise storage requirements: using an elimination
technique, a so-called ``product form'' is computed for each strongly connected component, whilst the
process of ``forward substitution'' is applied to the acyclic-graph structure.
}%
\MPendBox\end{Example}

%It is important to note the very strict requirement (\ref{def:Block-Ordered Relationfg}) on the functionals $f$
%and $g$.  Were this requirement to be omitted (retaining
%only that $f$ and $g$ are functional relations \emph{into} ---not onto---  the domain of $T$), 
%there would be no guarantee of non-redundancy.  
%As we shall see, our
%definition of block-ordering does guarantee the existence of a non-redundant polar covering (theorem 
%\ref{BD.provorder.onlyif}) but not vice-versa (corollary \ref{BD.provorder.if}).  This suggests that the requirement
%may be too strong.  
%Add later???? *****See section \ref{BD: Imperfect Block-Orderings} and  the conclusions for further discussion.

Theorem \ref{BD.provorder.theo} makes precise the statement that block orderings ---where they exist--- 
are unique ``up to isomorphism''.  
\begin{Theorem}\label{BD.provorder.theo}{\rm \ \ \ Suppose $T$ is  a provisional ordering.  
Suppose also that $f$ and $g$ are functional and onto the domain of $T$.  That is, suppose \begin{displaymath}f\ms{1}{\MPcomp}\ms{1}f^{\MPrev}\ms{4}{=}\ms{4}f{\MPldom{}}\ms{4}{=}\ms{4}T\ms{1}{\cap}\ms{1}T^{\MPrev}\ms{4}{=}\ms{4}g{\MPldom{}}\ms{4}{=}\ms{4}g\ms{1}{\MPcomp}\ms{1}g^{\MPrev}~~.\end{displaymath}Suppose further\footnote{The types of $T$ and $S$ may be different.  The types of $f$ and $h$, and of $g$ and $k$ will then also be
different.  As in lemma \ref{BD.provorder}, the 
requirement is that the types are compatible with the type restrictions on the
operators in all assumed properties.  
The symbol ``$I$'' in (\ref{BD.provorder.iso}) is overloaded: if the type of $T$ is $A{\sim}A$ and the type of $S$ is
$B{\sim}B$,  $\phi\ms{1}{\MPcomp}\ms{1}\phi^{\MPrev}$ has type $A{\sim}A$ and $\phi^{\MPrev}\ms{1}{\MPcomp}\ms{1}\phi$ has type $B{\sim}B$.} 
that $S$,  $h$  and $k$  satisfy the same properties as  $T$, $f$ and $g$ (respectively) and that \begin{equation}\label{BD.fRSg}
f^{\MPrev}\ms{1}{\MPcomp}\ms{1}T\ms{1}{\MPcomp}\ms{1}g\ms{4}{=}\ms{4}h^{\MPrev}\ms{1}{\MPcomp}\ms{1}S\ms{1}{\MPcomp}\ms{1}k~~.
\end{equation}Then \begin{equation}\label{BD.provorder.dom.theo}
f{\MPrdom{}}\ms{1}{=}\ms{1}h{\MPrdom{}}\ms{5}{\wedge}\ms{5}g{\MPrdom{}}\ms{1}{=}\ms{1}k{\MPrdom{}}~~,
\end{equation}\begin{equation}\label{BD.provorder.fghk}
f^{\MPrev}\ms{1}{\MPcomp}\ms{1}g\ms{2}{=}\ms{2}h^{\MPrev}\ms{1}{\MPcomp}\ms{1}k~~,
\end{equation}\begin{equation}\label{BD.provorder.fgRShk}
f^{\MPrev}\ms{1}{\MPcomp}\ms{1}T^{\MPrev}\ms{1}{\MPcomp}\ms{1}g\ms{4}{=}\ms{4}h^{\MPrev}\ms{1}{\MPcomp}\ms{1}S^{\MPrev}\ms{1}{\MPcomp}\ms{1}k\mbox{~~, and}
\end{equation}\begin{equation}\label{BD.provorder.fhgk}
f\ms{1}{\MPcomp}\ms{1}h^{\MPrev}\ms{4}{=}\ms{4}g\ms{1}{\MPcomp}\ms{1}k^{\MPrev}~~.
\end{equation}Also,  letting $\phi$ denote $f\ms{1}{\MPcomp}\ms{1}h^{\MPrev}$ (equally, by  (\ref{BD.provorder.fhgk}),  $g\ms{1}{\MPcomp}\ms{1}k^{\MPrev}$),\begin{equation}\label{BD.provorder.iso}
\phi\ms{1}{\MPcomp}\ms{1}\phi^{\MPrev}\ms{2}{=}\ms{2}T\ms{1}{\cap}\ms{1}T^{\MPrev}\ms{5}{\wedge}\ms{5}\phi^{\MPrev}\ms{1}{\MPcomp}\ms{1}\phi\ms{2}{=}\ms{2}S\ms{1}{\cap}\ms{1}S^{\MPrev}\ms{5}{\wedge}\ms{5}\phi{\MPcomp}T\ms{1}{=}\ms{1}S{\MPcomp}\phi~~.
\end{equation}In words, $\phi$ is an order isomorphism of the domains of $T$ and $S$.
}
\end{Theorem}
{\bf Proof}~~~In combination with the assumption (\ref{BD.fRSg}), properties (\ref{BD.provorder.dom.theo}), 
(\ref{BD.provorder.fgRShk}) and (\ref{BD.provorder.fghk}) are immediate from
(\ref{BD.provorder.dom}), (\ref{BD.provorder.Rwok}) and (\ref{BD.provorder.fwokg}), respectively.

Proof of (\ref{BD.provorder.fhgk}) is a step on the way to proving (\ref{BD.provorder.iso}). From
symmetry considerations, it is an obvious first step.
\begin{mpdisplay}{0.15em}{6.5mm}{0mm}{2}
	$f\ms{1}{\MPcomp}\ms{1}h^{\MPrev}$\push\-\\
	$=$	\>	\>$\{$	\>\+\+\+assumption:~ $k\ms{1}{\MPcomp}\ms{1}k^{\MPrev}\ms{2}{=}\ms{2}h{\MPldom{}}$\-\-$~~~ \}$\pop\\
	$f\ms{1}{\MPcomp}\ms{1}h^{\MPrev}\ms{1}{\MPcomp}\ms{1}k\ms{1}{\MPcomp}\ms{1}k^{\MPrev}$\push\-\\
	$=$	\>	\>$\{$	\>\+\+\+(\ref{BD.provorder.fghk})\-\-$~~~ \}$\pop\\
	$f\ms{1}{\MPcomp}\ms{1}f^{\MPrev}\ms{1}{\MPcomp}\ms{1}g\ms{1}{\MPcomp}\ms{1}k^{\MPrev}$\push\-\\
	$=$	\>	\>$\{$	\>\+\+\+assumption:~ $f\ms{1}{\MPcomp}\ms{1}f^{\MPrev}\ms{2}{=}\ms{2}g{\MPldom{}}$\-\-$~~~ \}$\pop\\
	$g\ms{1}{\MPcomp}\ms{1}k^{\MPrev}~~.$
\end{mpdisplay}
Now,
\begin{mpdisplay}{0.15em}{6.5mm}{0mm}{2}
	$\phi\ms{1}{\MPcomp}\ms{1}\phi^{\MPrev}$\push\-\\
	$=$	\>	\>$\{$	\>\+\+\+definition of $\phi$, converse\-\-$~~~ \}$\pop\\
	$f\ms{1}{\MPcomp}\ms{1}h^{\MPrev}\ms{1}{\MPcomp}\ms{1}h\ms{1}{\MPcomp}\ms{1}f^{\MPrev}$\push\-\\
	$=$	\>	\>$\{$	\>\+\+\+(\ref{BD.provorder.fhgk})\-\-$~~~ \}$\pop\\
	$g\ms{1}{\MPcomp}\ms{1}k^{\MPrev}\ms{1}{\MPcomp}\ms{1}h\ms{1}{\MPcomp}\ms{1}f^{\MPrev}$\push\-\\
	$=$	\>	\>$\{$	\>\+\+\+ (\ref{BD.provorder.fghk}) and converse\-\-$~~~ \}$\pop\\
	$g\ms{1}{\MPcomp}\ms{1}g^{\MPrev}\ms{1}{\MPcomp}\ms{1}f\ms{1}{\MPcomp}\ms{1}f^{\MPrev}$\push\-\\
	$=$	\>	\>$\{$	\>\+\+\+assumption:~ $f\ms{1}{\MPcomp}\ms{1}f^{\MPrev}\ms{2}{=}\ms{2}T\ms{1}{\cap}\ms{1}T^{\MPrev}\ms{2}{=}\ms{2}g\ms{1}{\MPcomp}\ms{1}g^{\MPrev}$\-\-$~~~ \}$\pop\\
	$T\ms{1}{\cap}\ms{1}T^{\MPrev}~~.$
\end{mpdisplay}
Symmetrically,  $\phi^{\MPrev}\ms{1}{\MPcomp}\ms{1}\phi\ms{2}{=}\ms{2}T\ms{1}{\cap}\ms{1}T^{\MPrev}$.  Finally,
\begin{mpdisplay}{0.15em}{6.5mm}{0mm}{2}
	$T{\MPcomp}\phi$\push\-\\
	$=$	\>	\>$\{$	\>\+\+\+definition of $\phi$\-\-$~~~ \}$\pop\\
	$T\ms{1}{\MPcomp}\ms{1}f\ms{1}{\MPcomp}\ms{1}h^{\MPrev}$\push\-\\
	$=$	\>	\>$\{$	\>\+\+\+assumptions:~ $f\ms{1}{\MPcomp}\ms{1}f^{\MPrev}\ms{2}{=}\ms{2}T\ms{1}{\cap}\ms{1}T^{\MPrev}\ms{2}{=}\ms{2}g\ms{1}{\MPcomp}\ms{1}g^{\MPrev}$\\
	$T\ms{3}{=}\ms{3}(T\ms{1}{\cap}\ms{1}T^{\MPrev})\ms{1}{\MPcomp}\ms{1}T\ms{1}{\MPcomp}\ms{1}(T\ms{1}{\cap}\ms{1}T^{\MPrev})$\-\-$~~~ \}$\pop\\
	$f\ms{1}{\MPcomp}\ms{1}f^{\MPrev}\ms{1}{\MPcomp}\ms{1}T\ms{1}{\MPcomp}\ms{1}g\ms{1}{\MPcomp}\ms{1}g^{\MPrev}\ms{1}{\MPcomp}\ms{1}f\ms{1}{\MPcomp}\ms{1}h^{\MPrev}$\push\-\\
	$=$	\>	\>$\{$	\>\+\+\+assumption:~ $f^{\MPrev}\ms{1}{\MPcomp}\ms{1}T\ms{1}{\MPcomp}\ms{1}g\ms{4}{=}\ms{4}h^{\MPrev}\ms{1}{\MPcomp}\ms{1}S\ms{1}{\MPcomp}\ms{1}k$ ~, (\ref{BD.provorder.fghk}) and converse\-\-$~~~ \}$\pop\\
	$f\ms{1}{\MPcomp}\ms{1}h^{\MPrev}\ms{1}{\MPcomp}\ms{1}S\ms{1}{\MPcomp}\ms{1}k\ms{1}{\MPcomp}\ms{1}k^{\MPrev}\ms{1}{\MPcomp}\ms{1}h\ms{1}{\MPcomp}\ms{1}h^{\MPrev}$\push\-\\
	$=$	\>	\>$\{$	\>\+\+\+assumption:~ $h\ms{1}{\MPcomp}\ms{1}h^{\MPrev}\ms{2}{=}\ms{2}S\ms{1}{\cap}\ms{1}S^{\MPrev}\ms{2}{=}\ms{2}k\ms{1}{\MPcomp}\ms{1}k^{\MPrev}$\-\-$~~~ \}$\pop\\
	$f\ms{1}{\MPcomp}\ms{1}h^{\MPrev}\ms{1}{\MPcomp}\ms{1}S$\push\-\\
	$=$	\>	\>$\{$	\>\+\+\+definition of $\phi$\-\-$~~~ \}$\pop\\
	$\phi{\MPcomp}S~~.$
\end{mpdisplay}
\vspace{-7mm}
\MPendBox

\subsection{Pair Algebras and Galois Connections}\label{Pair Algebras and Galois Connections}

In order to find lots of examples of block-ordered relations one need look no further than the theory of
Galois connections (which are, of course, ubiquitous).   In this section, we briefly review the notion of a
``pair algebra'' ---due to Hartmanis and Stearns \cite{HS64a,HS66a}--- and its relation to Galois
connections.  

Hartmanis and Stearns limited their analysis to finite, complete posets, and  their
analysis was less general than is  possible.  This work was extended in \cite{Bac98} to non-finite posets and
the current section is a short extract.

A Galois connection involves two posets  $(\mathcal{A},\sqsubseteq)$ and ($\mathcal{B}\ms{1}{,}\ms{1}{\preceq}$) and
two functions, $F\ms{1}{\in}\ms{1}\mathcal{A}{\leftarrow}\mathcal{B}$ and $G\ms{1}{\in}\ms{1}\mathcal{B}{\leftarrow}\mathcal{A}$.  These four components together form a
\emph{Galois connection} iff for all $b{\in}\mathcal{B}$ and $a{\in}\mathcal{A}$ \begin{equation}\label{galois}
F{.}b\ms{1}{\sqsubseteq}\ms{1}a\ms{2}{\equiv}\ms{2}b\ms{1}{\preceq}\ms{1}G{.}a~~.
\end{equation}We  refer to $F$ as the {\em lower adjoint} and to $G$ as the {\em upper adjoint}.

A Galois connection is thus a connection between two functions between posets.
Typical accounts of the properties of Galois connections (e.g.\  \cite{GHK80a}) focus
on the properties of these \emph{functions}. For example, given a function $F$, one may ask 
whether  $F$ is a lower adjoint in a Galois connection.  The question posed
by Hartmanis and Stearns was, however, rather different.

To motivate their  question, note that the statement $F{.}b\ms{1}{\sqsubseteq}\ms{1}a$ defines a 
\emph{relation} between $\mathcal{B}$  and $\mathcal{A}$.    So too does  $b\ms{1}{\preceq}\ms{1}G{.}a$.  The existence of a Galois
connection states that these two relations are equal.  A natural question is therefore:
under which conditions does an arbitrary (binary) relation between  two posets 
define a Galois connection between the sets?

Exploring the question in more detail leads to two separate questions.
The first is:  suppose $R$ is a relation between posets
$(\mathcal{A},\sqsubseteq)$ and  ($\mathcal{B}\ms{1}{,}\ms{1}{\preceq}$).  What is a necessary and sufficient condition that there exist
a function  $F$ such that \begin{displaymath}(a,b)\ms{1}{\in}\ms{1}R\ms{4}{\equiv}\ms{4}F{.}b\ms{1}{\sqsubseteq}\ms{1}a\mbox{~~~~?}\end{displaymath}The second is the dual of the first: given relation $R$,
what is a necessary and sufficient condition that there exist a function $G$ such
that \begin{displaymath}(a,b)\ms{1}{\in}\ms{1}R\ms{4}{\equiv}\ms{4}b\ms{1}{\preceq}\ms{1}G{.}a\mbox{~~~~?}\end{displaymath}The conjunction of these two conditions is a necessary and  sufficient condition for a
relation $R$ to define a Galois connection.  Such a relation is called  a \emph{pair algebra}\mpindexnew{pair algebra}{\mpT }{\mpT }{\mpT }{\mpT }{\mpT }{\mpT }{}.

\begin{Example}\label{pair.sep.ex}{\rm \ \ \ It is easy to demonstrate that the two questions are separate.  To this end,  
fig.\ \ref{fig:pairalg} depicts two posets and a relation between them.  The posets are $\{\alpha{,}\beta\}$ and $\{\mathsf{A}{,}\mathsf{B}\}$; both are
ordered lexicographically: the reflexive-transitive reduction of the lexicographic ordering is depicted by
the directed  edges.  The relation of type $\{\alpha{,}\beta\}{\sim}\{\mathsf{A}{,}\mathsf{B}\}$ is depicted by the undirected edges.    

\begin{figure}[h]
\centering \includegraphics{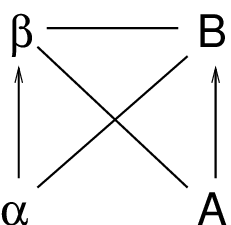}
 
\caption{A Relation on Two Posets}\label{fig:pairalg}
\end{figure}

Let the relation be denoted by $R$.  Define the function $F$ of type $\{\alpha{,}\beta\}\ms{1}{\leftarrow}\ms{1}\{\mathsf{A}{,}\mathsf{B}\}$ by $F{.}\mathsf{B}\ms{1}{=}\ms{1}\alpha$ and $F{.}\mathsf{A}\ms{1}{=}\ms{1}\beta$.  Then it
is easy to check that. for $a{\in}\{\alpha{,}\beta\}$ and $b{\in}\{\mathsf{A}{,}\mathsf{B}\}$,\begin{displaymath}(a,b)\ms{1}{\in}\ms{1}R\ms{4}{\equiv}\ms{4}F{.}b\ms{1}{\sqsubseteq}\ms{1}a~~.\end{displaymath}(There are just four cases to be considered.)  On the other hand,  there is no function $G$  of type 
 $\{\mathsf{A}{,}\mathsf{B}\}\ms{1}{\leftarrow}\ms{1}\{\alpha{,}\beta\}$  such that \begin{displaymath}(a,b)\ms{1}{\in}\ms{1}R\ms{4}{\equiv}\ms{4}b\ms{1}{\preceq}\ms{1}G{.}a~~.\end{displaymath}To check that this is indeed the case, it suffices to check that the assignment $G{.}\mathsf{A}\ms{1}{=}\ms{1}\alpha$ is invalid (because $\alpha\ms{1}{\sqsubseteq}\ms{1}\alpha$ but $(\alpha,\mathsf{A})\ms{1}{\not\in}\ms{1}R$)  and 
the assignment $G{.}\mathsf{A}\ms{1}{=}\ms{1}\beta$ is also invalid (because $\alpha\ms{1}{\sqsubseteq}\ms{1}\beta$  but $(\alpha,\mathsf{A})\ms{1}{\not\in}\ms{1}R$).   
%\vspace{-7mm}
}%
\MPendBox\end{Example}

\begin{Example}\label{pair.sep.gen}{\rm \ \ \ A less artificial, general way to demonstrate that the two questions are
separate is to consider the membership relation.  Specifically, suppose $\mathcal{S}$ is a set.  Then the membership
relation, denoted as usual by the ---overloaded---  symbol  ``${\in}$'', is a heterogeneous relation of type $\mathcal{S}\ms{1}{\sim}\ms{1}2^{\mathcal{S}}$ 
 (where $2^{\mathcal{S}}$  denotes the type of subsets of $\mathcal{S}$).   Now,  for all  $x$  of type $\mathcal{S}$ and $X$ of type $2^{\mathcal{S}}$,  \begin{displaymath}x\ms{1}{\in}\ms{1}X\ms{3}{\equiv}\ms{3}\{x\}\ms{1}{\subseteq}\ms{1}X~~.\end{displaymath}The right side of this equation has the form $F{.}b\ms{1}{\sqsubseteq}\ms{1}a$ where $F$ is the function that maps an element into a
singleton set and the ordering is the subset ordering.  Also, its  left side has the form $(a,b)\ms{1}{\in}\ms{1}R$, where the
relation $R$ is the membership relation and $a$ and $b$ are $x$ and $X$, respectively.  (This is where the
overloading of notation can become confusing, for which our apologies!)  It is, however, not possible to
express $x\ms{1}{\in}\ms{1}X$  in the form $x\ms{1}{\preceq}\ms{1}G{.}X$ (except in the trivial cases where $\mathcal{S}$ has cardinality at most one).   We
leave the proof to the reader.
  %See also example \ref{mem.analogie}.
%\vspace{-7mm}
}%
\MPendBox\end{Example}

\begin{Example}\label{heterogeneous.atmost}{\rm \ \ \ An example of a Galois connection is the definition
of the ceiling function on real numbers:  for all real numbers $x$,  $\lceil{}x\rceil$ is an integer such that, for all
integers $m$, \begin{displaymath}x\ms{1}{\leq}\ms{1}m\ms{4}{\equiv}\ms{4}\lceil{}x\rceil\ms{1}{\leq}\ms{1}m~~.\end{displaymath}To properly fit the definition of a Galois connection, it is necessary to make explicit the implicit coercion 
from  integers to real numbers in the left side of this equation.  Specifically,  we have, for all real numbers $x$
and integers $m$,  \begin{displaymath}x~{\leq}_{\MPReal}~\mathsf{real}{.}m\ms{8}{\equiv}\ms{8}\lceil{}x\rceil~{\leq}_{\MPInt}~m\end{displaymath}where $\mathsf{real}$ denotes the function that ``coerces'' an integer to a real, and $\leq_{\MPReal}$ and $\leq_{\MPInt}$ denote
the (homogeneous) at-most relations on, respectively,  real numbers and integers.  If, however, we
consider the symbol ``${\leq}$'' on the left side of the equation to denote the heterogeneous  at-most
relation of type $\MPReal\ms{1}{\sim}\ms{1}\MPInt$,  the fact that  \begin{displaymath}x\ms{1}{\leq}\ms{1}m\ms{7}{\equiv}\ms{7}\lceil{}x\rceil~{\leq}_{\MPInt}~m\end{displaymath}gives a  representation of the (heterogeneous) ``${\leq}$'' relation of type $\MPReal\ms{1}{\sim}\ms{1}\MPInt$ as a block-ordered
relation:  referring to definition \ref{def:Block-Ordered Relation},  the provisional 
ordering is ${\leq}_{\MPInt}$,  $f$ is the ceiling function and $g$ is the identity function.

More interesting is if we take the contrapositive.  We have, for all real numbers $x$ and integers $m$,  \begin{displaymath}m\ms{1}{<}\ms{1}x\ms{5}{\equiv}\ms{5}m\ms{1}{\leq}\ms{1}\lceil{}x\rceil{-}1~~.\end{displaymath}On the right of this equation is the (homogeneous) at-most relation on integers.  On the left is the
(heterogeneous) less-than relation of type $\MPInt\ms{1}{\sim}\ms{1}\MPReal$.  The equation demonstrates that this relation is
block-ordered; the ``blocks'' of real numbers being all the numbers that have the same ceiling.  (The
functional  $f$ is the identity function, the functional $g$ maps real number $x$ to $\lceil{}x\rceil{-}1$ and the provisional
ordering is the ordering $\leq_{\MPInt}$.)   The example is interesting because the (homogeneous) less-than relation 
on real numbers is \emph{not} block-ordered.  This is because  its diagonal is empty.  See examples 
\ref{diagonal.lessthan}  and \ref{less.than.staircase}. 
% **** Revise if staircase relations included.
%The example is interesting because we show in theorem 
%\ref{staircase.neq.blockordered} that the (homogeneous) less-than relation on real numbers is \emph{not}
%block-ordered.
}%
\MPendBox\end{Example}

Returning to the discussion immediately preceding example \ref{pair.sep.ex},  the two separate questions are each of interest in their own right: a
positive answer to either question may predict that  a given relation has a block-ordering of a specific
form: in the case of the first question, where the functional $g$ in definition \ref{def:Block-Ordered Relation} is
the identity function, and, in the case of the second question,   where the functional $f$ in definition
\ref{def:Block-Ordered Relation}  is the identity function.  In both cases, a further step is to check the
requirement on $f$ and $g$: in the first case, one has to check that the function $F$ is surjective and in the
second case that the function $G$ is surjective.  (A Galois connection is said to be ``perfect'' if both $F$ and $G$ are 
surjective.) For example,  the fact that \begin{displaymath}x\ms{1}{\leq}\ms{1}m\ms{8}{\equiv}\ms{8}x~{\leq}_{\MPReal}~\mathsf{real}{.}m\end{displaymath}does  \emph{not} define a block-ordering because the function $\mathsf{real}$ is not surjective.  

The relevant theory predicting exactly when the first of the two questions has 
a positive answer is as follows.  Suppose  $(\mathcal{B},\sqsubseteq)$ is a complete poset. 
Let ${\bigsqcap}$ denote the infimum operator for $\mathcal{B}$ and suppose $p$ is a predicate on $\mathcal{B}$.  Then we
define \emph{inf-preserving}   by \begin{equation}\label{p.inf-preserving}
p\mbox{ is  inf-preserving\ms{6}}{\equiv}\ms{6}{\left\langle\forall{}g\ms{3}{:}{:}\ms{3}p{.}({\sqcap}g)\ms{2}{\equiv}\ms{2}{\left\langle\forall{}x\ms{2}{:}{:}\ms{2}p{.}(g{.}x)\right\rangle}\right\rangle}~~.
\end{equation}So, for a given $a$, the predicate  $\langle{}b{:}{:}\ms{2}(a,b)\ms{1}{\in}\ms{1}R\rangle$ is  inf-preserving equivales\begin{displaymath}{\left\langle\forall{}g\ms{5}{:}{:}\ms{5}(a\ms{1},\ms{1}{\sqcap}g)\ms{1}{\in}\ms{1}R\ms{4}{\equiv}\ms{4}{\left\langle\forall{}x\ms{2}{:}{:}\ms{2}(a\ms{1},\ms{1}g{.}x)\ms{1}{\in}\ms{1}R\right\rangle}\right\rangle}~~.\end{displaymath}Then we have:
\begin{Theorem}\label{pair.alg.ex}{\rm \ \ \ Suppose $\mathcal{A}$ is a  set  and $(\mathcal{B},\sqsubseteq)$ is a complete poset.
Suppose  $R\ms{1}{\subseteq}\ms{1}\mathcal{A}{\MPtimes}\mathcal{B}$ is a relation between the two sets.  
Define $F$  by \begin{equation}\label{defF}
F{.}a\ms{7}{=}\ms{7}{\left\langle{\sqcap}b\ms{3}{:}\ms{3}(a,b)\ms{1}{\in}\ms{1}R\ms{3}{:}\ms{3}b\right\rangle}~~.
\end{equation}Then the following two statements are equivalent.
\begin{itemize}
\item   ${\left\langle\forall\ms{1}{}a{,}b\ms{4}{:}\ms{4}a{\in}\mathcal{A}\ms{1}{\wedge}\ms{1}b{\in}\mathcal{B}\ms{4}{:}\ms{4}(a,b)\ms{1}{\in}\ms{1}R\ms{4}{\equiv}\ms{4}F{.}a\ms{1}{\sqsubseteq}\ms{1}b\right\rangle}$.
\item For all $a$, the predicate $\langle{}b{:}{:}\ms{2}(a,b)\ms{1}{\in}\ms{1}R\rangle$ is  inf-preserving.
\end{itemize}
\vspace{-7mm}
}
\MPendBox\end{Theorem}

The answer to the second question is, of course, obtained by formulating the dual of theorem \ref{pair.alg.ex}.

In general, for most relations occurring in practical information systems the answer to the pair-algebra
questions will be negative: the required inf- and sup-preserving properties just do not hold.  
However, a  common way to define a pair algebra is to extend a given  
relation  to a relation between sets in such a way that the infimum and supremum preserving
properties are automatically satisfied.  Hartmanis and Stearns'  \cite{HS64a,HS66a} solution to the state
assignment problem was to consider the lattice of partitions of a given set; in so-called ``concept analysis'',
the technique is to extend a given relation to a relation between rectangles.  

An important property of Galois connections is the  theorem we call the  ``unity of opposites'':
 if $F$ and $G$ are  the adjoint functions in a Galois connection of the posets $(\mathcal{A},\sqsubseteq)$ and  ($\mathcal{B}{,}{\preceq}$),
 then there is an isomorphism between the posets $(F{.}\mathcal{B}\ms{1},\ms{1}\sqsubseteq)$ and  ($G{.}\mathcal{A}\ms{1}{,}\ms{1}{\preceq}$).  ($F{.}\mathcal{B}$ denotes the ``image''  of the
function $F$, and similarly for $G{.}\mathcal{A}$.)  Knowledge of the unity-of-opposites theorem suggests theorem 
\ref{BD.provorder.theo}, which expresses an isomorphism between different  representations of
block-ordered relations.

\subsection{Analogie Frappante}\label{BD: Equivalent Formulations}

In this section, we relate block-orderings to diagonals.     The main result is  theorem  \ref{BD.diagdom}; we
call theorem  \ref{BD.diagdom} the ``analogie frappante'' because it generalises Riguet's  ``analogie
frappante'' connecting ``relation de Ferrers'' to diagonals.   

Some elements of the following lemma have been recorded earlier by Winter \cite{Winter2004}.   We think
the overlap is justified because Winter's calculations make very heavy use of complementation whereas our
calculations avoid its use altogether.
\begin{Lemma}\label{BD.provorder}{\rm \ \ \ Suppose $T$ is  a provisional   ordering of type $C{\sim}C$.  
Suppose also that $f$ and $g$ are functional and onto the domain of $T$.  
That is,  suppose\footnote{The ordering  $T$ must be homogeneous 
but $f$ and $g$ may be heterogeneous and of different type, so long as both have target $C$.}  that \begin{displaymath}f\ms{1}{\MPcomp}\ms{1}f^{\MPrev}\ms{4}{=}\ms{4}f{\MPldom{}}\ms{4}{=}\ms{4}T\ms{1}{\cap}\ms{1}T^{\MPrev}\ms{4}{=}\ms{4}g{\MPldom{}}\ms{4}{=}\ms{4}g\ms{1}{\MPcomp}\ms{1}g^{\MPrev}~~.\end{displaymath}Let $R$ denote $f^{\MPrev}\ms{1}{\MPcomp}\ms{1}T\ms{1}{\MPcomp}\ms{1}g$.  Then\begin{equation}\label{BD.provorder.dom}
R{\MPldom{}}\ms{2}{=}\ms{2}f{\MPrdom{}}\ms{5}{\wedge}\ms{5}R{\MPrdom{}}\ms{1}{=}\ms{1}g{\MPrdom{}}~~,
\end{equation}\begin{equation}\label{BD.provorder.Rwok}
f^{\MPrev}\ms{1}{\MPcomp}\ms{1}T^{\MPrev}\ms{1}{\MPcomp}\ms{1}g\ms{4}{=}\ms{4}R{\MPldom{}}\ms{1}{\MPcomp}\ms{1}(R{\setminus}R{/}R)^{\MPrev}\ms{1}{\MPcomp}\ms{1}R{\MPrdom{}}\mbox{~, and}
\end{equation}\begin{equation}\label{BD.provorder.fwokg}
f^{\MPrev}\ms{1}{\MPcomp}\ms{1}g\ms{4}{=}\ms{4}\Delta{}R~~,
\end{equation}\begin{equation}\label{BD.provorder.domdelta}
R{\MPldom{}}\ms{2}{=}\ms{2}(\Delta{}R){\MPldom{}}\ms{5}{\wedge}\ms{5}R{\MPrdom{}}\ms{2}{=}\ms{2}(\Delta{}R){\MPrdom{}}~~,
\end{equation}\begin{equation}\label{BD.provorder.perdom}
R{\MPperldom{}}\ms{3}{=}\ms{3}\Delta{}R\ms{1}{\MPcomp}\ms{1}\Delta{}R^{\MPrev}\ms{3}{=}\ms{3}f^{\MPrev}\ms{1}{\MPcomp}\ms{1}f\ms{6}{\wedge}\ms{6}R{\MPperrdom{}}\ms{3}{=}\ms{3}\Delta{}R^{\MPrev}\ms{1}{\MPcomp}\ms{1}\Delta{}R\ms{3}{=}\ms{3}g^{\MPrev}\ms{1}{\MPcomp}\ms{1}g~~.
\end{equation}
}%
\end{Lemma}%
{\bf Proof}~~~Property (\ref{BD.provorder.dom}) is a straightforward application of domain calculus:  
\begin{mpdisplay}{0.15em}{6.5mm}{0mm}{2}
	$R{\MPrdom{}}$\push\-\\
	$=$	\>	\>$\{$	\>\+\+\+definition:   $R\ms{2}{=}\ms{2}f^{\MPrev}\ms{1}{\MPcomp}\ms{1}T\ms{1}{\MPcomp}\ms{1}g$\-\-$~~~ \}$\pop\\
	$(f^{\MPrev}\ms{1}{\MPcomp}\ms{1}T\ms{1}{\MPcomp}\ms{1}g){\MPrdom{}}$\push\-\\
	$=$	\>	\>$\{$	\>\+\+\+domains (specifically,  $\left[\ms{2}(U{\MPcomp}V){\MPrdom{}}\ms{1}{=}\ms{1}(U{\MPrdom{}}\ms{1}{\MPcomp}\ms{1}V){\MPrdom{}}\ms{2}\right]$  and $\left[\ms{2}(U^{\MPrev}){\MPrdom{}}\ms{1}{=}\ms{1}U{\MPldom{}}\ms{2}\right]$)\-\-$~~~ \}$\pop\\
	$(f{\MPldom{}}\ms{1}{\MPcomp}\ms{1}T\ms{1}{\MPcomp}\ms{1}g){\MPrdom{}}$\push\-\\
	$=$	\>	\>$\{$	\>\+\+\+ assumption:   $T\ms{2}{=}\ms{2}f{\MPldom{}}\ms{1}{\MPcomp}\ms{1}T\ms{1}{\MPcomp}\ms{1}g{\MPldom{}}$  (so $T\ms{2}{=}\ms{2}f{\MPldom{}}\ms{1}{\MPcomp}\ms{1}T$)\-\-$~~~ \}$\pop\\
	$(T{\MPcomp}g){\MPrdom{}}$\push\-\\
	$=$	\>	\>$\{$	\>\+\+\+domains (specifically,  $\left[\ms{2}(U{\MPcomp}V){\MPrdom{}}\ms{1}{=}\ms{1}(U{\MPrdom{}}\ms{1}{\MPcomp}\ms{1}V){\MPrdom{}}\ms{2}\right]$)\-\-$~~~ \}$\pop\\
	$(T{\MPrdom{}}\ms{1}{\MPcomp}\ms{1}g){\MPrdom{}}$\push\-\\
	$=$	\>	\>$\{$	\>\+\+\+lemma \ref{p.ordering.dom} and assumption: $T\ms{1}{\cap}\ms{1}T^{\MPrev}\ms{2}{=}\ms{2}g{\MPldom{}}$  \-\-$~~~ \}$\pop\\
	$g{\MPrdom{}}~~.$
\end{mpdisplay}
By a symmetric argument, $(f^{\MPrev}\ms{1}{\MPcomp}\ms{1}T\ms{1}{\MPcomp}\ms{1}g){\MPldom{}}\ms{2}{=}\ms{2}f{\MPrdom{}}$.  

Now we consider (\ref{BD.provorder.Rwok}).  The raison d'\^{e}tre of (\ref{BD.provorder.Rwok}) is that it expresses the
left side as a function of $f^{\MPrev}\ms{1}{\MPcomp}\ms{1}T\ms{1}{\MPcomp}\ms{1}g$.   In a pointwise
calculation a natural step is to use indirect ordering.  In a point-free calculation, this corresponds to using
factors.  That is, we exploit  lemma \ref{ppreorder.perleft}: 
\begin{mpdisplay}{0.15em}{6.5mm}{0mm}{2}
	$f^{\MPrev}\ms{1}{\MPcomp}\ms{1}T^{\MPrev}\ms{1}{\MPcomp}\ms{1}g$\push\-\\
	$=$	\>	\>$\{$	\>\+\+\+assumption: $T$ is a provisional ordering \\
	lemmas \ref{provisionalpreorder.conv}, \ref{ppreorder.perdomain} and  \ref{ppreorder.perleft}\-\-$~~~ \}$\pop\\
	$f^{\MPrev}\ms{2}{\MPcomp}\ms{2}(T\ms{1}{\cap}\ms{1}T^{\MPrev})\ms{2}{\MPcomp}\ms{2}T^{\MPrev}\ms{1}{\setminus}\ms{1}T^{\MPrev}\ms{1}{/}\ms{1}T^{\MPrev}\ms{2}{\MPcomp}\ms{2}(T\ms{1}{\cap}\ms{1}T^{\MPrev})\ms{2}{\MPcomp}\ms{2}g$\push\-\\
	$=$	\>	\>$\{$	\>\+\+\+assumption:   $f{\MPldom{}}\ms{2}{=}\ms{2}T\ms{1}{\cap}\ms{1}T^{\MPrev}\ms{2}{=}\ms{2}g{\MPldom{}}$ \-\-$~~~ \}$\pop\\
	$f^{\MPrev}\ms{2}{\MPcomp}\ms{2}T^{\MPrev}\ms{1}{\setminus}\ms{1}T^{\MPrev}\ms{1}{/}\ms{1}T^{\MPrev}\ms{2}{\MPcomp}\ms{2}g$\push\-\\
	$=$	\>	\>$\{$	\>\+\+\+lemma \ref{f.under.over.g} and assumption:   $T\ms{2}{=}\ms{2}f{\MPldom{}}\ms{1}{\MPcomp}\ms{1}T\ms{1}{\MPcomp}\ms{1}g{\MPldom{}}$\-\-$~~~ \}$\pop\\
	$f{\MPrdom{}}\ms{2}{\MPcomp}\ms{2}(g^{\MPrev}\ms{1}{\MPcomp}\ms{1}T^{\MPrev}\ms{1}{\MPcomp}\ms{1}f)\ms{1}{\setminus}\ms{1}(g^{\MPrev}\ms{1}{\MPcomp}\ms{1}T^{\MPrev}\ms{1}{\MPcomp}\ms{1}f)\ms{1}{/}\ms{1}(g^{\MPrev}\ms{1}{\MPcomp}\ms{1}T^{\MPrev}\ms{1}{\MPcomp}\ms{1}f)\ms{2}{\MPcomp}\ms{2}g{\MPrdom{}}$\push\-\\
	$=$	\>	\>$\{$	\>\+\+\+(\ref{BD.provorder.dom}) and definition of $R$\-\-$~~~ \}$\pop\\
	$R{\MPldom{}}\ms{2}{\MPcomp}\ms{2}R^{\MPrev}\ms{1}{\setminus}\ms{1}R^{\MPrev}\ms{1}{/}\ms{1}R^{\MPrev}\ms{2}{\MPcomp}\ms{2}R{\MPrdom{}}$\push\-\\
	$=$	\>	\>$\{$	\>\+\+\+factors\-\-$~~~ \}$\pop\\
	$R{\MPldom{}}\ms{1}{\MPcomp}\ms{1}(R{\setminus}R{/}R)^{\MPrev}\ms{1}{\MPcomp}\ms{1}R{\MPrdom{}}~~.$
\end{mpdisplay}
Note the use of lemma \ref{f.under.over.g}.   The discovery of this lemma is driven by the goal of the calculation.

The pointwise interpretation of $f^{\MPrev}\ms{1}{\MPcomp}\ms{1}g$ is a relation expressing  equality between values of $f$ and $g$.  This
suggests that, in order to prove (\ref{BD.provorder.fwokg}),  we begin by exploiting the anti-symmetry of $T$:
\begin{mpdisplay}{0.15em}{6.5mm}{0mm}{2}
	$f^{\MPrev}\ms{1}{\MPcomp}\ms{1}g$\push\-\\
	$=$	\>	\>$\{$	\>\+\+\+$f{\MPldom{}}\ms{3}{=}\ms{3}T\ms{1}{\cap}\ms{1}T^{\MPrev}\ms{3}{=}\ms{3}g{\MPldom{}}$ and domains\-\-$~~~ \}$\pop\\
	$f^{\MPrev}\ms{1}{\MPcomp}\ms{1}(T\ms{1}{\cap}\ms{1}T^{\MPrev})\ms{1}{\MPcomp}\ms{1}g$\push\-\\
	$=$	\>	\>$\{$	\>\+\+\+distributivity (valid because $f$ and $g$ are functional)\-\-$~~~ \}$\pop\\
	$f^{\MPrev}\ms{1}{\MPcomp}\ms{1}T\ms{1}{\MPcomp}\ms{1}g\ms{5}{\cap}\ms{5}f^{\MPrev}\ms{1}{\MPcomp}\ms{1}T^{\MPrev}\ms{1}{\MPcomp}\ms{1}g$\push\-\\
	$=$	\>	\>$\{$	\>\+\+\+definition of $R$ and (\ref{BD.provorder.Rwok})\-\-$~~~ \}$\pop\\
	$f^{\MPrev}\ms{1}{\MPcomp}\ms{1}T\ms{1}{\MPcomp}\ms{1}g\ms{5}{\cap}\ms{5}f{\MPrdom{}}\ms{1}{\MPcomp}\ms{1}((f^{\MPrev}\ms{1}{\MPcomp}\ms{1}T\ms{1}{\MPcomp}\ms{1}g)\ms{1}{\setminus}\ms{1}(f^{\MPrev}\ms{1}{\MPcomp}\ms{1}T\ms{1}{\MPcomp}\ms{1}g)\ms{1}{/}\ms{1}(f^{\MPrev}\ms{1}{\MPcomp}\ms{1}T\ms{1}{\MPcomp}\ms{1}g))^{\MPrev}\ms{1}{\MPcomp}\ms{1}g{\MPrdom{}}$\push\-\\
	$=$	\>	\>$\{$	\>\+\+\+(\ref{ARSB})  (see below)\-\-$~~~ \}$\pop\\
	$f{\MPrdom{}}\ms{1}{\MPcomp}\ms{1}f^{\MPrev}\ms{1}{\MPcomp}\ms{1}T\ms{1}{\MPcomp}\ms{1}g\ms{1}{\MPcomp}\ms{1}g{\MPrdom{}}\ms{5}{\cap}\ms{5}((f^{\MPrev}\ms{1}{\MPcomp}\ms{1}T\ms{1}{\MPcomp}\ms{1}g)\ms{1}{\setminus}\ms{1}(f^{\MPrev}\ms{1}{\MPcomp}\ms{1}T\ms{1}{\MPcomp}\ms{1}g)\ms{1}{/}\ms{1}(f^{\MPrev}\ms{1}{\MPcomp}\ms{1}T\ms{1}{\MPcomp}\ms{1}g))^{\MPrev}$\push\-\\
	$=$	\>	\>$\{$	\>\+\+\+domains  (specifically,  $f{\MPrdom{}}\ms{1}{\MPcomp}\ms{1}f^{\MPrev}\ms{2}{=}\ms{2}f^{\MPrev}$ and $g\ms{1}{\MPcomp}\ms{1}g{\MPrdom{}}\ms{2}{=}\ms{2}g$)\-\-$~~~ \}$\pop\\
	$f^{\MPrev}\ms{1}{\MPcomp}\ms{1}T\ms{1}{\MPcomp}\ms{1}g\ms{5}{\cap}\ms{5}((f^{\MPrev}\ms{1}{\MPcomp}\ms{1}T\ms{1}{\MPcomp}\ms{1}g)\ms{1}{\setminus}\ms{1}(f^{\MPrev}\ms{1}{\MPcomp}\ms{1}T\ms{1}{\MPcomp}\ms{1}g)\ms{1}{/}\ms{1}(f^{\MPrev}\ms{1}{\MPcomp}\ms{1}T\ms{1}{\MPcomp}\ms{1}g))^{\MPrev}$\push\-\\
	$=$	\>	\>$\{$	\>\+\+\+definition of $R$ and $\Delta{}R$\-\-$~~~ \}$\pop\\
	$\Delta{}R~~.$
\end{mpdisplay}
A crucial step in the above calculation is the use of the property\begin{equation}\label{ARSB}
U\ms{2}{\cap}\ms{2}p{\MPcomp}V{\MPcomp}q\ms{6}{=}\ms{6}p{\MPcomp}(U{\cap}V){\MPcomp}q\ms{6}{=}\ms{6}p{\MPcomp}U{\MPcomp}q\ms{2}{\cap}\ms{2}V
\end{equation}for all relations $U$ and $V$ and coreflexive relations $p$ and $q$.   This is a frequently used property of domain
restriction.  

The remaining equations (\ref{BD.provorder.domdelta}) and (\ref{BD.provorder.perdom}) are straightforward.  
First 
\begin{mpdisplay}{0.15em}{6.5mm}{0mm}{2}
	$(\Delta{}R){\MPldom{}}$\push\-\\
	$=$	\>	\>$\{$	\>\+\+\+(\ref{BD.provorder.fwokg})\-\-$~~~ \}$\pop\\
	$(f^{\MPrev}\ms{1}{\MPcomp}\ms{1}g){\MPldom{}}$\push\-\\
	$=$	\>	\>$\{$	\>\+\+\+domains and assumption:  $f{\MPldom{}}\ms{2}{=}\ms{2}g{\MPldom{}}$\-\-$~~~ \}$\pop\\
	$f{\MPrdom{}}$\push\-\\
	$=$	\>	\>$\{$	\>\+\+\+assumption:  $f{\MPldom{}}\ms{3}{=}\ms{3}T\ms{1}{\cap}\ms{1}T^{\MPrev}$\-\-$~~~ \}$\pop\\
	$((T\ms{1}{\cap}\ms{1}T^{\MPrev})\ms{1}{\MPcomp}\ms{1}f){\MPrdom{}}$\push\-\\
	$=$	\>	\>$\{$	\>\+\+\+domains and converse\-\-$~~~ \}$\pop\\
	$(f^{\MPrev}\ms{1}{\MPcomp}\ms{1}(T\ms{1}{\cap}\ms{1}T^{\MPrev})){\MPldom{}}$\push\-\\
	$=$	\>	\>$\{$	\>\+\+\+lemma \ref{p.ordering.dom} and domains\-\-$~~~ \}$\pop\\
	$(f^{\MPrev}\ms{1}{\MPcomp}\ms{1}T){\MPldom{}}$\push\-\\
	$=$	\>	\>$\{$	\>\+\+\+domains and assumption:  $g{\MPldom{}}\ms{2}{=}\ms{2}T\ms{1}{\cap}\ms{1}T^{\MPrev}$\\
	 and lemma \ref{p.ordering.dom}\-\-$~~~ \}$\pop\\
	$(f^{\MPrev}\ms{1}{\MPcomp}\ms{1}T\ms{1}{\MPcomp}\ms{1}g){\MPldom{}}~~.$
\end{mpdisplay}
That is  $(\Delta{}R){\MPldom{}}\ms{3}{=}\ms{3}R{\MPldom{}}$.    The dual equation  $(\Delta{}R){\MPrdom{}}\ms{3}{=}\ms{3}R{\MPrdom{}}$ is immediate from the fact that $(\Delta{}R)^{\MPrev}\ms{1}{=}\ms{1}\Delta(R^{\MPrev})$ and
properties of the domain operators.     For the per  domains, we have:
\begin{mpdisplay}{0.15em}{6.5mm}{0mm}{2}
	$R{\MPperldom{}}$\push\-\\
	$=$	\>	\>$\{$	\>\+\+\+$R{\MPldom{}}\ms{2}{=}\ms{2}(\Delta{}R){\MPldom{}}$ and $R{\MPrdom{}}\ms{2}{=}\ms{2}(\Delta{}R){\MPrdom{}}$ (above);  lemma \ref{diag.dom.perdom} \-\-$~~~ \}$\pop\\
	$(\Delta{}R){\MPperldom{}}$\push\-\\
	$=$	\>	\>$\{$	\>\+\+\+$\Delta{}R$ is difunctional,  theorem \ref{difunctional.strongdef} (with $R\ms{1}{:=}\ms{1}\Delta{}R$)\-\-$~~~ \}$\pop\\
	$\Delta{}R\ms{1}{\MPcomp}\ms{1}\Delta{}R^{\MPrev}$\push\-\\
	$=$	\>	\>$\{$	\>\+\+\+lemma \ref{BD.provorder} and definition of $\Delta{}R$ \-\-$~~~ \}$\pop\\
	$f^{\MPrev}\ms{1}{\MPcomp}\ms{1}g\ms{1}{\MPcomp}\ms{1}(f^{\MPrev}\ms{1}{\MPcomp}\ms{1}g)^{\MPrev}$\push\-\\
	$=$	\>	\>$\{$	\>\+\+\+converse and $f{\MPldom{}}\ms{4}{=}\ms{4}g{\MPldom{}}\ms{4}{=}\ms{4}g\ms{1}{\MPcomp}\ms{1}g^{\MPrev}$\-\-$~~~ \}$\pop\\
	$f^{\MPrev}\ms{1}{\MPcomp}\ms{1}f~~.$
\end{mpdisplay}
Again, the dual equation is immediate.  
%\vspace{-7mm}
\MPendBox

We now prove the converse of lemma  \ref{BD.provorder}.
\begin{Lemma}\label{BD.diagdom.if}{\rm \ \ \ A relation $R$ is block-ordered if $R{\MPldom{}}\ms{2}{=}\ms{2}(\Delta{}R){\MPldom{}}$ and $R{\MPrdom{}}\ms{2}{=}\ms{2}(\Delta{}R){\MPrdom{}}$.
}%
\end{Lemma}%
{\bf Proof}~~~Suppose $R{\MPldom{}}\ms{2}{=}\ms{2}(\Delta{}R){\MPldom{}}$ and $R{\MPrdom{}}\ms{2}{=}\ms{2}(\Delta{}R){\MPrdom{}}$.  Our task is to construct relations  $f$,  $g$ and  $T$ such that\begin{displaymath}R\ms{2}{=}\ms{2}f^{\MPrev}\ms{1}{\MPcomp}\ms{1}T\ms{1}{\MPcomp}\ms{1}g~~,\end{displaymath}\begin{displaymath}T\ms{1}{\cap}\ms{1}T^{\MPrev}\ms{3}{\subseteq}\ms{3}I\ms{7}{\wedge}\ms{7}T\ms{3}{=}\ms{3}(T\ms{1}{\cap}\ms{1}T^{\MPrev})\ms{1}{\MPcomp}\ms{1}T\ms{1}{\MPcomp}\ms{1}(T\ms{1}{\cap}\ms{1}T^{\MPrev})\ms{7}{\wedge}\ms{7}T{\MPcomp}T\ms{1}{\subseteq}\ms{1}T\mbox{~~~and}\end{displaymath} \begin{displaymath}f\ms{1}{\MPcomp}\ms{1}f^{\MPrev}\ms{4}{=}\ms{4}f{\MPldom{}}\ms{4}{=}\ms{4}T\ms{1}{\cap}\ms{1}T^{\MPrev}\ms{4}{=}\ms{4}g{\MPldom{}}\ms{4}{=}\ms{4}g\ms{1}{\MPcomp}\ms{1}g^{\MPrev}~~.\end{displaymath}Since $\Delta{}R$ is difunctional,  theorem \ref{fwokg} is the obvious place to start.  Applying the theorem, 
we can  construct $f$ and $g$ such that  \begin{displaymath}\Delta{}R\ms{2}{=}\ms{2}f^{\MPrev}\ms{1}{\MPcomp}\ms{1}g\ms{5}{\wedge}\ms{5}f\ms{1}{\MPcomp}\ms{1}f^{\MPrev}\ms{3}{=}\ms{3}f{\MPldom{}}\ms{3}{=}\ms{3}g\ms{1}{\MPcomp}\ms{1}g^{\MPrev}\ms{3}{=}\ms{3}g{\MPldom{}}~~.\end{displaymath}Using standard properties of the domain operators together with the assumption that 
$R{\MPldom{}}\ms{2}{=}\ms{2}(\Delta{}R){\MPldom{}}$ and $R{\MPrdom{}}\ms{2}{=}\ms{2}(\Delta{}R){\MPrdom{}}$, it follows that \begin{displaymath}R{\MPldom{}}\ms{2}{=}\ms{2}f{\MPrdom{}}\ms{5}{\wedge}\ms{5}R{\MPrdom{}}\ms{2}{=}\ms{2}g{\MPrdom{}}~~.\end{displaymath}It remains to construct the provisional ordering $T$.  The appropriate construction is suggested  by lemma
\ref{BD.provorder}, in particular (\ref{BD.provorder.Rwok}).  Specifically, we define $T$ by the equation\begin{equation}\label{def.prov.order}
T\ms{4}{=}\ms{4}g\ms{1}{\MPcomp}\ms{1}R{\setminus}R{/}R\ms{1}{\MPcomp}\ms{1}f^{\MPrev}~~.
\end{equation}The proof that $R\ms{2}{=}\ms{2}f^{\MPrev}\ms{1}{\MPcomp}\ms{1}T\ms{1}{\MPcomp}\ms{1}g$ is by mutual inclusion.  First note that \begin{equation}\label{prov.order.delta}
f^{\MPrev}\ms{1}{\MPcomp}\ms{1}T\ms{1}{\MPcomp}\ms{1}g\ms{4}{=}\ms{4}\Delta{}R\ms{1}{\MPcomp}\ms{1}R{\setminus}R{/}R\ms{1}{\MPcomp}\ms{1}\Delta{}R
\end{equation}since
\begin{mpdisplay}{0.15em}{6.5mm}{0mm}{2}
	$f^{\MPrev}\ms{1}{\MPcomp}\ms{1}T\ms{1}{\MPcomp}\ms{1}g$\push\-\\
	$=$	\>	\>$\{$	\>\+\+\+(\ref{def.prov.order})\-\-$~~~ \}$\pop\\
	$f^{\MPrev}\ms{1}{\MPcomp}\ms{1}g\ms{1}{\MPcomp}\ms{1}R{\setminus}R{/}R\ms{1}{\MPcomp}\ms{1}f^{\MPrev}\ms{1}{\MPcomp}\ms{1}g$\push\-\\
	$=$	\>	\>$\{$	\>\+\+\+$\Delta{}R\ms{2}{=}\ms{2}f^{\MPrev}\ms{1}{\MPcomp}\ms{1}g$\-\-$~~~ \}$\pop\\
	$\Delta{}R\ms{1}{\MPcomp}\ms{1}R{\setminus}R{/}R\ms{1}{\MPcomp}\ms{1}\Delta{}R~~.$
\end{mpdisplay}
So
\begin{mpdisplay}{0.15em}{6.5mm}{0mm}{2}
	$f^{\MPrev}\ms{1}{\MPcomp}\ms{1}T\ms{1}{\MPcomp}\ms{1}g$\push\-\\
	$\subseteq$	\>	\>$\{$	\>\+\+\+(\ref{prov.order.delta}) and $\Delta{}R\ms{1}{\subseteq}\ms{1}R$\-\-$~~~ \}$\pop\\
	$R\ms{1}{\MPcomp}\ms{1}R{\setminus}R{/}R\ms{1}{\MPcomp}\ms{1}R$\push\-\\
	$\subseteq$	\>	\>$\{$	\>\+\+\+cancellation\-\-$~~~ \}$\pop\\
	$R~~.$
\end{mpdisplay}
Also,
\begin{mpdisplay}{0.15em}{6.5mm}{0mm}{2}
	$R\ms{2}{\subseteq}\ms{2}f^{\MPrev}\ms{1}{\MPcomp}\ms{1}T\ms{1}{\MPcomp}\ms{1}g$\push\-\\
	$=$	\>	\>$\{$	\>\+\+\+(\ref{prov.order.delta})\-\-$~~~ \}$\pop\\
	$R\ms{3}{\subseteq}\ms{3}\Delta{}R\ms{1}{\MPcomp}\ms{1}R{\setminus}R{/}R\ms{1}{\MPcomp}\ms{1}\Delta{}R$\push\-\\
	$=$	\>	\>$\{$	\>\+\+\+per domains\-\-$~~~ \}$\pop\\
	$R{\MPperldom{}}\ms{1}{\MPcomp}\ms{1}R\ms{1}{\MPcomp}\ms{1}R{\MPperrdom{}}\ms{3}{\subseteq}\ms{3}\Delta{}R\ms{1}{\MPcomp}\ms{1}R{\setminus}R{/}R\ms{1}{\MPcomp}\ms{1}\Delta{}R$\push\-\\
	$=$	\>	\>$\{$	\>\+\+\+assumption:  $R{\MPldom{}}\ms{2}{=}\ms{2}(\Delta{}R){\MPldom{}}$ and $R{\MPrdom{}}\ms{2}{=}\ms{2}(\Delta{}R){\MPrdom{}}$,    lemma \ref{diag.dom.perdom}\-\-$~~~ \}$\pop\\
	$(\Delta{}R){\MPperldom{}}\ms{1}{\MPcomp}\ms{1}R\ms{1}{\MPcomp}\ms{1}(\Delta{}R){\MPperrdom{}}\ms{3}{\subseteq}\ms{3}\Delta{}R\ms{1}{\MPcomp}\ms{1}R{\setminus}R{/}R\ms{1}{\MPcomp}\ms{1}\Delta{}R$\push\-\\
	$=$	\>	\>$\{$	\>\+\+\+ $\Delta{}R$ is difunctional,  theorem \ref{difunctional.strongdef} (with $R\ms{1}{:=}\ms{1}\Delta{}R$)\-\-$~~~ \}$\pop\\
	$\Delta{}R\ms{1}{\MPcomp}\ms{1}\Delta{}R^{\MPrev}\ms{1}{\MPcomp}\ms{1}R\ms{1}{\MPcomp}\ms{1}\Delta{}R^{\MPrev}\ms{1}{\MPcomp}\ms{1}\Delta{}R\ms{4}{\subseteq}\ms{4}\Delta{}R\ms{1}{\MPcomp}\ms{1}R{\setminus}R{/}R\ms{1}{\MPcomp}\ms{1}\Delta{}R$\push\-\\
	$\Leftarrow$	\>	\>$\{$	\>\+\+\+monotonicity\-\-$~~~ \}$\pop\\
	$\Delta{}R^{\MPrev}\ms{1}{\MPcomp}\ms{1}R\ms{1}{\MPcomp}\ms{1}\Delta{}R^{\MPrev}\ms{4}{\subseteq}\ms{4}R{\setminus}R{/}R$\push\-\\
	$\Leftarrow$	\>	\>$\{$	\>\+\+\+$\Delta{}R^{\MPrev}\ms{1}{\subseteq}\ms{1}R{\setminus}R{/}R$, monotonicity\-\-$~~~ \}$\pop\\
	$R{\setminus}R{/}R\ms{1}{\MPcomp}\ms{1}R\ms{1}{\MPcomp}\ms{1}R{\setminus}R{/}R\ms{4}{\subseteq}\ms{4}R{\setminus}R{/}R$\push\-\\
	$=$	\>	\>$\{$	\>\+\+\+factors\-\-$~~~ \}$\pop\\
	$R\ms{1}{\MPcomp}\ms{1}R{\setminus}R{/}R\ms{1}{\MPcomp}\ms{1}R\ms{1}{\MPcomp}\ms{1}R{\setminus}R{/}R\ms{1}{\MPcomp}\ms{1}R\ms{4}{\subseteq}\ms{4}R$\push\-\\
	$=$	\>	\>$\{$	\>\+\+\+cancellation\-\-$~~~ \}$\pop\\
	$\mathsf{true}~~.$
\end{mpdisplay}
Combining the two inclusions we conclude that indeed $R\ms{2}{=}\ms{2}f^{\MPrev}\ms{1}{\MPcomp}\ms{1}T\ms{1}{\MPcomp}\ms{1}g$.  

We now establish the requirements on $T$.  First,
\begin{mpdisplay}{0.15em}{6.5mm}{0mm}{2}
	$T\ms{1}{\cap}\ms{1}T^{\MPrev}$\push\-\\
	$=$	\>	\>$\{$	\>\+\+\+definition and converse\-\-$~~~ \}$\pop\\
	$g\ms{1}{\MPcomp}\ms{1}R{\setminus}R{/}R\ms{1}{\MPcomp}\ms{1}f^{\MPrev}\ms{5}{\cap}\ms{5}f\ms{1}{\MPcomp}\ms{1}(R{\setminus}R{/}R)^{\MPrev}\ms{1}{\MPcomp}\ms{1}g^{\MPrev}$\push\-\\
	$\subseteq$	\>	\>$\{$	\>\+\+\+modular law\-\-$~~~ \}$\pop\\
	$f\ms{1}{\MPcomp}\ms{1}(f^{\MPrev}\ms{1}{\MPcomp}\ms{1}g\ms{1}{\MPcomp}\ms{1}R{\setminus}R{/}R\ms{1}{\MPcomp}\ms{1}f^{\MPrev}\ms{1}{\MPcomp}\ms{1}g\ms{4}{\cap}\ms{4}(R{\setminus}R{/}R)^{\MPrev})\ms{1}{\MPcomp}\ms{1}g^{\MPrev}$\push\-\\
	$=$	\>	\>$\{$	\>\+\+\+$\Delta{}R\ms{2}{=}\ms{2}f^{\MPrev}\ms{1}{\MPcomp}\ms{1}g$\-\-$~~~ \}$\pop\\
	$f\ms{1}{\MPcomp}\ms{1}(\Delta{}R\ms{1}{\MPcomp}\ms{1}R{\setminus}R{/}R\ms{1}{\MPcomp}\ms{1}\Delta{}R\ms{4}{\cap}\ms{4}(R{\setminus}R{/}R)^{\MPrev})\ms{1}{\MPcomp}\ms{1}g^{\MPrev}$\push\-\\
	$\subseteq$	\>	\>$\{$	\>\+\+\+$\Delta{}R\ms{1}{\subseteq}\ms{1}R$, monotonicity and cancellation\-\-$~~~ \}$\pop\\
	$f\ms{1}{\MPcomp}\ms{1}(R\ms{3}{\cap}\ms{3}(R{\setminus}R{/}R)^{\MPrev})\ms{1}{\MPcomp}\ms{1}g^{\MPrev}$\push\-\\
	$=$	\>	\>$\{$	\>\+\+\+$\Delta{}R\ms{2}{=}\ms{2}R\ms{3}{\cap}\ms{3}(R{\setminus}R{/}R)^{\MPrev}$\-\-$~~~ \}$\pop\\
	$f\ms{1}{\MPcomp}\ms{1}\Delta{}R\ms{1}{\MPcomp}\ms{1}g^{\MPrev}$\push\-\\
	$=$	\>	\>$\{$	\>\+\+\+$\Delta{}R\ms{2}{=}\ms{2}f^{\MPrev}\ms{1}{\MPcomp}\ms{1}g$\-\-$~~~ \}$\pop\\
	$f\ms{1}{\MPcomp}\ms{1}f^{\MPrev}\ms{1}{\MPcomp}\ms{1}g\ms{1}{\MPcomp}\ms{1}g^{\MPrev}$\push\-\\
	$=$	\>	\>$\{$	\>\+\+\+$f\ms{1}{\MPcomp}\ms{1}f^{\MPrev}\ms{3}{=}\ms{3}f{\MPldom{}}\ms{3}{=}\ms{3}g\ms{1}{\MPcomp}\ms{1}g^{\MPrev}\ms{3}{=}\ms{3}g{\MPldom{}}$\-\-$~~~ \}$\pop\\
	$f{\MPldom{}}~~.$
\end{mpdisplay}
Thus $T\ms{1}{\cap}\ms{1}T^{\MPrev}\ms{2}{\subseteq}\ms{2}f{\MPldom{}}$.   So $T\ms{1}{\cap}\ms{1}T^{\MPrev}\ms{2}{\subseteq}\ms{2}I$.   Now  
\begin{mpdisplay}{0.15em}{6.5mm}{0mm}{2}
	$f{\MPldom{}}\ms{2}{\subseteq}\ms{2}T\ms{1}{\cap}\ms{1}T^{\MPrev}$\push\-\\
	$=$	\>	\>$\{$	\>\+\+\+infima and $f{\MPldom{}}$ is coreflexive\-\-$~~~ \}$\pop\\
	$f{\MPldom{}}\ms{2}{\subseteq}\ms{2}T$\push\-\\
	$\Leftarrow$	\>	\>$\{$	\>\+\+\+domains\-\-$~~~ \}$\pop\\
	$f\ms{1}{\MPcomp}\ms{1}f^{\MPrev}\ms{3}{\subseteq}\ms{3}T$\push\-\\
	$\Leftarrow$	\>	\>$\{$	\>\+\+\+definition of $T$ and monotonicity\-\-$~~~ \}$\pop\\
	$f\ms{3}{\subseteq}\ms{3}g\ms{1}{\MPcomp}\ms{1}R{\setminus}R{/}R$\push\-\\
	$\Leftarrow$	\>	\>$\{$	\>\+\+\+$f{\MPldom{}}\ms{3}{=}\ms{3}g\ms{1}{\MPcomp}\ms{1}g^{\MPrev}$, domains and monotonicity\-\-$~~~ \}$\pop\\
	$g^{\MPrev}\ms{1}{\MPcomp}\ms{1}f\ms{3}{\subseteq}\ms{3}R{\setminus}R{/}R$\push\-\\
	$=$	\>	\>$\{$	\>\+\+\+$f^{\MPrev}\ms{1}{\MPcomp}\ms{1}g\ms{2}{=}\ms{2}\Delta{}R$\-\-$~~~ \}$\pop\\
	$\Delta{}R^{\MPrev}\ms{2}{\subseteq}\ms{2}R{\setminus}R{/}R$\push\-\\
	$=$	\>	\>$\{$	\>\+\+\+$\Delta{}R\ms{3}{=}\ms{3}R\ms{2}{\cap}\ms{2}(R{\setminus}R{/}R)^{\MPrev}$, converse\-\-$~~~ \}$\pop\\
	$\mathsf{true}~~.$
\end{mpdisplay}
So, by anti-symmetry we have established that $T\ms{1}{\cap}\ms{1}T^{\MPrev}\ms{2}{=}\ms{2}f{\MPldom{}}$.    Since also $f{\MPldom{}}\ms{1}{=}\ms{1}g{\MPldom{}}$, we conclude from the
definition of $T$ and properties of domains that \begin{displaymath}T\ms{3}{=}\ms{3}(T\ms{1}{\cap}\ms{1}T^{\MPrev})\ms{1}{\MPcomp}\ms{1}T\ms{1}{\MPcomp}\ms{1}(T\ms{1}{\cap}\ms{1}T^{\MPrev})~~.\end{displaymath}The final task is to show that $T$ is transitive:
\begin{mpdisplay}{0.15em}{6.5mm}{0mm}{2}
	$T{\MPcomp}T$\push\-\\
	$=$	\>	\>$\{$	\>\+\+\+definition\-\-$~~~ \}$\pop\\
	$g\ms{1}{\MPcomp}\ms{1}R{\setminus}R{/}R\ms{1}{\MPcomp}\ms{1}f^{\MPrev}\ms{1}{\MPcomp}\ms{1}g\ms{1}{\MPcomp}\ms{1}R{\setminus}R{/}R\ms{1}{\MPcomp}\ms{1}f^{\MPrev}$\push\-\\
	$=$	\>	\>$\{$	\>\+\+\+$\Delta{}R\ms{2}{=}\ms{2}f^{\MPrev}\ms{1}{\MPcomp}\ms{1}g$\-\-$~~~ \}$\pop\\
	$g\ms{1}{\MPcomp}\ms{1}R{\setminus}R{/}R\ms{1}{\MPcomp}\ms{1}\Delta{}R\ms{1}{\MPcomp}\ms{1}R{\setminus}R{/}R\ms{1}{\MPcomp}\ms{1}f^{\MPrev}$\push\-\\
	$\subseteq$	\>	\>$\{$	\>\+\+\+$\Delta{}R\ms{1}{\subseteq}\ms{1}R$\-\-$~~~ \}$\pop\\
	$g\ms{1}{\MPcomp}\ms{1}R{\setminus}R{/}R\ms{1}{\MPcomp}\ms{1}R\ms{1}{\MPcomp}\ms{1}R{\setminus}R{/}R\ms{1}{\MPcomp}\ms{1}f^{\MPrev}$\push\-\\
	$\subseteq$	\>	\>$\{$	\>\+\+\+factors\-\-$~~~ \}$\pop\\
	$g\ms{1}{\MPcomp}\ms{1}R{\setminus}R{/}R\ms{1}{\MPcomp}\ms{1}f^{\MPrev}$\push\-\\
	$=$	\>	\>$\{$	\>\+\+\+definition\-\-$~~~ \}$\pop\\
	$T~~.$
\end{mpdisplay}
\vspace{-7mm}
\MPendBox

It is interesting to reflect on the proof of lemma \ref{BD.diagdom.if}.    The hardest part is to find appropriate
definitions of $f$, $g$ and $T$ such that $R\ms{2}{=}\ms{2}f^{\MPrev}\ms{1}{\MPcomp}\ms{1}T\ms{1}{\MPcomp}\ms{1}g$.  The key to constructing $f$ and $g$ is Riguet's ``analogie
frappante'' \cite{Riguet51} whereby he introduced the  ``diff\a'{e}rence'' ---the diagonal $\Delta{}R$---  of the relation $R$.  
Expressing the diagonal in terms of factors as we have done makes many parts of the calculations very
straightforward.  One much less straightforward step is the use of lemma \ref{diag.dom.perdom} in the proof that 
$R\ms{2}{\subseteq}\ms{2}f^{\MPrev}\ms{1}{\MPcomp}\ms{1}T\ms{1}{\MPcomp}\ms{1}g$.  Note how calculational needs drive the search for the lemma:  in order to simplify the
inclusion it is necessary to expose the term $R{\setminus}R{/}R$ on the right side, and that is precisely what the lemma
enables. 

We conclude with the theorem that we call the ``analogie frappante''.  It is not the theorem that Riguet
suggested  but we have chosen to give it this name in order to recognise Riguet's contribution.
\begin{Theorem}[Analogie Frappante]\label{BD.diagdom}{\rm \ \ \ A relation $R$ is block-ordered if and only 
if  $R{\MPldom{}}\ms{2}{=}\ms{2}(\Delta{}R){\MPldom{}}$ and $R{\MPrdom{}}\ms{2}{=}\ms{2}(\Delta{}R){\MPrdom{}}$. 
}
\end{Theorem}
{\bf Proof} ~~~Lemma \ref{BD.provorder} establishes ``only-if'' and lemma \ref{BD.diagdom.if} establishes ``if''.
\MPendBox 

{} 

\begin{Example}\label{membership.analogie}{\rm \ \ \ A generic way to construct examples of relations that are not 
block-ordered is to exploit example  \ref{membership.diagonal}.  In order to avoid unnecessary repetition, we  
refer the reader to that example for the definition of the relation $\mathsf{in}$ given a finite set  $\mathcal{X}$ and a set $\mathcal{S}$ of 
subsets of $\mathcal{X}$.  

%(Example \ref{ex:notbd} is a slightly disguised instance of the generic construction:  the nodes $A$ and $B$ can be 
%identified with, respectively,   $\{\alpha{,}\beta\}$ and $\{\beta{,}\gamma\}$.)

Recall  that the diagonal  $\Delta\mathsf{in}$ of type $\mathcal{X}{\sim}\mathcal{S}$ is injective.  It follows that the
size of  $(\Delta\mathsf{in}){\MPldom{}}$  is at most  the size of  $\mathcal{S}$.    If, however, the set $\mathcal{S}$ has $\mathcal{X}$ as one of its elements, the size of $\mathsf{in}{\MPldom{}}$
equals the size of $\mathcal{X}$.   Theorem  \ref{BD.diagdom} thus predicts that,  if $\mathcal{X}$ is an element of $\mathcal{S}$,   a necessary
condition for $\mathsf{in}$ to be block-ordered is  that the sizes of  $\mathcal{X}$ and $\mathcal{S}$ must be equal;  conversely,  if $\mathcal{X}$ is an
element of $\mathcal{S}$,  $\mathsf{in}$ is not block-ordered if  the sizes of  $\mathcal{X}$ and $\mathcal{S}$ are  different.   

Fig.\ \ref{membershipdiagFig} (example \ref{membership.diagonal})  shows that, even if the sizes of $\mathcal{X}$ and $\mathcal{S}$ are
equal, the relation $\mathsf{in}$ may not be block-ordered: as remarked then,  for the choice of $\mathcal{S}$ shown in fig.\ 
\ref{membershipdiagFig},  $\mathsf{in}{\MPldom{}}$ and $(\Delta\mathsf{in}){\MPldom{}}$  are different since $0$ and $3$  are  elements  of  the former but not the 
latter.  

It is straightforward to construct instances of $\mathcal{X}$ and $\mathcal{S}$ such that the relation $\mathsf{in}$ is block-ordered.  It
suffices to ensure that three conditions are satisfied:   $\mathcal{X}$ is an element of $\mathcal{S}$,  the sizes of $\mathcal{X}$ and $\mathcal{S}$ are equal,
and, for each $x$ in $\mathcal{X}$,  the set of sets represented by $(x{\MPcomp}\mathsf{in}){\MPrdom{}}$ is totally ordered.   Fig.\ \ref{membershipBDFig} is
one such.  Referring to definition  \ref{def:Block-Ordered Relation},  the functional $f$ is $\Delta\mathsf{in}^{\MPrev}$ 
(depicted by rectangles)  and the functional $g$ is $I_{\mathcal{S}}$;  the ordering relation is 
the subset relation $\mathsf{in}{\setminus}\mathsf{in}$ (depicted by the directed graph).  

\begin{figure}[h]
\centering \includegraphics{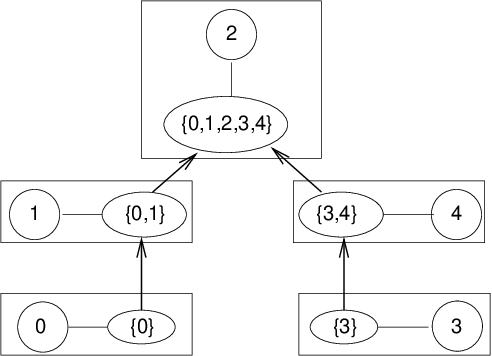}
 
\caption{A Block-Ordered Membership Relation}\label{membershipBDFig}
\end{figure}

\vspace{-7mm}
}%
\MPendBox\end{Example}

The following theorem is a corollary of theorem \ref{diagonal.core}.   In combination with theorem 
\ref{BD.diagdom} it states that a relation is block-ordered iff its core is block-ordered.  Testing whether or
not  a given relation is block-ordered can thus be decomposed into computing a core of the relation
and then testing whether or not that is block-ordered.   (For practical purposes computing an index of the
relation is to be preferred.)
\begin{Theorem}\label{diagonal.core.domain}{\rm \ \ \ Suppose $R$ is an arbitrary relation and suppose $C$ is a core of $R$ as
witnessed by $\lambda$ and $\rho$.  Then \begin{displaymath}R{\MPldom{}}\ms{3}{=}\ms{3}(\Delta{}R){\MPldom{}}\ms{7}{\equiv}\ms{7}C{\MPldom{}}\ms{3}{=}\ms{3}(\Delta{}C){\MPldom{}}~~.\end{displaymath}Dually,  \begin{displaymath}R{\MPrdom{}}\ms{3}{=}\ms{3}(\Delta{}R){\MPrdom{}}\ms{7}{\equiv}\ms{7}C{\MPrdom{}}\ms{3}{=}\ms{3}(\Delta{}C){\MPrdom{}}~~.\end{displaymath}
}
\end{Theorem}
{\bf Proof}~~~Suppose $R$, $C$,   $\lambda$ and $\rho$ are as in definition \ref{core}.  Then
\begin{mpdisplay}{0.15em}{6.5mm}{0mm}{2}
	$C{\MPldom{}}\ms{3}{=}\ms{3}(\Delta{}C){\MPldom{}}$\push\-\\
	$=$	\>	\>$\{$	\>\+\+\+definition \ref{core} and theorem  \ref{diagonal.core} \-\-$~~~ \}$\pop\\
	$(\lambda\ms{1}{\MPcomp}\ms{1}R\ms{1}{\MPcomp}\ms{1}\rho^{\MPrev}){\MPldom{}}\ms{3}{=}\ms{3}(\lambda\ms{1}{\MPcomp}\ms{1}\Delta{}R\ms{1}{\MPcomp}\ms{1}\rho^{\MPrev}){\MPldom{}}$\push\-\\
	$\Rightarrow$	\>	\>$\{$	\>\+\+\+Leibniz\-\-$~~~ \}$\pop\\
	$(\lambda^{\MPrev}\ms{1}{\MPcomp}\ms{1}(\lambda\ms{1}{\MPcomp}\ms{1}R\ms{1}{\MPcomp}\ms{1}\rho^{\MPrev}){\MPldom{}}){\MPldom{}}\ms{3}{=}\ms{3}(\lambda^{\MPrev}\ms{1}{\MPcomp}\ms{1}(\lambda\ms{1}{\MPcomp}\ms{1}\Delta{}R\ms{1}{\MPcomp}\ms{1}\rho^{\MPrev}){\MPldom{}}){\MPldom{}}$\push\-\\
	$=$	\>	\>$\{$	\>\+\+\+domains\-\-$~~~ \}$\pop\\
	$(\lambda^{\MPrev}\ms{1}{\MPcomp}\ms{1}\lambda\ms{1}{\MPcomp}\ms{1}R\ms{1}{\MPcomp}\ms{1}\rho^{\MPrev}){\MPldom{}}\ms{3}{=}\ms{3}(\lambda^{\MPrev}\ms{1}{\MPcomp}\ms{1}\lambda\ms{1}{\MPcomp}\ms{1}\Delta{}R\ms{1}{\MPcomp}\ms{1}\rho^{\MPrev}){\MPldom{}}$\push\-\\
	$=$	\>	\>$\{$	\>\+\+\+ $\lambda^{\MPrev}\ms{1}{\MPcomp}\ms{1}\lambda\ms{1}{\MPcomp}\ms{1}R\ms{4}{=}\ms{4}R{\MPperldom{}}\ms{1}{\MPcomp}\ms{1}R\ms{4}{=}\ms{4}R$,\\
	  $(\rho^{\MPrev}){\MPldom{}}\ms{3}{=}\ms{3}(\rho^{\MPrev}\ms{1}{\MPcomp}\ms{1}\rho){\MPldom{}}\ms{3}{=}\ms{3}(R{\MPperrdom{}}){\MPldom{}}\ms{3}{=}\ms{3}R{\MPrdom{}}$,  and domains \-\-$~~~ \}$\pop\\
	$R{\MPldom{}}\ms{3}{=}\ms{3}(\lambda^{\MPrev}\ms{1}{\MPcomp}\ms{1}\lambda\ms{1}{\MPcomp}\ms{1}\Delta{}R\ms{1}{\MPcomp}\ms{1}\rho^{\MPrev}){\MPldom{}}$\push\-\\
	$=$	\>	\>$\{$	\>\+\+\+$(\rho^{\MPrev}){\MPldom{}}\ms{3}{=}\ms{3}(\rho^{\MPrev}\ms{1}{\MPcomp}\ms{1}\rho){\MPldom{}}$  and   domains\-\-$~~~ \}$\pop\\
	$R{\MPldom{}}\ms{3}{=}\ms{3}(\lambda^{\MPrev}\ms{1}{\MPcomp}\ms{1}\lambda\ms{1}{\MPcomp}\ms{1}\Delta{}R\ms{1}{\MPcomp}\ms{1}\rho^{\MPrev}\ms{1}{\MPcomp}\ms{1}\rho){\MPldom{}}$\push\-\\
	$=$	\>	\>$\{$	\>\+\+\+theorem \ref{diagonal.core}\-\-$~~~ \}$\pop\\
	$R{\MPldom{}}\ms{3}{=}\ms{3}(\lambda^{\MPrev}\ms{1}{\MPcomp}\ms{1}\Delta{}C\ms{1}{\MPcomp}\ms{1}\rho){\MPldom{}}$\push\-\\
	$=$	\>	\>$\{$	\>\+\+\+theorem \ref{diagonal.core}\-\-$~~~ \}$\pop\\
	$R{\MPldom{}}\ms{3}{=}\ms{3}(\Delta{}R){\MPldom{}}~~.$
\end{mpdisplay}
Similarly,
\begin{mpdisplay}{0.15em}{6.5mm}{0mm}{2}
	$R{\MPldom{}}\ms{3}{=}\ms{3}(\Delta{}R){\MPldom{}}$\push\-\\
	$=$	\>	\>$\{$	\>\+\+\+definition \ref{core}, theorem  \ref{diagonal.core} and Leibniz\-\-$~~~ \}$\pop\\
	$(\lambda^{\MPrev}\ms{1}{\MPcomp}\ms{1}C\ms{1}{\MPcomp}\ms{1}\rho){\MPldom{}}\ms{3}{=}\ms{3}(\lambda^{\MPrev}\ms{1}{\MPcomp}\ms{1}\Delta{}C\ms{1}{\MPcomp}\ms{1}\rho){\MPldom{}}$\push\-\\
	$\Rightarrow$	\>	\>$\{$	\>\+\+\+Leibniz and domains\-\-$~~~ \}$\pop\\
	$(\lambda\ms{1}{\MPcomp}\ms{1}\lambda^{\MPrev}\ms{1}{\MPcomp}\ms{1}C\ms{1}{\MPcomp}\ms{1}\rho){\MPldom{}}\ms{3}{=}\ms{3}(\lambda\ms{1}{\MPcomp}\ms{1}\lambda^{\MPrev}\ms{1}{\MPcomp}\ms{1}\Delta{}C\ms{1}{\MPcomp}\ms{1}\rho){\MPldom{}}$\push\-\\
	$=$	\>	\>$\{$	\>\+\+\+$\rho{\MPldom{}}\ms{3}{=}\ms{3}(\rho\ms{1}{\MPcomp}\ms{1}\rho^{\MPrev}){\MPldom{}}$  and   domains\-\-$~~~ \}$\pop\\
	$(\lambda\ms{1}{\MPcomp}\ms{1}\lambda^{\MPrev}\ms{1}{\MPcomp}\ms{1}C\ms{1}{\MPcomp}\ms{1}\rho\ms{1}{\MPcomp}\ms{1}\rho^{\MPrev}){\MPldom{}}\ms{3}{=}\ms{3}(\lambda\ms{1}{\MPcomp}\ms{1}\lambda^{\MPrev}\ms{1}{\MPcomp}\ms{1}\Delta{}C\ms{1}{\MPcomp}\ms{1}\rho\ms{1}{\MPcomp}\ms{1}\rho^{\MPrev}){\MPldom{}}$\push\-\\
	$=$	\>	\>$\{$	\>\+\+\+theorem \ref{diagonal.core} (applied twice)\-\-$~~~ \}$\pop\\
	$C{\MPldom{}}\ms{3}{=}\ms{3}(\Delta{}C){\MPldom{}}~~.$
\end{mpdisplay}
The property \begin{displaymath}R{\MPldom{}}\ms{3}{=}\ms{3}(\Delta{}R){\MPldom{}}\ms{7}{\equiv}\ms{7}C{\MPldom{}}\ms{3}{=}\ms{3}(\Delta{}C){\MPldom{}}\end{displaymath}follows by mutual implication.  The dual follows by instantiating  $R$ to  $R^{\MPrev}$ and applying the properties of
converse.
\MPendBox

By combining the definition of block-ordering with theorem \ref{diagonal.core}, it is immediately clear that 
$R$ is block-ordered if  its core $C$ is a provisional ordering.  In general, a core of a block-ordered relation will not
be a provisional ordering.  This is  because the types of the targets of the  components $\lambda$ and $\rho$ in the
definition of a core are  not required to be the same; on the other hand, orderings are  required to be 
homogeneous relations.    However by carefully restricting the choice of core,  it is possible to construct a
core that is indeed a provisional ordering. 
\begin{Theorem}\label{diagonal.core.ordering}{\rm \ \ \ Suppose $R$ is an arbitrary relation.  Then  if  $R$ is block-ordered it
has a core  that is a provisional ordering.
}
\end{Theorem}
{\bf Proof}~~~Suppose $R$ is block-ordered. That is, suppose that $f$, $g$ and $T$ are relations such that  $T$ is a
provisional ordering, \begin{displaymath}R\ms{2}{=}\ms{2}f^{\MPrev}\ms{1}{\MPcomp}\ms{1}T\ms{1}{\MPcomp}\ms{1}g\end{displaymath}and \begin{displaymath}f\ms{1}{\MPcomp}\ms{1}f^{\MPrev}\ms{4}{=}\ms{4}f{\MPldom{}}\ms{4}{=}\ms{4}T\ms{1}{\cap}\ms{1}T^{\MPrev}\ms{4}{=}\ms{4}g{\MPldom{}}\ms{4}{=}\ms{4}g\ms{1}{\MPcomp}\ms{1}g^{\MPrev}~~.\end{displaymath}Then,  by lemma \ref{BD.provorder},    $R{\MPperldom{}}\ms{2}{=}\ms{2}f^{\MPrev}\ms{1}{\MPcomp}\ms{1}f$ and ,  $R{\MPperrdom{}}\ms{2}{=}\ms{2}g^{\MPrev}\ms{1}{\MPcomp}\ms{1}g$.   Thus $f$ and $g$ satisfy the conditions
for witnessing  a core  $C$ of $R$.  (Cf.\ definition \ref{core} with $\lambda{,}\rho\ms{1}{:=}\ms{1}f{,}g$.)  Consequently,
\begin{mpdisplay}{0.15em}{6.5mm}{0mm}{2}
	$C$\push\-\\
	$=$	\>	\>$\{$	\>\+\+\+definition \ref{core}\-\-$~~~ \}$\pop\\
	$f\ms{1}{\MPcomp}\ms{1}R\ms{1}{\MPcomp}\ms{1}g^{\MPrev}$\push\-\\
	$=$	\>	\>$\{$	\>\+\+\+$R\ms{2}{=}\ms{2}f^{\MPrev}\ms{1}{\MPcomp}\ms{1}T\ms{1}{\MPcomp}\ms{1}g$\-\-$~~~ \}$\pop\\
	$f\ms{1}{\MPcomp}\ms{1}f^{\MPrev}\ms{1}{\MPcomp}\ms{1}T\ms{1}{\MPcomp}\ms{1}g\ms{1}{\MPcomp}\ms{1}g^{\MPrev}$\push\-\\
	$=$	\>	\>$\{$	\>\+\+\+$f\ms{1}{\MPcomp}\ms{1}f^{\MPrev}\ms{4}{=}\ms{4}f{\MPldom{}}\ms{4}{=}\ms{4}T\ms{1}{\cap}\ms{1}T^{\MPrev}\ms{4}{=}\ms{4}g{\MPldom{}}\ms{4}{=}\ms{4}g\ms{1}{\MPcomp}\ms{1}g^{\MPrev}$\-\-$~~~ \}$\pop\\
	$(T\ms{1}{\cap}\ms{1}T^{\MPrev})\ms{1}{\MPcomp}\ms{1}T\ms{1}{\MPcomp}\ms{1}(T\ms{1}{\cap}\ms{1}T^{\MPrev})$\push\-\\
	$=$	\>	\>$\{$	\>\+\+\+$T$ is a provisional ordering, lemma \ref{p.ordering.dom} and domains\-\-$~~~ \}$\pop\\
	$T~~.$
\end{mpdisplay}
We conclude that $C$ is the  provisional ordering $T$. 
%\vspace{-7mm}
\MPendBox

Combining theorem \ref{diagonal.core.ordering} with the theorem that all cores of a given relation are
isomorphic, we conclude that any core of a
block-ordered relation is isomorphic to a provisional ordering.  Loosely speaking, block-ordered
relations are   provisional orderings  up to isomorphism and reduction to the core.
\begin{Example}\label{diagonal.lessthan.again}{\rm \ \ \ From the Galois connection, for all reals $x$ and integers  $m$,  \begin{displaymath}\lceil{}x\rceil\ms{1}{\leq}\ms{1}m\ms{4}{\equiv}\ms{4}x\ms{1}{\leq}\ms{1}m\end{displaymath}defining the ceiling function, we deduce  that the heterogeneous relation  ${}_{\MPReal}{\leq}_{\MPInt}$  has core the provisional
ordering ${\leq}_{\MPInt}$.  This is because  the ceiling function is surjective.   Its core in \emph{not} the ordering ${\leq}_{\MPReal}$ because
the  coercion $\mathsf{real}$  from integers to reals is  not surjective.  (See also  example  \ref{heterogeneous.atmost}.)

On the other hand, if a  Galois connection  \begin{displaymath}F{.}b\ms{1}{\sqsubseteq}\ms{1}a\ms{2}{\equiv}\ms{2}b\ms{1}{\preceq}\ms{1}G{.}a\end{displaymath}of posets  $(\mathcal{A},\sqsubseteq)$ and ($\mathcal{B}\ms{1}{,}\ms{1}{\preceq}$)  is ``perfect'' (i.e.\ both $F$ and $G$ are surjective),  both the orderings ${\sqsubseteq}$ and ${\preceq}$ are 
cores of the defined heterogeneous relation.  That the orderings are isomorphic is an instance of  the
unity-of-opposites theorem \cite{Bac02}.
}%
\MPendBox\end{Example}

\section{Staircase Relations}\label{sec:Staircase Relations}

For any binary  relation $R$, the relations $R{\setminus}R$ and $R{/}R$ are preorders.  That is, both are transitive and
reflexive.   (If $R$ has type $A{\sim}B$ then $R{\setminus}R$ has type $B{\sim}B$ and $R{/}R$ has type $A{\sim}A$.)
That  relation $R$ is a ``staircase'' relation  means formally 
that the preorder $R{\setminus}R$ is linear\footnote{An
ordering  $S$ ---of any sort--- is said to be \emph{linear}  if $S\ms{1}{\cup}\ms{1}S^{\MPrev}\ms{3}{=}\ms{3}{\MPplattop}$.  Sometimes the word ``total'' is
used instead of linear.  For example,  Riguet \cite{Riguet51} uses the term ``totalement
ordonn\a'{e}es''.}.  For brevity, we denote the property of being a staircase relation by $\mathsf{SC}$.  That is:
\begin{Definition}\label{BD}{\rm \ \ \ The predicate $\mathsf{SC}$ on (binary) relations is defined by, for all $R$,\begin{displaymath}\mathsf{SC}{.}R\ms{9}{\equiv}\ms{9}R{\setminus}R\ms{1}{\cup}\ms{1}(R{\setminus}R)^{\MPrev}\ms{3}{=}\ms{3}{\MPplattop}~~.\end{displaymath}A relation that satisfies the predicate $\mathsf{SC}$ is called a \emph{staircase} relation.
%\vspace{-7mm}
}
\MPendBox\end{Definition}

The pointwise formulation of the relation $R{\setminus}R$ is \begin{displaymath}b0\MPdopen{}R{\setminus}R\MPdclose{}b1\ms{7}{\equiv}\ms{7}{\left\langle\forall{}a\ms{2}{:}\ms{2}a\MPdopen{}R\MPdclose{}b0\ms{2}{:}\ms{2}a\MPdopen{}R\MPdclose{}b1\right\rangle}~~.\end{displaymath}In terms of the mental picture shown in fig.\ \ref{fig:staircase},  a vertical line through a
point $b$  depicts the set of points $a$ such that $a\MPdopen{}R\MPdclose{}b$;  a staircase relation is one such that the points of type $B$
can be ordered in such a way that the preorder $R{\setminus}R$ is depicted by the left-to-right ordering of points on
the  $B$-axis,   and the length of the  vertical lines increases monotonically (although not always
strictly) as one proceeds from left to right in the diagram.  

%In terms of the mental picture  of a staircase relation shown  in fig. \ref{fig:staircase},  the relation $R{\setminus}R$ is a
%relation between vertical lines: a relation is a staircase relation if the elements of the type $B$ in the figure
%can be ordered in such a way that the vertical lines are increasing  in length; the lengths increase  in
%``steps''.    (Similarly, the relation $R{/}R$  is a relation between horizontal lines.)

Various equivalent definitions of the predicate $\mathsf{SC}$ are given below.  Riguet \cite{Riguet51} defined a ``relation
de Ferrers'' to be a relation satisfying (\ref{BDneg}).    The equivalence of (\ref{BDunder}) and (\ref{BDover})
corresponds in the mental picture of a staircase relation to the property that the 
 vertical lines being increasing in length is equivalent to the horizontal lines being decreasing  in length. 
(Cf.\  the statement of Riguet's theorem quoted in the introduction.)
\begin{Lemma}\label{BDequivalents}{\rm \ \ \ The following are all equivalent formulations of $\mathsf{SC}{.}R$:\begin{equation}\label{BDunder}
R{\setminus}R\ms{2}{\cup}\ms{2}(R{\setminus}R)^{\MPrev}\ms{4}{=}\ms{4}{\MPplattop}~~,
\end{equation}\begin{equation}\label{BDover}
R{/}R\ms{2}{\cup}\ms{2}(R{/}R)^{\MPrev}\ms{4}{=}\ms{4}{\MPplattop}~~,
\end{equation}\begin{equation}\label{BDboth}
R\ms{2}{\cup}\ms{2}(R{\setminus}R{/}R)^{\MPrev}\ms{4}{=}\ms{4}{\MPplattop}~~,
\end{equation}\begin{equation}\label{BDneg}
R\ms{1}{\MPcomp}\ms{1}{\neg}R^{\MPrev}\ms{1}{\MPcomp}\ms{1}R\ms{3}{\subseteq}\ms{3}R~~.
\end{equation}
}%
\end{Lemma}%
{\bf Proof}~~~We prove first that (\ref{BDover}) and (\ref{BDneg}) are equivalent:
\begin{mpdisplay}{0.15em}{6.5mm}{0mm}{2}
	$R\ms{1}{\MPcomp}\ms{1}{\neg}R^{\MPrev}\ms{1}{\MPcomp}\ms{1}R\ms{3}{\subseteq}\ms{3}R$\push\-\\
	$=$	\>	\>$\{$	\>\+\+\+factors\-\-$~~~ \}$\pop\\
	$R\ms{1}{\MPcomp}\ms{1}{\neg}R^{\MPrev}\ms{4}{\subseteq}\ms{4}R{/}R$\push\-\\
	$=$	\>	\>$\{$	\>\+\+\+complements\-\-$~~~ \}$\pop\\
	${\MPplattop}\ms{4}{\subseteq}\ms{4}R{/}R\ms{2}{\cup}\ms{2}{\neg}(R\ms{1}{\MPcomp}\ms{1}{\neg}R^{\MPrev})$\push\-\\
	$=$	\>	\>$\{$	\>\+\+\+ (\ref{underovernot}) with $R{,}S\ms{2}{:=}\ms{2}R^{\MPrev}\ms{1}{,}\ms{1}R^{\MPrev}$ (and $R\ms{1}{=}\ms{1}(R^{\MPrev})^{\MPrev}$)\-\-$~~~ \}$\pop\\
	${\MPplattop}\ms{5}{\subseteq}\ms{5}R{/}R\ms{3}{\cup}\ms{3}R^{\MPrev}\ms{1}{\setminus}\ms{1}R^{\MPrev}$\push\-\\
	$=$	\>	\>$\{$	\>\+\+\+(\ref{factor.props.under}) with $R{,}S\ms{1}{:=}\ms{1}R{,}R$\-\-$~~~ \}$\pop\\
	${\MPplattop}\ms{4}{\subseteq}\ms{4}R{/}R\ms{2}{\cup}\ms{2}(R{/}R)^{\MPrev}$\push\-\\
	$=$	\>	\>$\{$	\>\+\+\+$\left[\ms{1}S\ms{1}{\subseteq}\ms{1}{\MPplattop}\ms{1}\right]$ with $S\ms{2}{:=}\ms{2}R{/}R\ms{2}{\cup}\ms{2}(R{/}R)^{\MPrev}$ and anti-symmetry\-\-$~~~ \}$\pop\\
	${\MPplattop}\ms{4}{=}\ms{4}R{/}R\ms{2}{\cup}\ms{2}(R{/}R)^{\MPrev}~~.$
\end{mpdisplay}
A symmetric argument establishes the equivalence of (\ref{BDunder}) and (\ref{BDneg}):
\begin{mpdisplay}{0.15em}{6.5mm}{0mm}{2}
	$R\ms{1}{\MPcomp}\ms{1}{\neg}R^{\MPrev}\ms{1}{\MPcomp}\ms{1}R\ms{3}{\subseteq}\ms{3}R$\push\-\\
	$=$	\>	\>$\{$	\>\+\+\+factors\-\-$~~~ \}$\pop\\
	${\neg}R^{\MPrev}\ms{1}{\MPcomp}\ms{1}R\ms{4}{\subseteq}\ms{4}R{\setminus}R$\push\-\\
	$=$	\>	\>$\{$	\>\+\+\+complements\-\-$~~~ \}$\pop\\
	${\MPplattop}\ms{4}{\subseteq}\ms{4}R{\setminus}R\ms{2}{\cup}\ms{2}{\neg}({\neg}R^{\MPrev}\ms{1}{\MPcomp}\ms{1}R)$\push\-\\
	$=$	\>	\>$\{$	\>\+\+\+ (\ref{underovernot}) with $S{,}T\ms{2}{:=}\ms{2}R^{\MPrev}\ms{1}{,}\ms{1}R^{\MPrev}$ \-\-$~~~ \}$\pop\\
	${\MPplattop}\ms{5}{\subseteq}\ms{5}R{\setminus}R\ms{3}{\cup}\ms{3}R^{\MPrev}\ms{1}{/}\ms{1}R^{\MPrev}$\push\-\\
	$=$	\>	\>$\{$	\>\+\+\+(\ref{factor.props.over}) with $R{,}S\ms{1}{:=}\ms{1}R{,}R$ (and $R\ms{1}{=}\ms{1}(R^{\MPrev})^{\MPrev}$)\-\-$~~~ \}$\pop\\
	${\MPplattop}\ms{4}{\subseteq}\ms{4}R{\setminus}R\ms{2}{\cup}\ms{2}(R{\setminus}R)^{\MPrev}$\push\-\\
	$=$	\>	\>$\{$	\>\+\+\+$\left[\ms{1}S\ms{1}{\subseteq}\ms{1}{\MPplattop}\ms{1}\right]$ with $S\ms{2}{:=}\ms{2}R{\setminus}R\ms{2}{\cup}\ms{2}(R{\setminus}R)^{\MPrev}$ and anti-symmetry\-\-$~~~ \}$\pop\\
	${\MPplattop}\ms{4}{=}\ms{4}R{\setminus}R\ms{2}{\cup}\ms{2}(R{\setminus}R)^{\MPrev}~~.$
\end{mpdisplay}
  Finally,
\begin{mpdisplay}{0.15em}{6.5mm}{0mm}{2}
	$R\ms{1}{\MPcomp}\ms{1}{\neg}R^{\MPrev}\ms{1}{\MPcomp}\ms{1}R\ms{3}{\subseteq}\ms{3}R$\push\-\\
	$=$	\>	\>$\{$	\>\+\+\+factors\-\-$~~~ \}$\pop\\
	${\neg}R^{\MPrev}\ms{3}{\subseteq}\ms{3}R{\setminus}R{/}R$\push\-\\
	$=$	\>	\>$\{$	\>\+\+\+converse and complements \-\-$~~~ \}$\pop\\
	${\MPplattop}\ms{4}{\subseteq}\ms{4}R\ms{2}{\cup}\ms{2}(R{\setminus}R{/}R)^{\MPrev}$\push\-\\
	$=$	\>	\>$\{$	\>\+\+\+$\left[\ms{1}S\ms{1}{\subseteq}\ms{1}{\MPplattop}\ms{1}\right]$ with $S\ms{2}{:=}\ms{2}R\ms{2}{\cup}\ms{2}(R{\setminus}R{/}R)^{\MPrev}$ and anti-symmetry\-\-$~~~ \}$\pop\\
	${\MPplattop}\ms{4}{=}\ms{4}R\ms{2}{\cup}\ms{2}(R{\setminus}R{/}R)^{\MPrev}~~.$
\end{mpdisplay}
\vspace{-7mm}
\MPendBox

An example of a staircase relation predicted by lemma \ref{BD.ord} is the at-most relation --- on
natural numbers, integers, rational numbers or reals.  

Two general methods for identifying examples of staircase relations are given in lemmas 
\ref{BD.ord} and \ref{BD.strictord}.  
\begin{Lemma}\label{BD.ord}{\rm \ \ \ A linear preorder  is a staircase relation.  That is,  for all (homogeneous)  $R$,\begin{displaymath}\mathsf{SC}{.}R\ms{7}{\Leftarrow}\ms{7}R{\MPcomp}R\ms{1}{\subseteq}\ms{1}R\ms{4}{\wedge}\ms{4}I\ms{1}{\subseteq}\ms{1}R\ms{4}{\wedge}\ms{4}R\ms{1}{\cup}\ms{1}R^{\MPrev}\ms{2}{=}\ms{2}{\MPplattop}~~.\end{displaymath}
}%
\end{Lemma}%
{\bf Proof}~~~We have\begin{displaymath}R\ms{1}{=}\ms{1}R{\setminus}R{/}R\ms{6}{\Leftarrow}\ms{6}R{\MPcomp}R\ms{1}{\subseteq}\ms{1}R\ms{3}{\wedge}\ms{3}I\ms{1}{\subseteq}\ms{1}R\end{displaymath}since
\begin{mpdisplay}{0.15em}{6.5mm}{0mm}{2}
	$R\ms{2}{\subseteq}\ms{2}R{\setminus}R{/}R$\push\-\\
	$=$	\>	\>$\{$	\>\+\+\+factors\-\-$~~~ \}$\pop\\
	$R{\MPcomp}R{\MPcomp}R\ms{2}{\subseteq}\ms{2}R$\push\-\\
	$\Leftarrow$	\>	\>$\{$	\>\+\+\+monontonicity and transitivity\-\-$~~~ \}$\pop\\
	$R{\MPcomp}R\ms{1}{\subseteq}\ms{1}R$
\end{mpdisplay}
and
\begin{mpdisplay}{0.15em}{6.5mm}{0mm}{2}
	$R{\setminus}R{/}R\ms{2}{\subseteq}\ms{2}R$\push\-\\
	$=$	\>	\>$\{$	\>\+\+\+$\left[\ms{2}R\ms{1}{=}\ms{1}I{\setminus}R{/}I\ms{2}\right]$\-\-$~~~ \}$\pop\\
	$R{\setminus}R{/}R\ms{2}{\subseteq}\ms{2}I{\setminus}R{/}I$\push\-\\
	$\Leftarrow$	\>	\>$\{$	\>\+\+\+(anti)monotonicity\-\-$~~~ \}$\pop\\
	$I\ms{1}{\subseteq}\ms{1}R~~.$
\end{mpdisplay}
Also,\begin{displaymath}R^{\MPrev}\ms{1}{\MPcomp}\ms{1}R^{\MPrev}\ms{2}{\subseteq}\ms{2}R^{\MPrev}\ms{4}{\wedge}\ms{4}I\ms{1}{\subseteq}\ms{1}R^{\MPrev}\ms{7}{\equiv}\ms{7}R{\MPcomp}R\ms{1}{\subseteq}\ms{1}R\ms{3}{\wedge}\ms{3}I\ms{1}{\subseteq}\ms{1}R~~.\end{displaymath}(The converse of a preorder is a preorder.) So
\begin{mpdisplay}{0.15em}{6.5mm}{0mm}{2}
	$\mathsf{SC}{.}R$\push\-\\
	$=$	\>	\>$\{$	\>\+\+\+lemma \ref{BDequivalents}, in particular (\ref{BDboth})\-\-$~~~ \}$\pop\\
	$R\ms{2}{\cup}\ms{2}(R{\setminus}R{/}R)^{\MPrev}\ms{4}{=}\ms{4}{\MPplattop}$\push\-\\
	$=$	\>	\>$\{$	\>\+\+\+assumption:  $R$ is a preorder \\
	(hence,  $R^{\MPrev}$ is a preorder and $R^{\MPrev}\ms{2}{=}\ms{2}R^{\MPrev}\ms{1}{\setminus}\ms{1}R^{\MPrev}\ms{1}{/}\ms{1}R^{\MPrev}$)\\
	lemma \ref{factor.props}, in particular  (\ref{factor.props.both})\-\-$~~~ \}$\pop\\
	$R\ms{2}{\cup}\ms{2}R^{\MPrev}\ms{4}{=}\ms{4}{\MPplattop}$\push\-\\
	$=$	\>	\>$\{$	\>\+\+\+assumption:  $R$ is linear (i.e.  $R\ms{1}{\cup}\ms{1}R^{\MPrev}\ms{2}{=}\ms{2}{\MPplattop}$)\-\-$~~~ \}$\pop\\
	$\mathsf{true}~~.$
\end{mpdisplay}
\vspace{-7mm}
\MPendBox

The second way of constructing a staircase relation is to reduce a linear preorder by eliminating its
reflexive part (making it so-called ``strict'').  For example, the less-than relation (on  
natural numbers, integers, rational numbers or reals) is a staircase relation.    
Formally, we have:
\begin{Lemma}\label{BD.strictord}{\rm \ \ \ For all (homogeneous)  $R$,\begin{displaymath}\mathsf{SC}{.}R\ms{7}{\Leftarrow}\ms{7}R{\MPcomp}R\ms{1}{\subseteq}\ms{1}R\ms{4}{\wedge}\ms{4}R\ms{1}{\cup}\ms{1}I\ms{1}{\cup}\ms{1}R^{\MPrev}\ms{2}{=}\ms{2}{\MPplattop}~~.\end{displaymath}
}%
\end{Lemma}%
{\bf Proof}~~~
\begin{mpdisplay}{0.15em}{6.5mm}{0mm}{2}
	$\mathsf{SC}{.}R$\push\-\\
	$=$	\>	\>$\{$	\>\+\+\+ (\ref{BDboth})\-\-$~~~ \}$\pop\\
	$R\ms{2}{\cup}\ms{2}(R{\setminus}R{/}R)^{\MPrev}\ms{4}{=}\ms{4}{\MPplattop}$\push\-\\
	$=$	\>	\>$\{$	\>\+\+\+$\left[\ms{1}X\ms{1}{\subseteq}\ms{1}{\MPplattop}\ms{1}\right]$ and antisymmetry\-\-$~~~ \}$\pop\\
	${\MPplattop}\ms{2}{\subseteq}\ms{2}R\ms{2}{\cup}\ms{2}(R{\setminus}R{/}R)^{\MPrev}$\push\-\\
	$\Leftarrow$	\>	\>$\{$	\>\+\+\+assumption:  $R\ms{1}{\cup}\ms{1}I\ms{1}{\cup}\ms{1}R^{\MPrev}\ms{2}{=}\ms{2}{\MPplattop}$, so ${\MPplattop}\ms{2}{\subseteq}\ms{2}R\ms{1}{\cup}\ms{1}I\ms{1}{\cup}\ms{1}R^{\MPrev}$\\
	monotonicity and transitivity\-\-$~~~ \}$\pop\\
	$I\ms{1}{\cup}\ms{1}R^{\MPrev}\ms{2}{\subseteq}\ms{2}(R{\setminus}R{/}R)^{\MPrev}$\push\-\\
	$=$	\>	\>$\{$	\>\+\+\+converse, factors and distributivity\-\-$~~~ \}$\pop\\
	$R{\MPcomp}I{\MPcomp}R\ms{1}{\cup}\ms{1}R{\MPcomp}R{\MPcomp}R\ms{3}{\subseteq}\ms{3}R$\push\-\\
	$=$	\>	\>$\{$	\>\+\+\+supremum and monotonicity\-\-$~~~ \}$\pop\\
	$R{\MPcomp}R\ms{2}{\subseteq}\ms{2}R$\push\-\\
	$=$	\>	\>$\{$	\>\+\+\+assumption\-\-$~~~ \}$\pop\\
	$\mathsf{true}~~.$
\end{mpdisplay}
\vspace{-7mm}
\MPendBox

\begin{Example}\label{less.than.staircase}{\rm \ \ \ The less-than relations on the integers, ${<}_{\MPInt}$, on the rationals,  ${<}_{\MPRational}$,
and on the reals, ${<}_{\MPReal}$,  
are all staircase relations since in each case $<{\setminus}<$  is the at-most relation, $\leq$.    
See example \ref{diagonal.lessthan} for details of the preorder  in each case.  
The less-than relation on the integers is a (linearly) block-ordered relation but the
less-than relation on the rationals and the less-than relation on the reals are not block-ordered.  This is
because, as shown in example \ref{diagonal.lessthan},  the less-than relations on the rationals and on the
reals both have empty diagonals.
}%
\MPendBox\end{Example}

That the less-than relation on the real
numbers is not block-ordered is a consequence of the fact that if $x\ms{1}{<}\ms{1}y$ the interval between $x$ and $y$ can
always be subdivided at will.  (That is, it is always possible to find a real number $z$ such that $x\ms{1}{<}\ms{1}z$ and 
$z\ms{1}{<}\ms{1}y$.)  The same is also true of the rationals.   Abstracting from the details of the less-than
relation, we get the following theorem.  (Winter \cite{Winter2004} proves a similar theorem. See section 
\ref{Diagonal:Conclusion} for further discussion.)
\begin{Theorem}\label{dense.SC}{\rm \ \ \ Suppose $R$ is a homogeneous relation such that \begin{displaymath}R\ms{1}{\neq}\ms{1}{\MPplatbottom}\ms{6}{\wedge}\ms{6}I{\cap}R\ms{1}{=}\ms{1}{\MPplatbottom}\ms{6}{\wedge}\ms{6}R\ms{1}{=}\ms{1}R{\MPcomp}R\ms{6}{\wedge}\ms{6}R\ms{1}{\cup}\ms{1}I\ms{1}{\cup}\ms{1}R^{\MPrev}\ms{2}{=}\ms{2}{\MPplattop}~~.\end{displaymath}Then $R$ is a staircase relation and $\Delta{}R\ms{1}{=}\ms{1}{\MPplatbottom}$.

It follows that any such relation is \emph{not} block-ordered.
}
\end{Theorem}
{\bf Proof}~~~Lemma \ref{BD.strictord} proves that $R$ is a staircase relation. 

Comparing the above conditions on $R$ with those in  lemma \ref{BD.strictord}, the additions are the 
non-emptiness property $R\ms{1}{\neq}\ms{1}{\MPplatbottom}$,   the ``strictness''  property   $I{\cap}R\ms{1}{=}\ms{1}{\MPplatbottom}$  and the  ``subdivision'' property
$R\ms{2}{\subseteq}\ms{2}R{\MPcomp}R$.    (The less-than relation on real numbers has the subdivision  property whereas the less-than
relation on the integers does not.) Applying lemma \ref{dense.diag} (below), the subdivision and strictness 
properties imply that  $\Delta{}R\ms{1}{=}\ms{1}{\MPplatbottom}$. That $R$ is not block-ordered follows from theorem \ref{BD.diagdom} and the 
assumption that $R\ms{1}{\neq}\ms{1}{\MPplatbottom}$.
%\vspace{-7mm}
\MPendBox

The lemma used to prove theorem \ref{dense.SC} is the following:
\begin{Lemma}\label{dense.diag}{\rm \ \ \ \begin{displaymath}R\ms{1}{\subseteq}\ms{1}R{\MPcomp}R\ms{2}{\Rightarrow}\ms{2}(\Delta{}R\ms{1}{=}\ms{1}{\MPplatbottom}\ms{2}{\equiv}\ms{2}I{\cap}R\ms{1}{\subseteq}\ms{1}{\MPplatbottom})~~.\end{displaymath}
}%
\end{Lemma}%
{\bf Proof}~~~
\begin{mpdisplay}{0.15em}{6.5mm}{0mm}{2}
	$R\ms{3}{\subseteq}\ms{3}R\ms{1}{\MPcomp}\ms{1}{\neg}R^{\MPrev}\ms{1}{\MPcomp}\ms{1}R$\push\-\\
	$\Rightarrow$	\>	\>$\{$	\>\+\+\+monotonicity\-\-$~~~ \}$\pop\\
	$I{\cap}R\ms{4}{\subseteq}\ms{4}I\ms{3}{\cap}\ms{3}R\ms{1}{\MPcomp}\ms{1}{\neg}R^{\MPrev}\ms{1}{\MPcomp}\ms{1}R$\push\-\\
	$\Rightarrow$	\>	\>$\{$	\>\+\+\+modular law\-\-$~~~ \}$\pop\\
	$I{\cap}R\ms{3}{\subseteq}\ms{3}R{\MPcomp}(R^{\MPrev}\ms{1}{\MPcomp}\ms{1}R^{\MPrev}\ms{2}{\cap}\ms{2}{\neg}R^{\MPrev}){\MPcomp}R$\push\-\\
	$=$	\>	\>$\{$	\>\+\+\+assumption:  $R\ms{1}{\subseteq}\ms{1}R{\MPcomp}R$\-\-$~~~ \}$\pop\\
	$I{\cap}R\ms{3}{\subseteq}\ms{3}R{\MPcomp}(R^{\MPrev}\ms{2}{\cap}\ms{2}{\neg}R^{\MPrev}){\MPcomp}R$\push\-\\
	$=$	\>	\>$\{$	\>\+\+\+complements\-\-$~~~ \}$\pop\\
	$I{\cap}R\ms{1}{\subseteq}\ms{1}{\MPplatbottom}$\push\-\\
	$=$	\>	\>$\{$	\>\+\+\+$I\ms{1}{=}\ms{1}I^{\MPrev}$,  converse and shunting\-\-$~~~ \}$\pop\\
	$I\ms{2}{\subseteq}\ms{2}{\neg}R^{\MPrev}$\push\-\\
	$\Rightarrow$	\>	\>$\{$	\>\+\+\+monotonicity\-\-$~~~ \}$\pop\\
	$R{\MPcomp}R\ms{3}{\subseteq}\ms{3}R\ms{1}{\MPcomp}\ms{1}{\neg}R^{\MPrev}\ms{1}{\MPcomp}\ms{1}R$\push\-\\
	$\Rightarrow$	\>	\>$\{$	\>\+\+\+assumption:  $R\ms{1}{\subseteq}\ms{1}R{\MPcomp}R$ and transitivity\-\-$~~~ \}$\pop\\
	$R\ms{3}{\subseteq}\ms{3}R\ms{1}{\MPcomp}\ms{1}{\neg}R^{\MPrev}\ms{1}{\MPcomp}\ms{1}R~~.$
\end{mpdisplay}
That is, \begin{equation}\label{dense.IandR}
R\ms{1}{\subseteq}\ms{1}R{\MPcomp}R\ms{4}{\Rightarrow}\ms{4}(R\ms{3}{\subseteq}\ms{3}R\ms{1}{\MPcomp}\ms{1}{\neg}R^{\MPrev}\ms{1}{\MPcomp}\ms{1}R\ms{4}{\equiv}\ms{4}I{\cap}R\ms{1}{\subseteq}\ms{1}{\MPplatbottom})~~.
\end{equation}So
\begin{mpdisplay}{0.15em}{6.5mm}{0mm}{2}
	$\Delta{}R\ms{1}{=}\ms{1}{\MPplatbottom}$\push\-\\
	$=$	\>	\>$\{$	\>\+\+\+$\left[\ms{1}{\MPplatbottom}\ms{1}{\subseteq}\ms{1}X\ms{1}\right]$ and antisymmetry, definition of $\Delta{}R$\-\-$~~~ \}$\pop\\
	$R\ms{2}{\cap}\ms{2}(R{\setminus}R{/}R)^{\MPrev}\ms{2}{\subseteq}\ms{2}{\MPplatbottom}$\push\-\\
	$=$	\>	\>$\{$	\>\+\+\+shunting \-\-$~~~ \}$\pop\\
	$R\ms{2}{\subseteq}\ms{2}{\neg}(R{\setminus}R{/}R)^{\MPrev}$\push\-\\
	$=$	\>	\>$\{$	\>\+\+\+(\ref{underover.to.not})\-\-$~~~ \}$\pop\\
	$R\ms{3}{\subseteq}\ms{3}R\ms{1}{\MPcomp}\ms{1}{\neg}R^{\MPrev}\ms{1}{\MPcomp}\ms{1}R$\push\-\\
	$=$	\>	\>$\{$	\>\+\+\+assumption:  $R\ms{1}{\subseteq}\ms{1}R{\MPcomp}R$,  (\ref{dense.IandR}) \-\-$~~~ \}$\pop\\
	$I{\cap}R\ms{1}{\subseteq}\ms{1}{\MPplatbottom}~~.$
\end{mpdisplay}
\vspace{-7mm}
\MPendBox

\section{Conclusion}\label{Diagonal:Conclusion}

The primary novel contribution of this paper is the combination of theorems \ref{diagonal.core}  and
\ref{diagonal.core.ordering}: essentially, a block-ordered relation is a relation whose core is a provisional
ordering.   The discovery of these theorems was inspired by Riguet's suggestion of an
 ``analogie frappante'' linking the notion of a ``relation de Ferrers'' and the (difunctional)  ``diff\a'{e}rence''
of a relation.  Theorem \ref{BD.diagdom} is a precise statement of an ``analogie'' linking the notions of a
block-ordered relation and  the diagonal of a relation.   

A secondary, but nevertheless important,  contribution of this paper is  our (almost) exclusive use of the 
properties of factors of a relation, particularly with respect to formulating and reasoning about the
notion of the diagonal of a relation (Riguet's ``diff\a'{e}rence''), as opposed to Riguet's use of nested
complements.    Indeed, our only use of complements is in section \ref{sec:Staircase Relations} where we
formulated the notion of a staircase relation ---in terms of factors--- and showed its equivalence to
Riguet's notion of a ``relation de Ferrers'' ---which he formulated in terms of nested complements--- . 

Our motivation for including section \ref{sec:Staircase Relations} is partly to give proper credit to Riguet's
contribution but also to rectify misleading/incorrect statements 
 in the extant literature\footnote{See  \cite{RCB2020}, in
particular the concluding section,  for references to the relevant literature.}.  Specifically, the claim that
a ``relation de Ferrers \ldots {} can be rewritten in staircase block form''  \cite[Definition 4.4.11]{SchmidtStrohlein} 
is, at best, confused: as shown in example \ref{less.than.staircase},  the less-than relation on real numbers is a
staircase relation but not block-ordered.  The lesson to be learnt is, in our view, that mental pictures, such 
as the one of a staircase relation shown in fig.\  \ref{fig:staircase} and informal natural-language statements,  
can never be relied on.  Ultimately,  it is vital that
informal notions are  formalised and the properties of the formal notions are explored in detail in order
to confirm that they do indeed conform to their intended meaning.

\paragraph{Acknowledgement}\label{difun:Acknowledgements}

Many thanks to Jules Desharnais for helping to locate Riguet's publications.

%\newpage

\bibliographystyle{alpha}
\bibliography{bibliogr}

\end{document}